\documentclass[a4paper,11pt]{article}
\pdfoutput=1 

\usepackage{jheppub} 

\usepackage[T1]{fontenc} 
\usepackage[font=footnotesize,labelfont=bf]{caption}
\usepackage{enumitem}
\usepackage{graphicx}
\usepackage{subcaption}
\usepackage{bm}
\usepackage{indentfirst}
\usepackage{amsmath}
\usepackage{dirtytalk}
\usepackage{ulem}
\usepackage{cancel}
\def\spose#1{\hbox to 0pt{#1\hss}}
\def\lta{\mathrel{\spose{\lower 3pt\hbox{$\mathchar"218$}}
		\raise 2.0pt\hbox{$\mathchar"13C$}}}
\def\gta{\mathrel{\spose{\lower 3pt\hbox{$\mathchar"218$}}
		\raise 2.0pt\hbox{$\mathchar"13E$}}}

\newcommand{\de}[2]{\kern - #1 em \mathrm{d} #2}

\def\Oc{{\mathcal O}}
\def\Cc{{\mathcal C}}
\def\Tc{{\mathcal T}}
\def\Pc{{\mathcal P}}

\def\Hc{{\mathcal H}}
\def\Lc{{\mathcal L}}
\def\Nc{{\mathcal N}}
\newcommand{\LF}{\left(}
\newcommand{\RF}{\right)}
\newcommand{\LT}{\left[}
\newcommand{\RT}{\right]}
\newcommand{\pd}{\partial}

\title{\boldmath Finding origins of CMB anomalies in the inflationary quantum fluctuations}

\author[a,b,c]{Enrique Gazta\~naga}
\author[a]{K. Sravan Kumar}

\affiliation[a]{Insitute of Cosmology and Gravitation, University of Portsmouth,
	Dennis Sciama Building, Burnaby Road,
	Portsmouth, PO1 3FX, United Kingdom}
\affiliation[b]
{Institute of Space Sciences (ICE, CSIC), 08193 Barcelona, Spain}
\affiliation[c]
{Institut d\'~Estudis Espacials de Catalunya (IEEC), 08034 Barcelona, Spain}

\emailAdd{enrique.gaztanaga@port.ac.uk}
\emailAdd{sravan.kumar@port.ac.uk}

\abstract{
In this paper, we present compelling evidence for the parity asymmetry (a discrete symmetry separate from isotropy) in the Cosmic Microwave Background (CMB) map, measured through two-point temperature correlations. 
Any asymmetry associated with discrete symmetries, such as parity,  challenges our understanding of quantum physics associated with primordial physics rather than LCDM ($\Lambda$ Cold-Dark-Matter) itself. We commence by conducting a comprehensive analysis of the Planck CMB, focusing on the distribution of power in low-multipoles and temperature anticorrelations at parity conjugate points in position space. We find tension with the near scale-invariant power-law power spectrum of Standard Inflation (SI), with p-values of the order $\Oc\LF 10^{-4}-10^{-3} \RF$. Alternatively, we explore the framework of direct-sum inflation (DSI), where a quantum fluctuation arises as a direct sum of two components evolving forward and backward in time at parity conjugate points in physical space. This mechanism results in a parity-asymmetric scale-dependent power spectrum, particularly prominent at low-multipoles, without any additional free model parameters. Our findings indicate that DSI is consistent with data on parity asymmetry, the absence of power at $\theta>60^{\circ}$, and power suppression at low-even-multipoles which are major data anomalies in the SI model.
Furthermore, we discover that the parameters characterizing the hemispherical power asymmetry anomaly become statistically insignificant when the large SI quadrupole amplitude is reduced to align with the data. DSI explains this low quadrupole with a p-value of 3.5\%, 39 times higher than SI. Combining statistics from parameters measuring parity and low-$\ell$ angular power spectrum, we find that DSI is 50-650 times more probable than SI. In summary, our investigation suggests that while CMB temperature fluctuations exhibit homogeneity and isotropy, they also display parity-asymmetric behavior consistent with predictions of DSI. This observation provides a tantalizing evidence for the quantum mechanical nature of gravity.
 }

\makeatother
\makeatletter
\gdef\@fpheader{}
\makeatother

\begin{document} 
	\maketitle
	\flushbottom

	\section{Introduction}

 The 20th century witnessed the emergence of two profoundly successful theories, General Relativity (GR) and Quantum Mechanics (QM), each excelling in explaining the macroscopic and microscopic realms, respectively. Additionally, the Standard Model (SM) of particle physics stands as a significant achievement, resulting from the successful integration of Quantum Mechanics with special relativity in the form of Quantum Field Theory (QFT).
  The discrete symmetries such as charge conjugation ($\Cc$), Parity ($\Pc$), and time-reversal ($\Tc$) played a vital role in understanding and testing SM of particle physics. For example, the Parity violation observed as asymmetric angular distribution of electron emission ($\beta-$ decay) in Cobalt-60 \cite{Lee:1956qn,Wu:1957my,Christenson:1964fg}, 
	the $\Cc\Pc$-violation in the weak interactions and the $\Cc\Pc\Tc$ invariance of scattering amplitudes \cite{Coleman:2018mew}. 
 Moreover, $\Cc\Pc$-violation has been identified as a potential explanation for the observed particle-antiparticle asymmetry in the Universe. This observation strongly suggests the necessity of extending beyond the SM of particle physics \cite{Sakharov:1967dj,Canetti:2012zc,Kaufman:2014rpa}. It's important to note that discrete operations like $\Cc,\Pc,\Tc$ are intricately linked to quantum rather than classical physics.
Gravity stands out as the weakest among all fundamental forces in nature. The exploration of the status of discrete symmetries in quantum gravity\footnote{Throughout the paper, our definition of quantum gravity pertains to classical spacetime with quantum mechanical fluctuations (i.e., linearized quantum gravity) or quantum fields in curved spacetime, applicable to physics far below the Planck scale.} or for quantum fields in curved spacetime naturally arises as a fundamental question. A definitive answer to this inquiry is pivotal for constructing a comprehensive theory of quantum gravity that remains valid up to Planck scales and beyond.

Parity is an inherent characteristic of elementary particles and antiparticles. The Parity transformation $\Pc$, combined with charge conjugation $\Cc$—both being unitary operations \cite{Coleman:2018mew}—transforms a particle into its corresponding antiparticle while flipping the sign of the momenta. In contrast, the time reversal operation is anti-unitary, counteracting the effects of $\Pc$ by reversing the sign of physical momenta.
In the realm of quantum gravity, particularly when addressing the early Universe's cosmology, such as the cosmic microwave background (CMB), the evolution of spacetime geometry transitions from quantum to classical. In this context, the concept of Parity asymmetry suggests differences in physics at Parity conjugate points. Classical GR portrays spacetime as a dynamical entity, and understanding time reversal for quantum fields propagating in classical spacetime poses a non-trivial challenge.

This paper delves into the foundational understanding of parity and time reversal in quantum gravity, revealing compelling observational evidence in the latest CMB data. Additionally, the paper rigorously reevaluates the status of CMB anomalies, specifically the so-called Hemispherical Power Asymmetry (HPA), often considered as a violation of the cosmological principle \cite{Land:2005ad,Hoftuft:2009rq,Mukherjee:2015mma,Akrami:2014eta,Jones:2023ncn}. The findings indicate no significant statistical evidence supporting the existence of HPA, thus challenging previous claims associated with cosmological principle violations.
	
	The paper is organized as follows. In Sec.~\ref{sec:visual-parity}, we present a visual picture of CMB's parity asymmetry, which has been consistently observed over the last 3 decades. We also show that a better way to characterize the parity asymmetry is through temperature anticorrelation maps of the CMB data. In Sec.~\ref{sec:priseed}, we discuss what could be the primordial (quantum) physics in the scope of inflationary cosmology that can lead to the parity asymmetric feature of CMB. We present an intuitive discussion on the idea of direct-sum inflation (DSI), where quantum fluctuations during inflation contain a parity asymmetric time evolution that leaves imprints in the CMB once they become classical. In Sec.~\ref{sec:SIvsDSI}, we study the probabilities of SI and DSI given the CMB data and vice versa using the observables that characterize only parity and not isotropy both in configuration space and also in the harmonic space. We record our results in terms of the p-values, which give the best description of the SI and DSI models in terms of fitting the data. In Sec.~\ref{sec:H60}, we establish a connection between lower power in the quadrupole and the observed lack of correlations at angular scales $\theta>60^\circ$. Then, we analyze the probability of DSI, explaining the lack of power on large scales compared to SI. In Sec.~\ref{sec:ruleoutHPA}, we examine the HPA and related observables with a new set of simulations that further test the statistical significance of HPA. Also, we assess if the lower quadrupole observed in the data can influence the significance of HPA. In Sec.~\ref{sec:theory}, we detail the SI and fundamental questions that lead to the DSI framework. We also calculate the parity-asymmetric angular power spectrum in DSI, which has been used to test the model with the Planck CMB data. We discuss the implications of DSI for stochastic inflation and vice versa. We also comment on how our observational tests of DSI can provide valuable input to the stochastic framework of inflationary quantum fluctuations and the issue of quantum-to-classical transition. We also elucidate how the DSI predictions for parity-asymmetric CMB remain stable concerning the tiny modifications in the coarse-graining scale of stochastic inflation. We end with Conclusions and outlook, followed by some appendices.
  
In Appendix.~\ref{sec:isotropy}, we present the spherical harmonic decomposition of CMB temperature fluctuation, define the angular power spectrum, and explain the distinguish parity with anisotropy. In Appendix~\ref{sec:mirror}, we show how we make the parity conjugate maps of CMB data with an illustration of the parity conjugate map of Earth. 
  In Appendix~\ref{sec:harmonicparity}, we explain and quantify the sampling variance related to the angular power spectra $C_{\ell}$ for both the data and the models. We explicitly demonstrate that the distribution of the sampling variance of $C_{\ell}$ is non-Gaussian, making the conversion of $\chi^2$ values to probabilities challenging. Additionally, we present the covariant matrix of $C_{\ell}$ for low $\ell$.
Appendix~\ref{sec:simulations} is dedicated to discussing our CMB simulations, noise considerations, and the impact of using masks on the data sets of the Planck satellite. In Appendix~\ref{sec:characterization}, we delve into the observables associated with the absence of correlations at large angular scales.
In Appendix~\ref{sec:visualcom}, we conduct a visual comparison of CMB maps with DSI and SI simulations. Furthermore, we present simulated maps for HPA to distinguish anisotropy and parity asymmetric CMB maps visually.

	\subsection{Conventions, Notations and Data analysis:}
 
 Throughout the paper, we follow the metric signature, which is mostly positive. We work in the units of $\hbar=c=1$ and $M_p^2=\frac{1}{\sqrt{8\pi G}}=1$.  Often in observations $\Tc(\hat{n})$  is given in units of Kelvin (K) or micro-Kelvin $\mu$K instead of units of $T_0$. In all the paper, we will therefore normalize  $\Tc(\hat{n})$  to  $\sigma_\Tc \equiv \sqrt{<\Tc(\hat{n})^2>}$ (where the average $<\dots >$ is overall sky directions) so that $\Tc(\hat{n})$ becomes a dimensionless scalar field of unit variance. This is because our study is not about the overall amplitude of fluctuations but only their symmetrical properties. In the whole paper, we only speak about point-parity, which is a discrete transformation $\textbf{x}\to -\textbf{x}$ or $\LF \theta,\, \varphi \RF\to \LF \pi-\theta,\, \pi+\varphi\RF$ in spherical coordinates. Whenever we say CMB angular power spectrum, it only means in the context of two-point temperature correlations. In this paper, we will use the data maps, masks, and best-fit LCDM model provided by Planck 2018 data release.\footnote{
		Available in the  Planck Archive webpage \url{ https://pla.esac.esa.int/\#maps}
		and described in the Planck Legacy Archive wiki \url{https://wiki.cosmos.esa.int/planck-legacy-archive/index.php/Simulation_data}.}
	We focus on the component-separated {\sc Commander} (COMM$_m$ from now on), {\sc SMICA} \footnote{{Named \sc COM-CMB-IQU-smica-2048-R3.00-full} .}
	(SMIC$_m$), {\sc SEVEM} (SEVE$_m$) and {\sc NILC} (NILC$_m$).
	The subscript '$m$' in the name indicates that we use the mask corresponding to each component separation, which is the default.
	The fraction of unmasked sky is: $f_{sky}=88.8, 84.2, 83.8$ and $78.6\%$ respectively for $N_{side}=2048$.
	The subscript '$c$' indicates that we use the common mask	\footnote{{\sc COM-Mask-CMB-common-Mask-Int-2048-R3.00}} which covers $77.9\%$ of the sky. We do not use the CMB measurement inside the mask as they are contaminated and have different noise properties. To study the parity asymmetry of the CMB map and visualizations of data and theoretical model realizations, we used HEALPix software\footnote{https://healpix.sourceforge.io/} with further details explained in Appendix.~\ref{sec:mirror}.
	
	\section{The parity asymmetric cosmic microwave background (CMB)}
	
	\label{sec:visual-parity}
	
	The observation of CMB from Cosmic Microwave Background Explorer (COBE) satellite in 1992 has revealed for the first time the blueprint of large-scale structure (LSS) of the Universe with temperature fluctuations of the order $\frac{\Delta T}{\bar{T}}\sim 10^{-5}$ with Gaussian distribution of two-point correlations revealing the large scale homogeneous and isotropic nature of primordial Universe \cite{COBEw2,Julien}. COBE Sattelite has mainly measured the CMB anisotropies at angular scales $\theta\gtrsim 6^\circ$, corresponding to the low-multipoles $\ell\lesssim 30$. The successors of COBE measurements in the later decades are carried by the Wilkinson Microwave Anisotropy Probe (WMAP) and Planck Satellites. They have confirmed the near-scale invariant spectrum of primordial fluctuations at small angular scales $\theta<1^\circ$ leaving the large-scale anomalies at low-multipoles \cite{Schwarz:2015cma}. 
	First of all, CMB sky is, on average, statistically homogeneous and isotropic, and the CMB temperature fluctuations being one part in $10^5$ is the first hint to speculate their origin must be spacetime fluctuations around the Friedman-Lema\^itre-Robertson-Walker (FLRW) Universe. 

\begin{figure}
		\centering
		\includegraphics[width=0.75\linewidth]{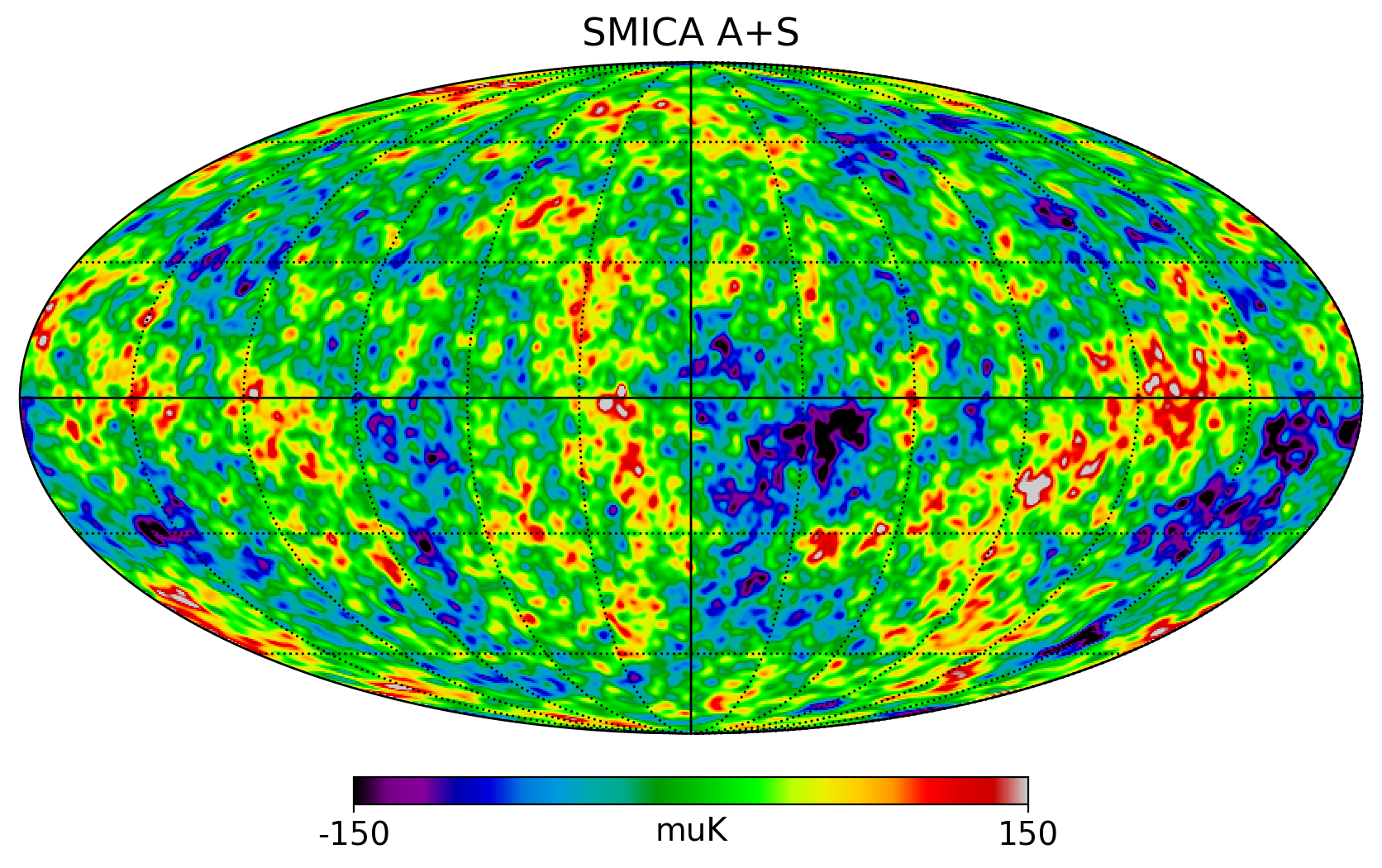}
		\vskip -1.cm
		\includegraphics[width=0.75\linewidth]{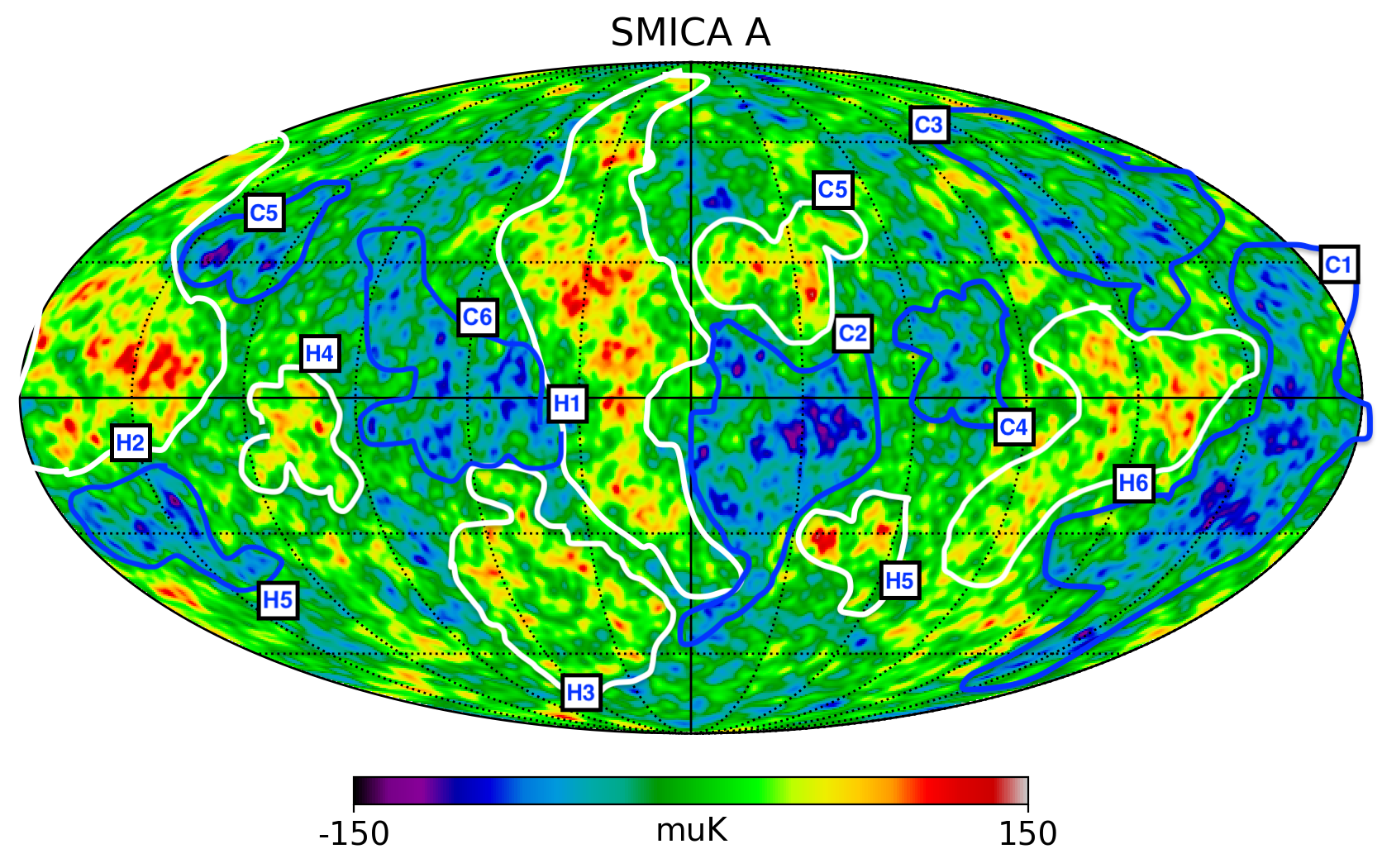}
		\vskip -1.cm
		\includegraphics[width=0.75\linewidth]{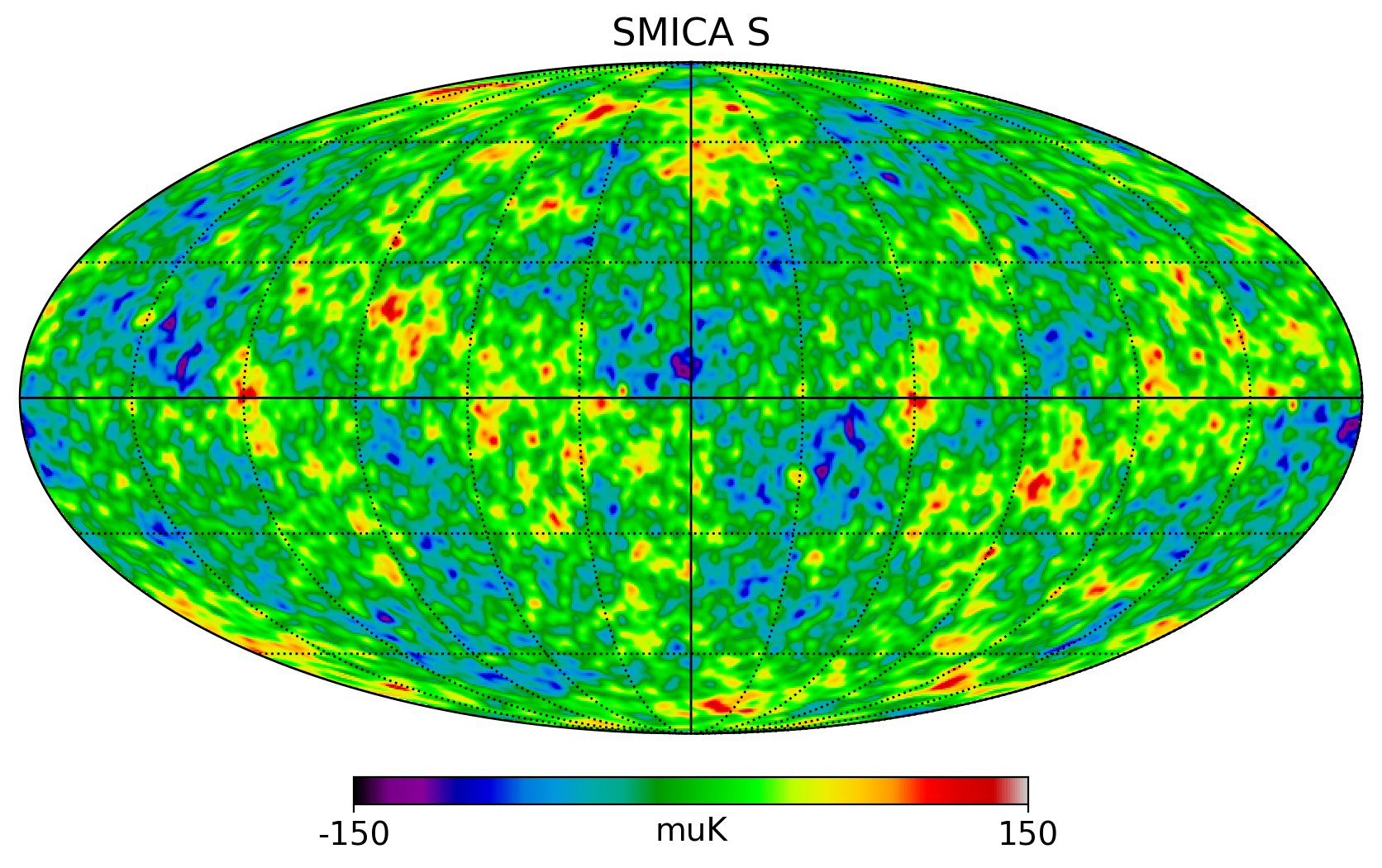}
		\caption{
			The Mollview projections above display the antisymmetric (A, middle) and symmetric (S, bottom) decomposition of the Planck 2018 map (SMICA, top). The mirror (odd/even parity) symmetry is readily apparent. Notably, large-scale cold structures crossing the galactic plane from the Southern Galactic Cap (SGC) to the North (labeled C1) have corresponding hot (antisymmetric) mirror structures (labeled H1) crossing the opposite side of the galactic plane from the Northern Galactic Cap (NGC) to the South. This symmetry holds for structures C2, C3, C4, and C5. These structures align closely with the original SMICA map (A+S) and are either entirely or partially absent in the $S$ even parity map, indicating that they are genuine odd parity components. In contrast, the very hot superstructure around H6, just above C1 in the original map, consists of both even and odd components, as expected for isotropic fluctuations without a well-defined parity. See also Fig.~\ref{fig:mirror} and Fig.~\ref{fig:Ctheta} for further illustration of parity odd preferred CMB.}
		\label{fig:smicaSA}
	\end{figure}

 In Appendix \ref{sec:isotropy}, we give a brief description of temperature fluctuations, isotropy, and parity.
	We can decompose the CMB temperature fluctuation  $\Tc(\hat{n})$
 as the sum of its symmetric (even parity)
	$S(\hat{n})$  and its antisymmetric (odd parity)
	$A(\hat{n})$ components:
	\begin{equation}
		\Tc(\hat{n}) = S(\hat{n}) + A(\hat{n})  
	\end{equation}
	where
	\begin{equation}
		\begin{aligned}
			& S(\hat{n}) \equiv   \frac{1}{2} \left[  \Tc(\hat{n}) +  \Tc(-\hat{n})   \right] = S(-\hat{n}) \\ &
			A(\hat{n}) \equiv   \frac{1}{2} \left[   \Tc(\hat{n}) -  \Tc(-\hat{n})  \right]   = -A(-\hat{n})
		\end{aligned}      
		\label{eq:parity}
	\end{equation}
	where $(-\hat{n})$ is the antipodal direction or parity $\Pc$ conjugate of $(\hat{n})$. From the SMICA map of Planck data, \cite{Planck:2019evm}, we create the projections of Symmetric (S) and antisymmetric (A) parts in Fig.~\ref{fig:smicaSA}.  
	Here, we can witness that the antisymmetric map (A) and the symmetric map (S) contain large-scale structures whose shapes are $\Pc$ conjugate images of each other. However, the remarkable revelation here is that the total map (A+S) appears very close to the antisymmetric map (A), meaning that the CMB is odd parity preferred than even \footnote{Which is indirectly known through the even-odd asymmetry in $C_\ell$ estimator $R_{TT}$ of Eq.~\ref{eq:RTT} which is measured to be different from unity  \cite{Muir:2018hjv,Schwarz:2015cma}. In Fig.~\ref{fig:smicaSA}, we present the position space representation of what it means CMB prefers odd parity. } This is exactly what we call in the rest of the paper parity asymmetry that we show that can emerge from quantum fluctuations during early Universe inflationary expansion.\footnote{It is worth noting that the global (geometrical) parity asymmetry considered here has nothing to do with parity-violating (i.e., pseudo scalar $\Pc \phi_{\rm pseudo} = -\phi_{\rm pseudo}$) terms in the GR Lagrangian which can result in 
 additional CMB polarizations, their cross-correlations, cosmic birefringence, and any higher-order temperature (chiral) correlations that were widely addressed in the literature  \cite{Lue:1998mq,Bartolo:2017szm,Minami:2020odp,Philcox:2023ffy}. To be more precise, in this paper, we are speaking about the inflationary (scalar) metric fluctuations differing in magnitude at the parity conjugate regions of physical space. 
 Note that a scalar field $ \Phi$ generally obeys $ \Phi (\textbf{x}) \neq  \pm\Phi(- \textbf{x})$. In our language, parity asymmetry means a correlation between the scalar field values at the antipodal points in physical space.
 Furthermore, our theoretical framework (DSI) would lead to parity asymmetry for tensor power spectrum (B-modes), which we defer for future investigation.} The odd parity favored CMB map in Fig.~\ref{fig:smicaSA} implies that the temperature fluctuation of the CMB contains an additional purely antisymmetric component 
	\begin{equation}
		\Tc\LF \hat{n} \RF = \tilde{\Tc}\LF \hat{n}\RF  + \Delta\Tc\LF \hat{n} \RF
		\label{eq:pureantisym}
	\end{equation}
	where $\bar{\Tc}(\hat{n})$ is neither symmetric nor antisymmetric. 
	This component $\Delta \Tc\LF \hat{n} \RF$ satisfies the purely antisymmetric property 
	\begin{equation}
		\Delta\Tc\LF \hat{n} \RF = -\Delta \Tc \LF -\hat{n}\RF\,. 
		\label{eq:propPA}
	\end{equation}
	It is this, additional component that makes the CMB map (A+S) look more like A than S as we see in Fig.~\ref{fig:smicaSA}. If $\Delta\Tc\LF \hat{n} \RF=0$, then there is no defined parity; we expect even and odd multipoles to be similar. However, this is not what we see in the data; the CMB is strongly aligned with the odd parity. 
	We carry out further analysis of this in the upcoming sections. 

	\subsection{Characterizing parity asymmetry with CMB temperature correlations}
		
	Temperature fluctuations $\Tc(\hat{n})$ in the sky correspond to local anisotropies, but we can still test if they are statistically consistent with isotropy. 
	Isotropic fluctuations are those for which $ \vert a_{\ell m}\vert^2$ in Eq.~\ref{eq:cl}
	do not depend strongly on $m$. This is a good pose question for a Gaussian field (which will be our assumption throughout this paper) as we can test the null hypothesis of whether a given direction is statistically consistent with any other one.  
	
	We can also define a similar measure in configuration space:
	\begin{equation}
		w[\hat{\theta}]   \equiv < \Tc(\hat{n}_1) \Tc(\hat{n}_2) >  
		\label{eq:C_theta_v}
	\end{equation}
	where $\hat{\theta}\equiv \hat{n}_2-\hat{n}_1$ and the average is over all the pairs in the map.
	Isotropic fluctuations are characterized by the property that this quantity does not exhibit significant dependence on the direction. In other words, we can average over all directions, collapsing them into a single angular distance $\theta = \vert \hat{\theta} \vert$, as demonstrated in Eq.~\ref{eq:cl} for $C_\ell$. 
	We then have: 
	$w[\theta]=<w[\hat{\theta}]>$, where the average is over all directions in $\hat{\theta}$. This results in:
	\begin{equation}
		w[\theta]  = \sum_\ell^{\ell_{max}} \frac{2\ell+1}{4\pi} C_\ell \, P_\ell[\cos{\theta}]
		\label{eq:C_theta}
	\end{equation}
	where $\cos{\theta}=\hat{n}_1 \cdot \hat{n}_2$.  Parity asymmetry comes from the anticorrelation at $\theta=180^\circ$:
	\begin{equation}
 w(180^\circ) =		<Z> =  < \Tc(\hat{n}) \Tc(-\hat{n}) > = \sum_\ell^{\ell_{max}} \frac{2\ell+1}{4\pi} \Big[ C_{\ell=even}-C_{\ell=odd}\Big]
	\end{equation}
	The $Z(\bar{n})$ map defined as $Z(\bar{n})=\Tc(\hat{n}) \Tc(-\hat{n})$ is an accurate measure of parity in all directions (regardless of isotropy). We will study here the Z-parity statistics, a significant anomaly in the Planck data away from the near scale-invariant primordial spectra within the $\Lambda$ (cosmological constant) Cold Dark Matter (CDM) model. Its mean value is:

 \begin{equation}
     Z^1  \equiv <Z(\bar{n})> = \sum_{i=1}^{i=12 N_{side}^2} P(Z_i) Z_i
     \label{eq:z1}
 \end{equation}
where $P(Z_i)$ is the distribution (or histogram) of $Z_i=Z(\bar{n}_i)$ values in the map pixels $\bar{n}_i$. We will also study the skewness: 
\begin{equation}
   Z^3 \equiv <Z^3>  = \sum_{i=1}^{i=12 N_{side}^2} P(Z_i) (Z_i-<Z>)^3
   \label{eq:skewness}
\end{equation}
 of the $P(Z)$ distribution.

	
	$Z$-parity variable is a more direct quantity to characterize the parity of the maps in comparison with the harmonic space variable $R^{TT}$ Eq.~\ref{eq:RTT}. 
	The Z-variable was also utilized in \cite{Creswell:2021eqi}, but with a key distinction—we center our analysis on temperatures normalized to unit variance. Such a normalized approach enables us to delve into parity asymmetry independently of pixel variance and map resolution.
	We also apply a 4-sigma clipping to mitigate the influence of rare values or artifacts. Such discrepancies are often evident in various Planck component separation maps around the Galactic plane and foreground sources. By implementing this clipping, our goal is to measure the parity of the entire distribution rather than focusing on the parity of rare or most extreme events.

\begin{figure}
		\centering
		\includegraphics[width=1.\linewidth]{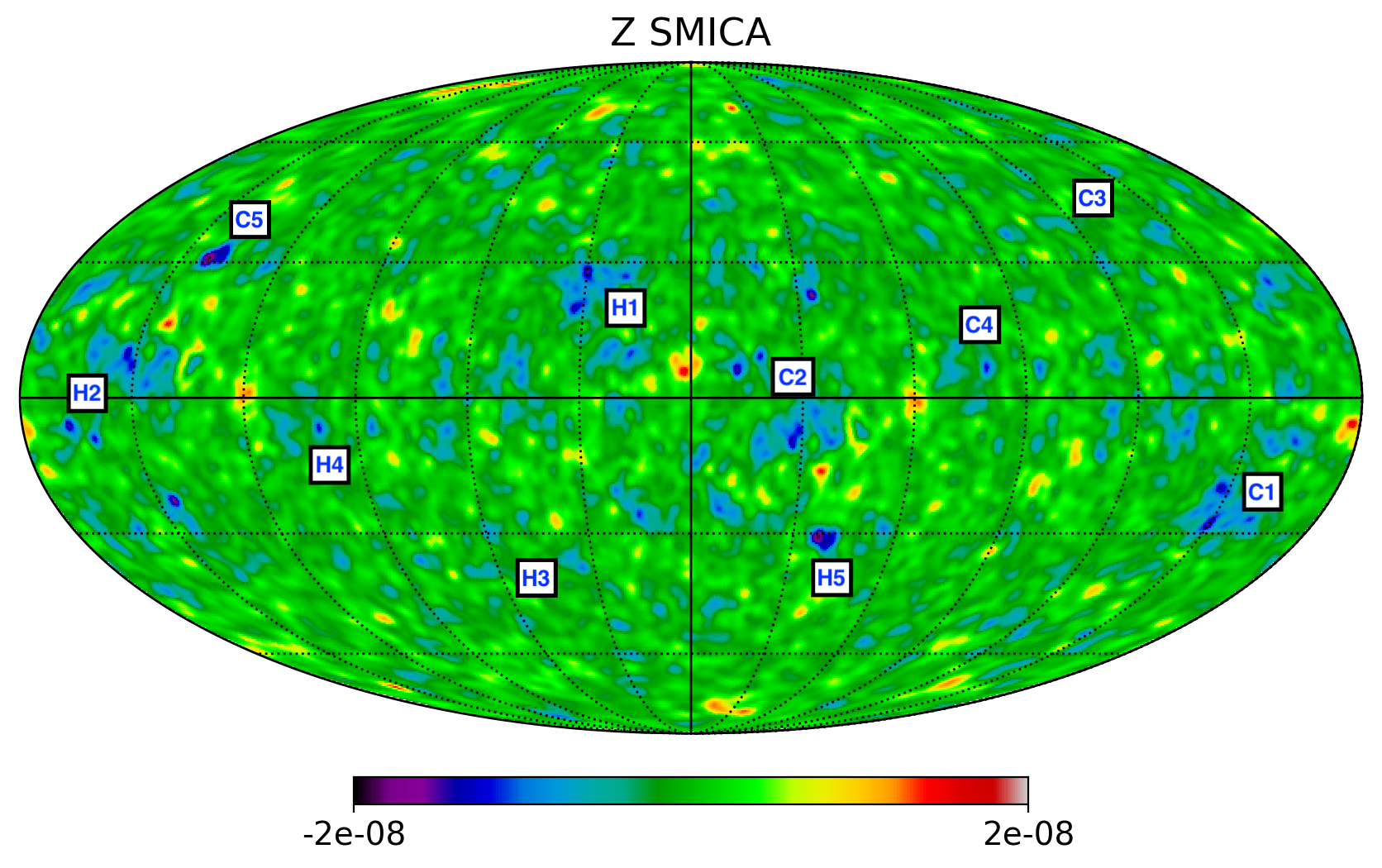}
		\caption{Un-normalised (units of Kelvin$^2$) parity product: $Z(\hat{n})=\Tc(\hat{n})\Tc(-\hat{n})$ for Planck (SMICA) map. This map is also a $\Pc$ conjugate image of itself. Positive values (red) correspond to the symmetric (even parity) component of $\Tc$, and negative values (blue) to the antisymmetric (odd parity) component. The $Z(\hat{n})$ map shows some extreme antisymmetric (i.e., blue) values that we already identified in the middle panel of Fig.~\ref{fig:smicaSA} (we use the same labels here). But if we focus on values around the mean (i.e., cyan-blue and yellow-orange), there is an excess of negative $Z$ values, indicating a preference for odd parity. We will quantify this here by comparing the 1-point $P(Z)$ distributions in data and simulations.} 
		\label{fig:Zsmica}
	\end{figure}

	It's interesting to note the close connection between these two parity measurements, even when one is conducted in harmonic space and the other in configuration space. 
	Harmonic space offers the advantage that different multipoles have uncorrelated errors in the ideal scenario of a full sky and Gaussian statistics. On the other hand, configuration space provides the benefit of easily accommodating any masking shape. Therefore, both measurements complement each other. But note that $Z(\hat{n})$ captures a wealth of information beyond just the mean of $P(Z)$, as depicted in Fig.~\ref{fig:Zsmica} as a full sky map. For a Gaussian field, the parity information lies in the shape of the 1-point $P(Z)$, as higher N-point correlations are all given in terms of the temperature $C_\ell$.

	\subsection{Even-odd discrete symmetry in the harmonic space}
	\label{sec:parity}
	
	Parity is a discrete symmetry separate from rotational invariance. To be very precise, parity transformation cannot be achieved by any continuous rotations. In other words, parity is not an operation that is part of the group $SO(3)$. 
	The temperature fluctuations of parity asymmetric CMB can be split into even-odd components of spherical harmonics as
	\begin{equation}
		\Tc \LF  \hat{n} \RF  =  \sum a_{\ell m} Y_{\ell m }\LF \hat{n} \RF 
		= S(\hat{n}) + A(\hat{n}) =
		\sum \LF a^{S}_{\ell m} + a^{A}_{\ell m} \RF Y_{\ell m }\LF \hat{n}   \RF
	\end{equation}
 From Eq.~\ref{eq:propPA} and  
	since  $Y_{\ell m}(-\hat{n}) = (-1)^\ell Y_{\ell m}(\hat{n})$ we immediately see that:

\begin{equation}
		\begin{aligned}
				a_{\ell m}= \Bigg\{
				\begin{matrix}
					\Tilde{a}_{\ell m}-\Delta a_{\ell m} \equiv a_{\ell m}^{S} \quad \rm{for}  \quad \ell=\rm{even} \\ \Tilde{a}_{\ell m}+\Delta a_{\ell m} \equiv a_{\ell m}^{A}   \quad  \rm{for}  \quad 
					\ell=\rm{odd} 
				\end{matrix}
		\end{aligned}
		\label{eq:Tparity}
	\end{equation}
 
	 Note how parity symmetry happens in the $\ell$ variable, and it is therefore separate from the local isotropic properties, which relate to the $m$ variable or the direction of a given multipole $\ell$ (see Fig.\ref{fig:alm}). So we can see here how parity and isotropy can and should be tested separately in data.
	
	In Fig.~\ref{fig:smicaSA}, the parity decomposition in the Planck 2018 maps (SMICA component separation) is illustrated. We can achieve the same parity split in harmonic space using Eq.~\ref{eq:Tparity} or in configuration space using Eq.~\ref{eq:parity}, where $ \Tc(-\hat{n})$  can be obtained from a parity transformation of $\Tc(\hat{n})$  (see Appendix \S\ref{sec:mirror}). Both approaches are mathematically equivalent. It is worth noting that all components $ \Tc(\hat{n})$, $ S(\hat{n}) $, and $ A(\hat{n}) $ can be perfectly isotropic and still can have a well-defined parity (a)symmetry. Parity is an additional discrete symmetry on top of isotropy, applying equally to all directions, and cannot be achieved by rotation.

 To measure the asymmetry in the distribution of power ($C_{\ell}$) in the even-odd $\ell$ we use the following quantity 
 \begin{equation}
		R^{TT} = \frac{D_{+}(\ell_{max})}{D_{-}(\ell_{max})} = \frac{\sum_{\ell=\rm{even}}^{\ell_{max}} \ell(\ell+1) C_\ell}
		{\sum_{\ell=\rm{odd}}^{\ell_{max}} \ell(\ell+1) C_\ell} .
		\label{eq:RTT}
	\end{equation}
 One issue with the $R^{TT}$ is that it depends strongly both on $\ell_{max}$ and on the values of the quadrupole and octopole \cite{Planck:2019evm,Schwarz:2015cma,Muir:2018hjv}, which could be affected by a break in the shape of the primordial power spectrum.

	\section{Primordial (quantum) seeds for parity asymmetric CMB}
	
	\label{sec:priseed}
	
	Cosmic inflation \cite{Starobinsky:1980te,Guth:1980zm,Linde:1981mu} is the most prominent paradigm consistent with near scale-invariant and Gaussian nature of CMB temperature fluctuations. From COBE to the latest Planck observations \cite{Planck:2018jri} the angular power spectrum $C_{\ell}$ \cite{Durrer:2020fza}
	\begin{equation}
		C_{\ell}  = \frac{2}{9\pi} \int \frac{dk}{k} \Pc_\zeta\LF k \RF j_{\ell}^2\LF k/k_s \RF
		\label{Cllstan}
	\end{equation}
	at $\ell>200$ or $\theta<1^\circ$ is found to be consistent with the power-law form of the primordial power spectrum (in convolution with the adequate thermal LCDM transfer function)
	\begin{equation}
		\Pc_\zeta  = A_s\LF \frac{k}{k_\ast} \RF^{n_s-1} 
		\label{HZpower}
	\end{equation}
with $A_s = 2.2\times 10^{-9}$ is called the primordial power spectrum amplitude, and the scalar spectral index $n_s = 0.9634\pm 0.0048$ (Planck TT+TE+EE) at $k_\ast = 0.05 \,\text{Mpc}^{-1}$ from the Planck data \cite{Planck:2018jri} and $k_s = \frac{1}{\tau-\tau_{LS}}= 7\times 10^{-5}\,\text{Mpc}^{-1}$ is given by the distance $(\tau-\tau_{LS})$ to the CMB surface of last scattering \cite{Durrer:2020fza}. The power spectrum Eq.~\ref{HZpower} has been understood as the outcome of inflationary quantum fluctuations generated around near exponential (quasi de Sitter) expansion of the Universe, which we call in the rest of the paper as Standard Inflation (SI) \cite{Mukhanov:1981xt}, which we review in greater detail SI in Sec.~\ref{sec:theory}.  Inflationary cosmology involves a non-perturbative modification of GR by introducing at least one new additional massive scalar (often called "inflaton") degree of freedom responsible for near de Sitter cosmic expansion followed by an inflaton-matter-dominated phase that results in particle production by reheating after the end of inflation. The success of the inflation is that its prediction of a slight departure from scale invariance is quantified by $n_s\lesssim 1$. Even though SI is largely successful, several fundamental questions do loom around regarding the nature of inflationary quantum fluctuations and their imprints in the CMB, which we discuss in detail in Sec.~\ref{sec:theory}. On the observational aspect as well, SI with Eq.~\ref{HZpower} is not consistent with large-scale features of CMB, more prominently at angular scales $\theta>6^\circ$ or $\ell< 30$) \cite{Schwarz:2015cma,Hansen:2008ym}. This was known even from the days of WMAP, and numerous phenomenological and theoretical attempts to explain it have been made either by modifying the inflationary potential or by proposing empirical modifications of SI power spectrum or by Planck-scale quantum gravity inspired (phenomenological) models Eq.~\ref{HZpower} (See for example \cite{Sinha:2005mn,Contaldi:2003zv,Iqbal:2015tta,Pedro:2013pba,Agullo:2020cvg} and the many papers that were followed). 

\begin{figure}
		\centering
		\includegraphics[width=0.7\linewidth]{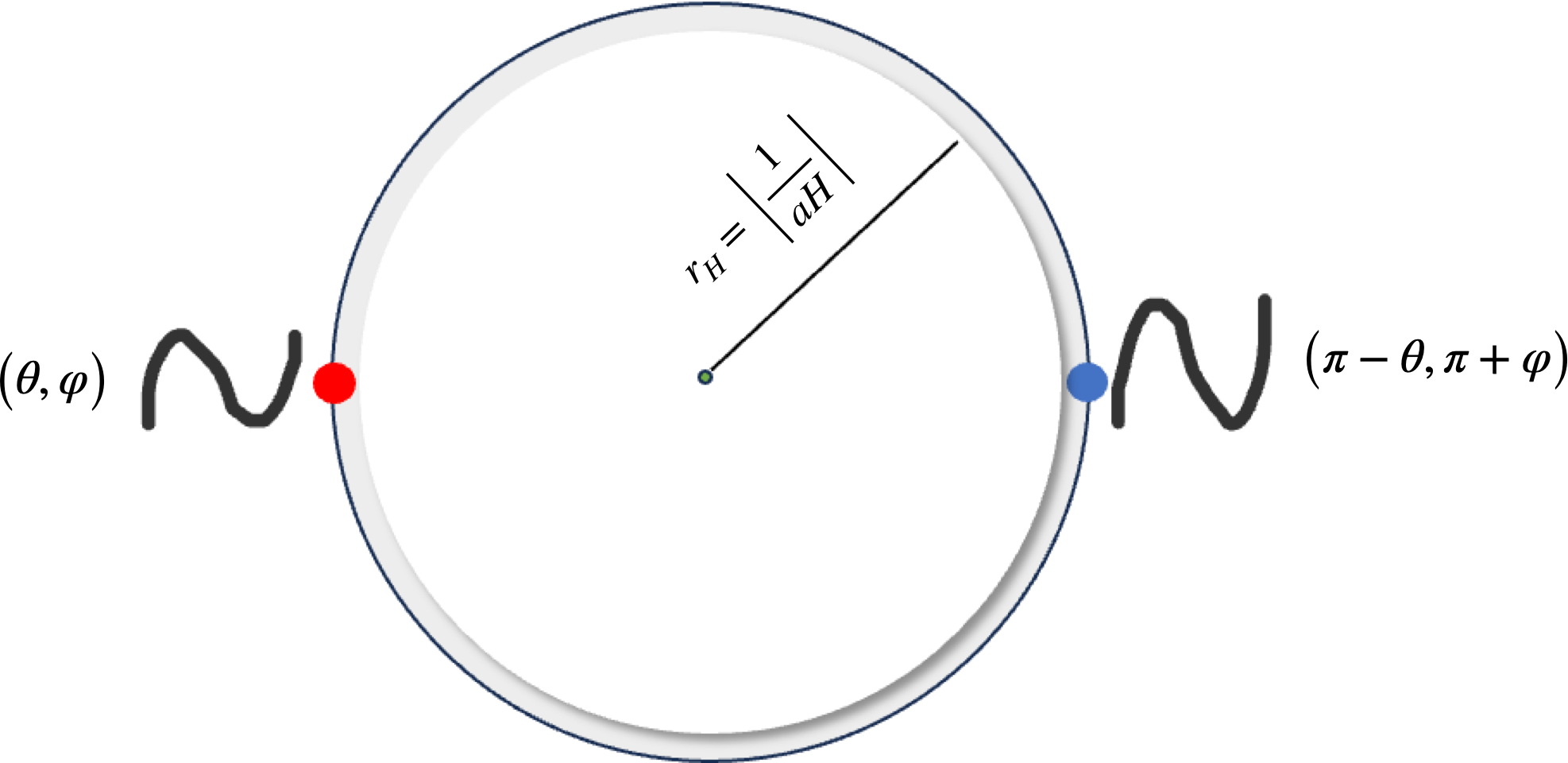}\quad \includegraphics[width=0.7\linewidth]{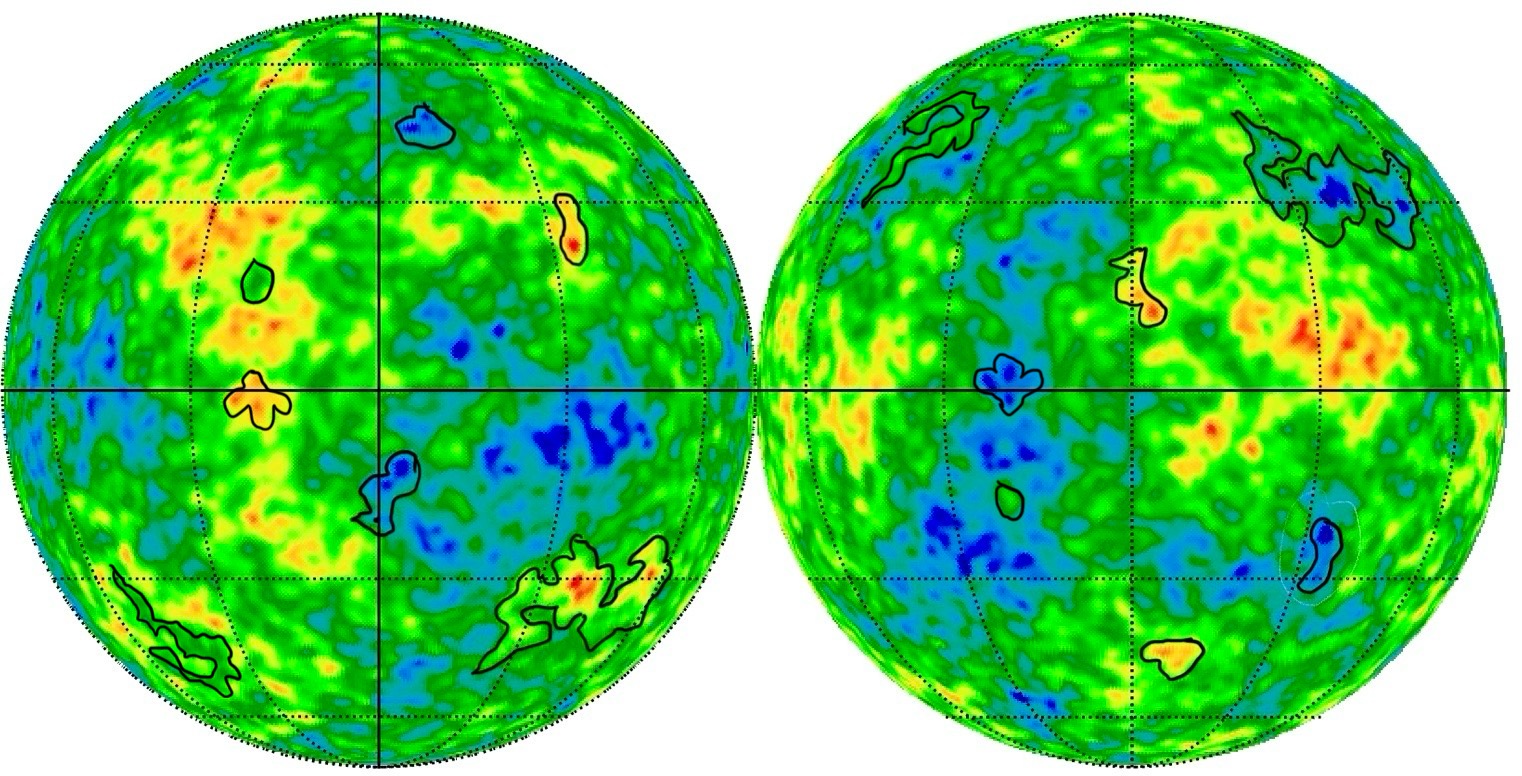}
		\caption{The top panel portrays the co-moving horizon during a particular moment of inflationary expansion, showcasing the evolution of a quantum fluctuation in DSI. This quantum fluctuation moves forward in time at $\LF \pi-\theta, \pi+\varphi \RF$ and backward at $\LF \theta,\varphi \RF$, leading to an odd parity excess due to the breaking of time-reversal symmetry by the background inflationary spacetime. In the bottom panel, you get a glimpse of the spherical distribution of the odd component in one such quantum fluctuation, displayed in two spherical caps centered at the red and blue points. The image is sourced from the odd component of the Planck 2018 (SMICA) map, which dominates over the even one (see also Fig.~\ref{fig:Ctheta} and Fig.~\ref{fig:smicaParity1}). We can schematically see how quantum fluctuations in DSI can leave their imprints in the CMB once they exit the horizon and become super-horizon fluctuations. 
		}
		\label{fig:mirror}
	\end{figure}

Inflationary expansion can be seen as an adiabatic departure from de Sitter spacetime \cite{Starobinsky:1980te} which can be seen as a spontaneous breaking of $\Pc\Tc$ symmetry.\footnote{Note that de Sitter spacetime is $\Pc\Tc$ symmetric, for example in the de Sitter metric in flat FLRW coordinates is $ds^2 = \frac{1}{\tau^2H^2}\LF -d\tau^2+d\textbf{x}^2 \RF$ with $H$ being the Hubble parameter and $\tau$ being the conformal time is invariant under $\tau\to -\tau$ and $\textbf{x}\to -\textbf{x}$ (see Sec.~\ref{sec:theory} for more detailed explanations).} An interesting question that can be asked here is what could be the signature of $\Pc\Tc$ symmetry breaking that inflationary quantum fluctuations generate which can be observed in the CMB. In the context of the standard treatment of inflationary quantum fluctuations \cite{Mukhanov:1990me}, the concepts of $\Pc$ and $\Tc$ do not play any role. Thus, standard inflation (SI) cannot address the parity asymmetry of the CMB map. A recently proposed direct-sum quantum field theory in curved spacetime (DQFT-CS) \cite{Kumar:2022zff,Kumar:2023ctp,Kumar:2023hbj} introduces the idea of inflationary quantum fluctuation as a direct-sum of a component evolving forward in time and another component that evolve backward in time at parity-conjugate points of physical space. Notably, the \say{time} (which is a parameter in quantum theory) is treated in this framework separately from the classical background inflationary metric, which gives a possibility for the quantum field to have both components having time opposite (quantum) evolution. This framework of treating inflationary quantum fluctuations is what we call Direct-Sum Inflation (DSI). In Fig.~\ref{fig:mirror}, we depict how the scheme of parity asymmetric nature of quantum fluctuation in DSI can resonate with what the CMB data conveys. 
	
A detailed explanation of the DSI theory is provided in Sec.~\ref{sec:theory}. The primary and noteworthy new result presented in this paper is the even-odd asymmetric angular power spectrum: 
	\begin{equation}
		\begin{aligned}
			&   C_{\ell}^{odd}  =  \frac{2}{9\pi} \int_0^{k_c} \frac{dk}{k} j_{\ell}^2\LF \frac{k}{k_s} \RF
			\Pc_\zeta(k) \LF 1+\Delta \Pc_v \RF 
			\\
			& C_{\ell}^{even}  = \frac{2}{9\pi}\int_0^{k_c} \frac{dk}{k} j_{\ell}^2\LF \frac{k}{k_s} \RF\Pc_\zeta(k) \LF 1-\Delta \Pc_v \RF  
		\end{aligned}
  \label{eq:even-oddcl}
	\end{equation}
It's essential to emphasize that this outcome introduces no new free parameters (with respect to the SI model in Eq.~\ref{HZpower}) and simply emerges from:
\begin{equation}
		\Delta \Pc_v= \LF 1-n_s \RF\operatorname{Re}\LT \frac{2}{H_{3/2}^{(1)} \LF  \frac{k}{k_\ast} \RF} \frac{\pd H^{(1)}_{\nu_s}\LF \frac{k}{k_\ast} \RF}{\pd\nu_s} \Bigg\vert_{\nu_s=\frac{3}{2}} \RT
		\label{eq:delpPR}
	\end{equation}	
Here, $H^{(1)}_{\nu_s}\LF z \RF, j_{\ell}\LF z \RF$ is the Hankel and  Bessel functions of the first kind and $k_c = 0.02 k_\ast$ is the cut-scale we impose as Eq.~\ref{eq:delpPR} is accurate enough for low-$\ell$ or large angular scales (See Sec.~\ref{sec:theory} for more details). 
In the next section, we estimate from Eq.~\ref{eq:even-oddcl} several observables related to parity asymmetric CMB temperature maps at low-$\ell$ and compare them with the predictions of SI.

	

	\section{Standard inflation versus Direct-Sum inflation}
\label{sec:SIvsDSI}

 In this section  we test two different theories of inflationary quantum fluctuations SI and DSI.  
 	We evaluate the significance of the observed parity asymmetry by comparing  statistics in the Planck maps with the corresponding distribution obtained from simulations of each model. The data and the simulated maps are processed in the same way.
 More details on the simulations and their validation are given in Appendix \ref{sec:simulations}.

	\begin{figure}
		\centering
		\includegraphics[width=.49\linewidth]{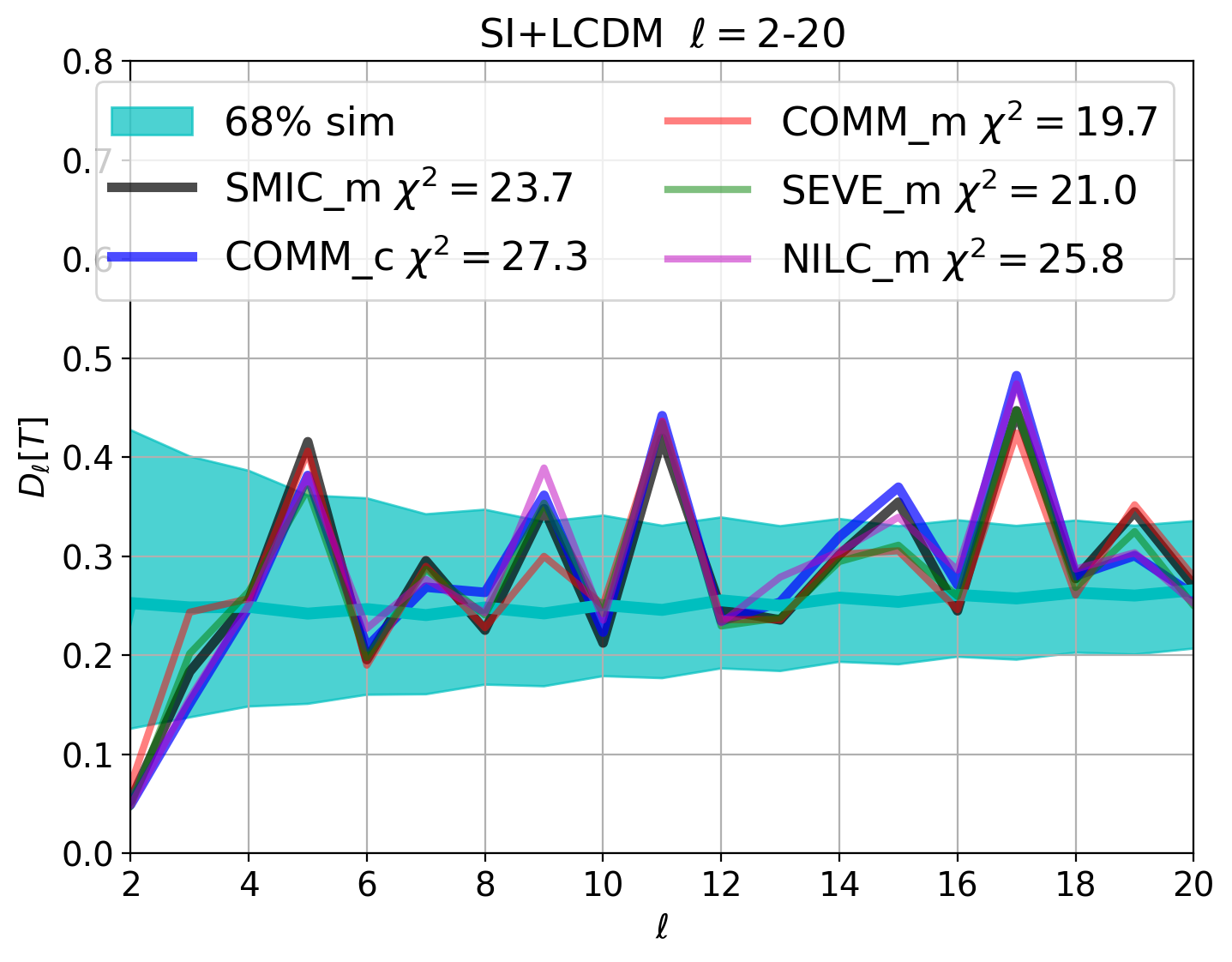} 
		\includegraphics[width=.49\linewidth]{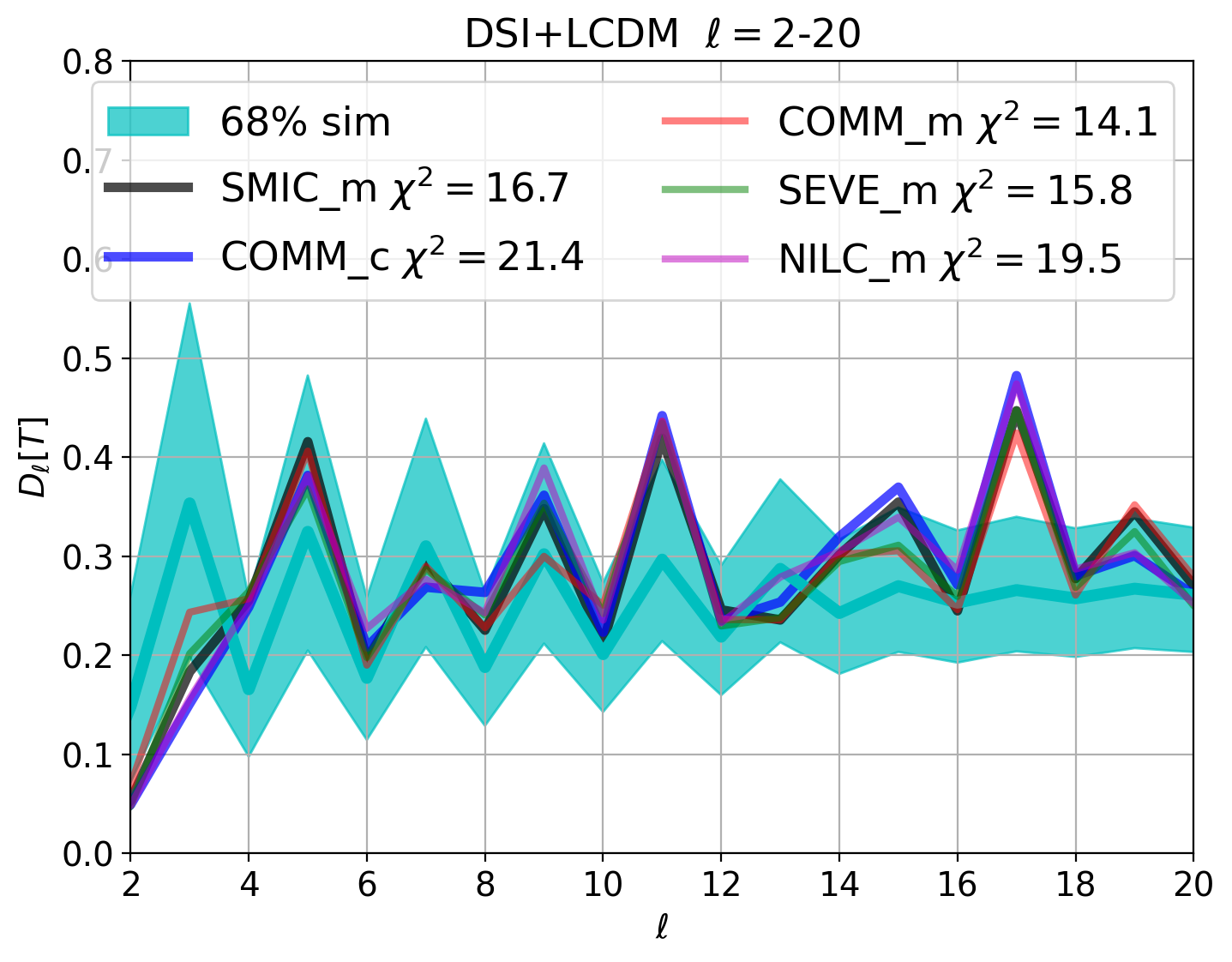} 
		\includegraphics[width=.49\linewidth]{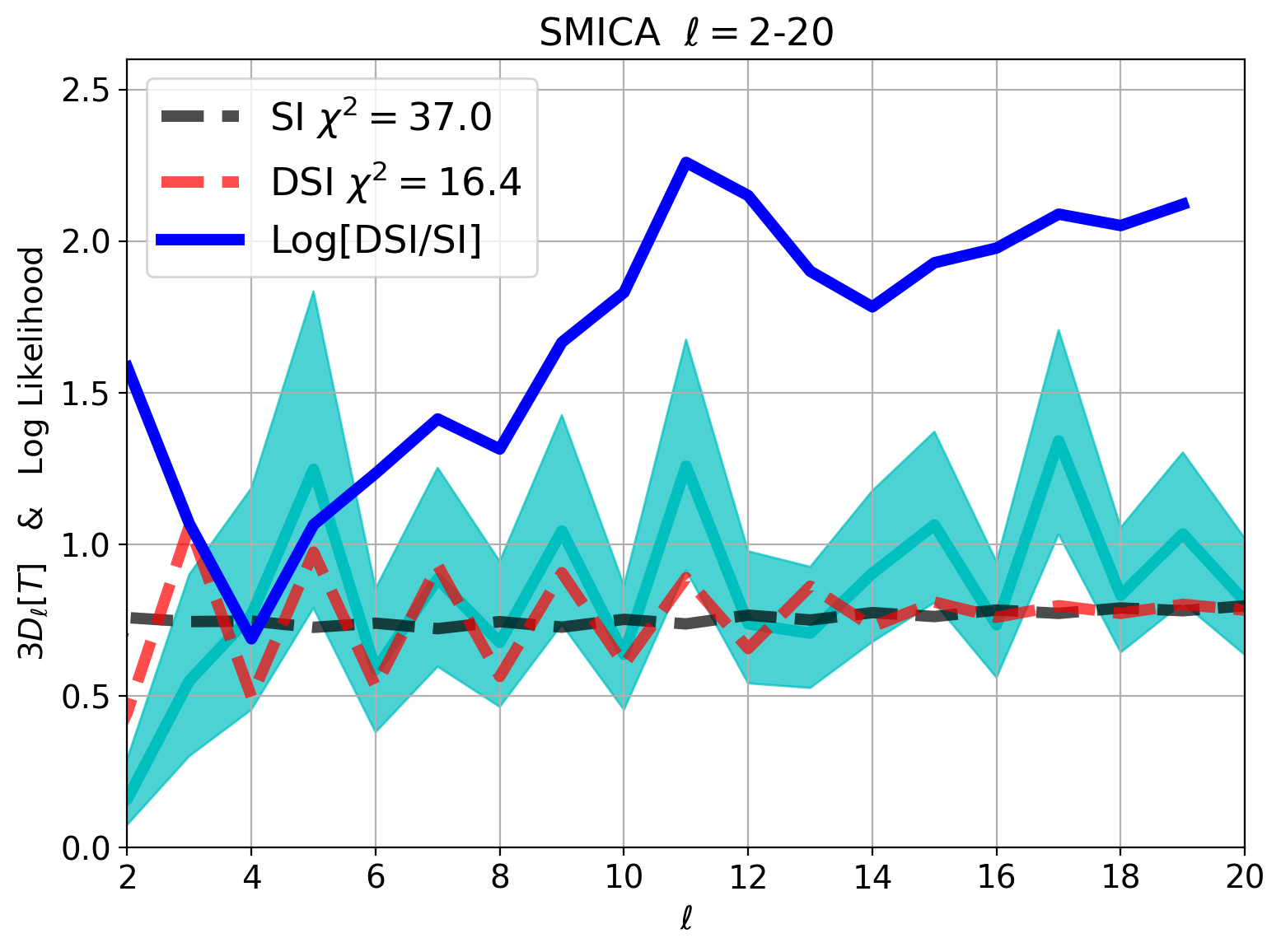}       
		\includegraphics[width=.49\linewidth]{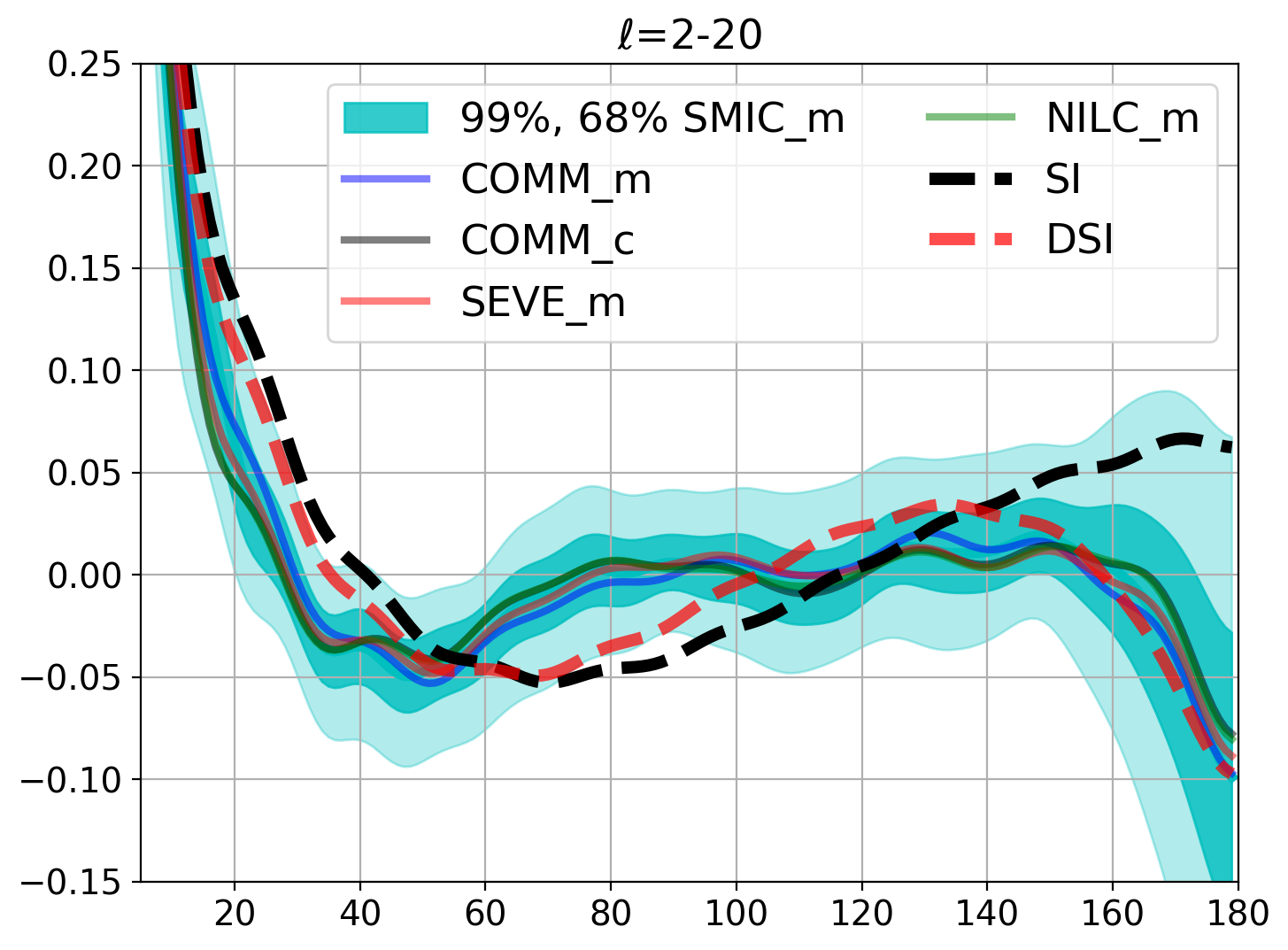} 
		
		\caption{Comparison of measured $C_\ell$ in several Planck (masked, smoothed and normalized) temperature maps (lines labeled SMIC$_m$, COMM$_m$,  COMM$_c$, SEVE$_m$ and NILC$_m$) against the corresponding $C_\ell$ in different simulations (shaded regions correspond to 68\% variations in the $10^6$ realizations). The top panels show simulations of SI (left) and DSI (right) LCDM models. The bottom left panel shows instead realizations of the observed Planck SMIC$_m$ map compared to the same models. The blue line shows the log of the ratio of the cumulative DSI to SI likelihoods according to SMICA$_m$  realizations:  DSI is up to 150 times more likely than SI. The bottom right panel compares the corresponding $w(\theta)$ with 68\% and 99\% errors in the  SMIC$_m$ data, which it is also aligned with odd symmetry (compare to Fig.~\ref{fig:Ctheta}).}
		\label{fig:clfit}
	\end{figure}
	
	Fig.~\ref{fig:clfit} compares the normalized $C_\ell$ in the SI and DSI models against the corresponding outputs in the different Planck maps.
	Given our focus on the largest possible scales, our analysis is dominated by sampling variance, rendering other sources of errors and uncertainties negligible.  Sampling variance errors are proportionate to the signal. Consequently, the errors depend on the assumptions we make regarding the model.  We explore two distinct approaches. The first involves calculating errors from different models and estimating $p[D|M]$, the probability of the data $D$ given the model $M$.  The top panels of Fig.~\ref{fig:clfit} are examples of this. According to the $\chi^2$ test, the SMIC$_m$ data in the range $\ell=2-20$ is closer to the DSI models (top right) than the SI model (top left), with a significance of $\Delta \chi^2 \simeq 7$. Similar results are obtained for $\ell=2-30$, as shown in the bottom panels of Fig.~\ref{fig:clLCDM}.
	
	In the second approach, we estimate errors in the data by using the data $C_\ell$ as input for simulations. In this case, we estimate the more common quantity: $p[M|D]$, the probability of the model given the data. The bottom left panel illustrates this second approach. The odd parity DSI model is also preferred, given the data, over standard LCDM (SI), but with a larger significance: $\Delta \chi^2=20.6$. This appears to indicate strong evidence for odd parity. However, as explained in Appendix~\ref{sec:harmonicparity}, it is not straightforward
	to convert such $\chi^2$ values into probabilities because, among other things, the $C_\ell$ distribution is not Gaussian for the lowest $\ell$.\footnote{For a Gaussian field $a_{\ell m}$ is also Gaussian, but $C_\ell$ in Eq,\ref{eq:cl} is not, as it is quadratic in $a_{\ell m}$.} This is illustrated in the left panel of Fig.~\ref{fig:CLpdf}.
	
	But we can directly estimate the likelihood ratio between DSI and SI using the data simulations, eliminating the need to presume a Gaussian distribution for $C_{\ell}$. We will still assume that the different $C_\ell$ values are approximately uncorrelated. This assumption seems reasonable, given that we have measured the covariance matrix and found that the Pearson cross-correlation coefficients are consistently smaller than 15\% (see right panel of Fig.~\ref{fig:CLpdf}) and have little impact ($<15\%$) on the likelihood ratio.
	
	In this scenario, the likelihood simplifies to the product of individual $C_\ell$ likelihoods. The blue line in the bottom left panel of Fig.~\ref{fig:clfit} shows the logarithm of the ratio of cumulative likelihoods for the two models. Examining individual multipoles, the quadrupole $C_2$ alone indicates that DSI is about 40 times more likely than SI. However, when including $C_3$ and $C_4$, this ratio decreases below 10. Subsequent multipoles then elevate the accumulated likelihood ratio to over 100 at $\ell \simeq 11$. Based solely on the $C_\ell$ measurements, we can confidently assert that DSI is approximately 150 times more likely than SI (using $\ell<20$ or $\ell<30$). This substantial difference will be subjected to further testing using alternative approaches that focus solely on the measurements of parity.

 The bottom right panel of Fig.~\ref{fig:clfit} shows 68\% and 99\% confidence regions in the 2-point correlation $w(\theta)$ of the SMIC$_m$ data realizations as compared to the DSI and SI mean model predictions using Eq.~\ref{eq:C_theta}. Even when there is a lot of covariance in these measurements, the data is at odds with SI data in a very significant way (e.g. see text around Eq.~\ref{eq:S12} and \cite{Camacho-Quevedo:2021bvt} and references therein). As we will show, odd parity is very related to this anomaly both because it predicts a low $C_2$, which has a large impact on $w(\theta)$ (see  Fig.~\ref{fig:Ctheta}), and also because of the strong observed negative correlation at $\theta=180^\circ$, connecting the antipodes at $\theta=180^\circ$ to $\theta=0^\circ$ with odd parity.
	
	\subsection{Results on parity asymmetry in SI versus DSI}
	\label{sec:parity}
	
	We next present two alternative tests to compare models to data. The first one is based on configuration space by doing a $\chi^2$ comparison of the full $P(Z)$ parity distribution. The second one shows a more direct comparison of the likelihood using extreme probabilities (p-values) of $R^{TT}$ in Eq.~\ref{eq:RTT} and the quadrupole $C_2$ (i.e. in harmonic space) as well as moments (mean $Z^1$ and skewness $Z^3$) of the $P(Z)$ distribution (in configuration space).
	
	\subsubsection{Z-parity statistics}

\begin{figure}
		\centering
		\includegraphics[width=.49\linewidth]{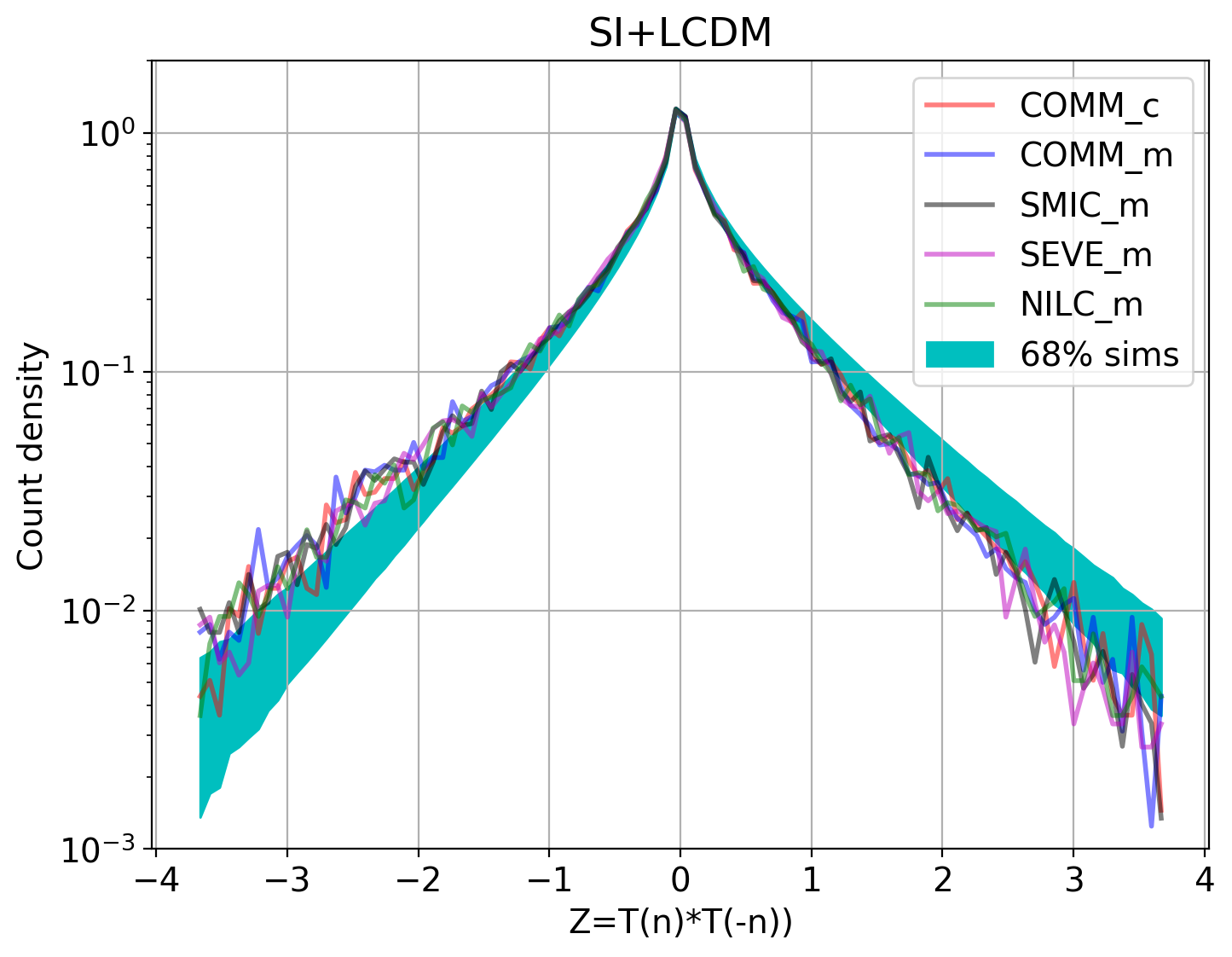}
		\includegraphics[width=.49\linewidth]{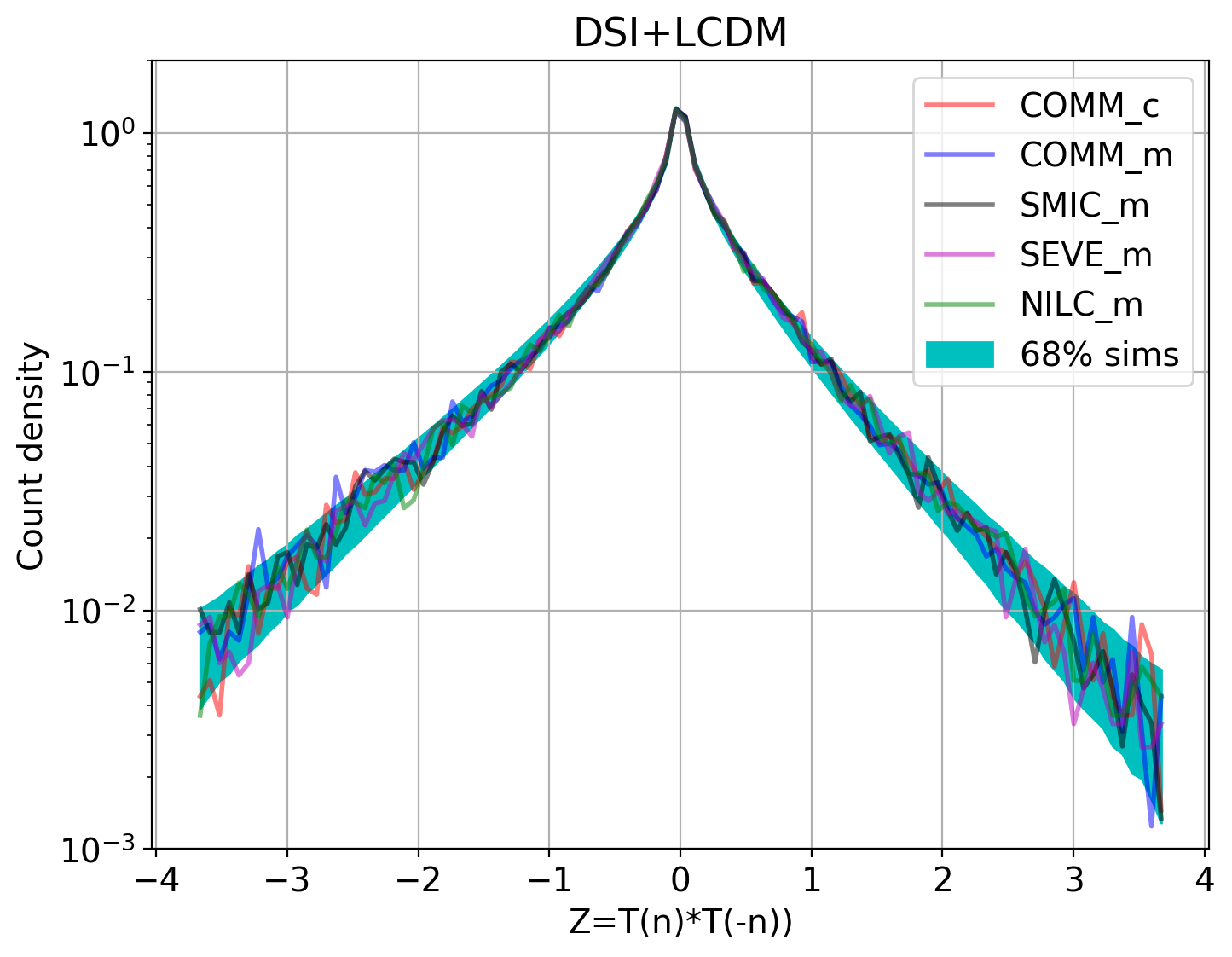} 
		\includegraphics[width=.49\linewidth]{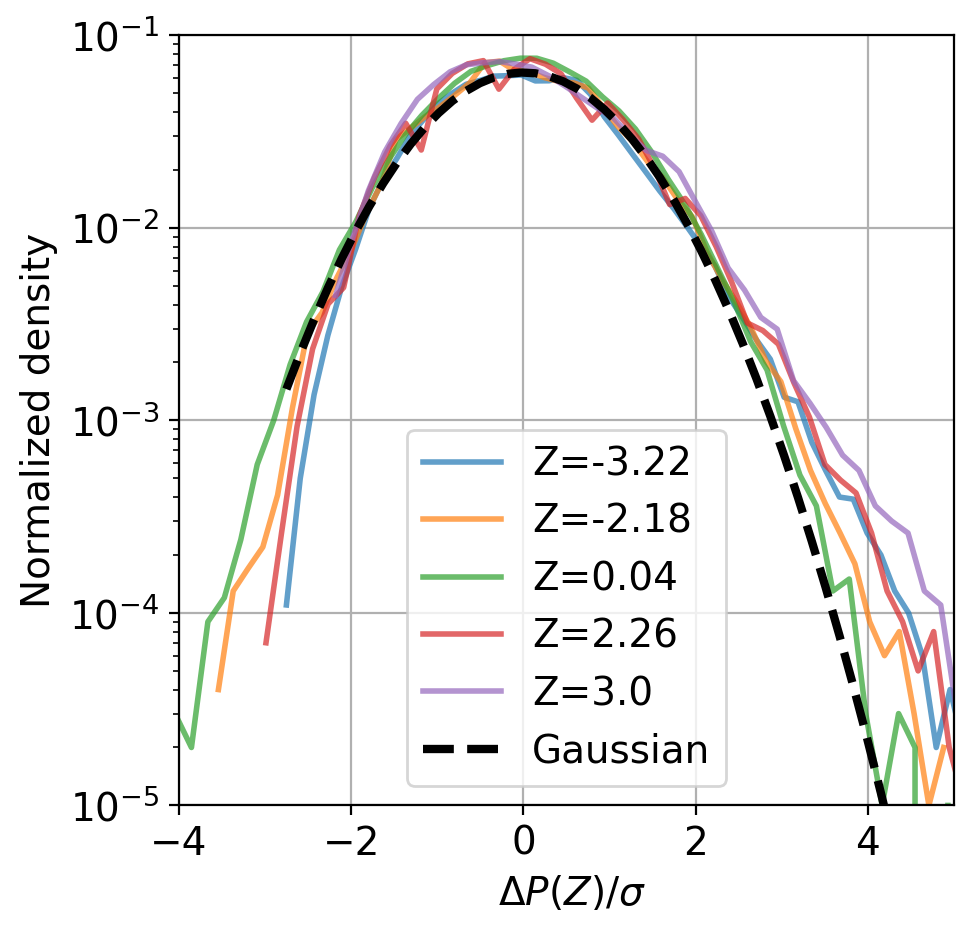} 
		\includegraphics[width=.49\linewidth]{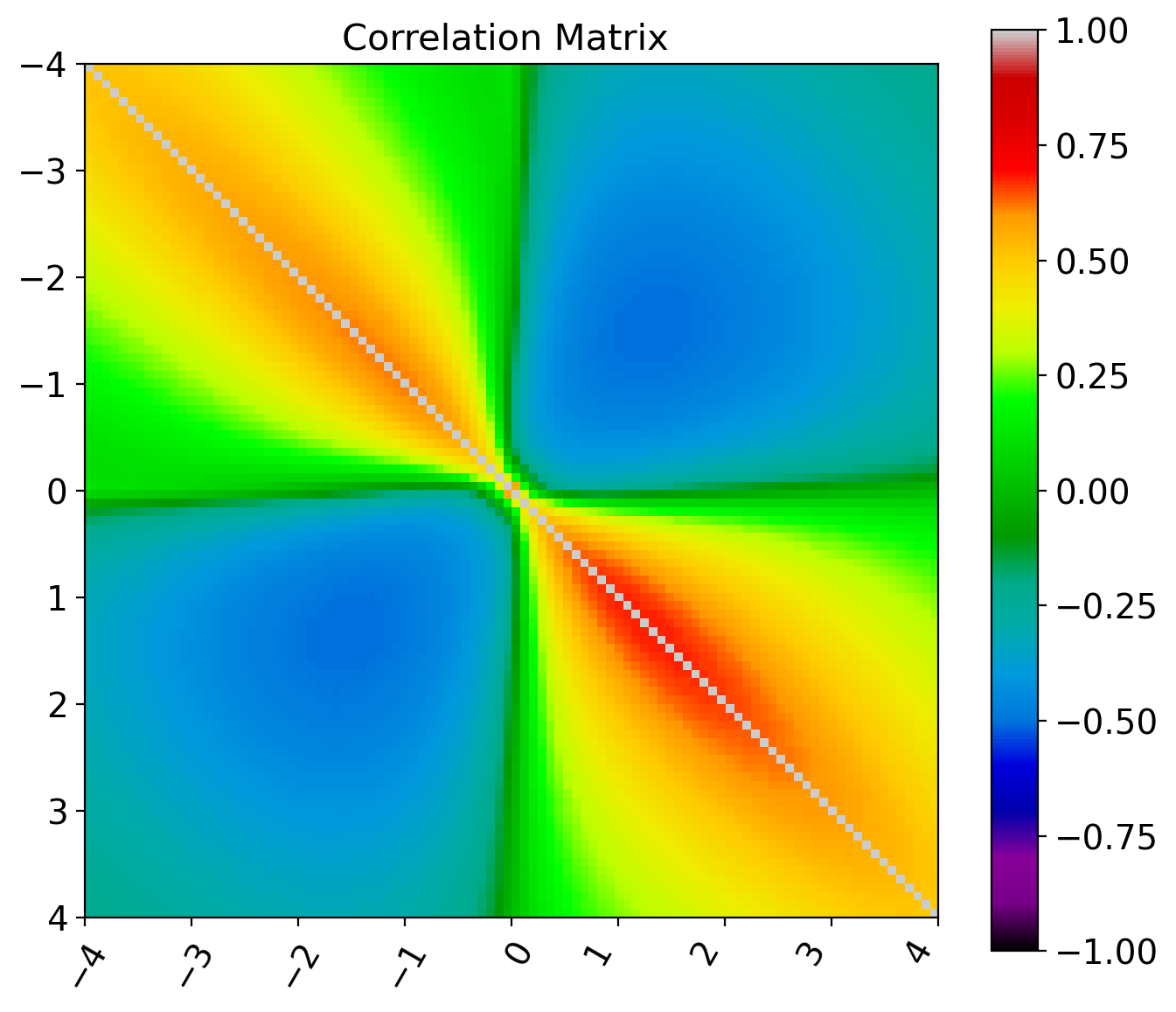}   	\caption{{\sc Top panels} show $P(Z)$ histogram of $Z(\hat{n}) \equiv \Tc(\hat{n}) \Tc(-\hat{n})$ CMB maps (for the different component separation and masking) as compared with the corresponding $P(Z)$ in simulations of the best-fit LCDM with SI (left) or  DSI (right). The shaded region corresponds to the 68\% interval around the median in the corresponding simulations. There is an excess of negative $Z$ values in comparison with the lack of positive ones in the SI predictions, which indicates a preference for odd parity, which is in agreement with DSI.  
			{\sc Bottom Left} shows the distribution of normalized count variations  $\hat{\Delta}_i$ in 
			Eq.~\ref{eq:Cij}. This distribution is not well approximated by a Gaussian (dashed line), making it hard to interpret the $\chi^2$ test.  {\sc Bottom Right} shows the covariance in $P(Z)$, which is very similar for the different models.
		}
		\label{fig:ZhistSMICA}
	\end{figure}

	Fig.~\ref{fig:Zsmica}
	shows a comparison of the $Z$ parity 1-point probability density distribution $P(Z)$, i.e. the probability for a given value of $Z$ to be observed,  in models and observations (see also $Z$ maps in Fig.~\ref{fig:ZhistSMICA} and Fig.~\ref{fig:smicaParity1}-\ref{fig:simParityDSI}).
	The figures indicate a preference for odd parity in the measured CMB data. This pattern holds across various component separations and masking conditions (as labeled in the Figure). Notably, there is an excess of negative $Z$ values and a deficiency of positive ones compared to SI, which inherently lacks a defined parity. The DSI model, characterized by 20\% excess in odd parity, appears to offer a much-improved fit to the data. Yet, how much improvement does it provide? 
	The considerable covariance $C_{ij}$ between bins in the $P(Z)$ histogram (as illustrated in the lower right panel of Fig.~\ref{fig:ZhistSMICA}) prompts the question of whether the observed differences are statistically significant. The $C_{ij}$ matrix shows 30-40\% anticorrelation between the antipodal fluctuations in $Z$ and some 20-70\% correlation within $Z$ fluctuations of the equal sign. This could, in principle, explain the observed parity asymmetry just as sampling variance fluctuations.

	To evaluate this, we perform a $\chi^2 = \Delta_i C_{ij}^{-1} \Delta_j$ test on the histogram $P_i=P(Z_i)$ presented in the top panels of Fig.~\ref{fig:ZhistSMICA}. Here, $\Delta_i \simeq [P^{data}_i-P^{model}_i]/\sigma_i$ represents the normalized differences between the model realizations and data at bin $i$, and $\sigma_i$ denotes the RMS scatter in the realizations. We employed $10^6$ realizations for each model to calculate the normalized covariance matrix as:
	\begin{equation}
		C_{ij} \equiv <\hat{\Delta}_i \hat{\Delta}_j> \quad ; \quad 
		\hat{\Delta}_i \equiv \frac{P_i^{sim}-<P_i^{sim}>}{\sigma_i} .
		\label{eq:Cij}
	\end{equation}
	In the expression above, $\hat{\Delta}_i$ represents the 
	fluctuations in the simulations. The right bottom panel of Fig.~\ref{fig:ZhistSMICA} displays $C_{ij}$ for the SI realizations, with the covariance for DSI being very similar.  The inversion of the covariance matrix $C_{ij}$ is performed using the (Moore-Penrose) pseudo-inverse based on Singular-Value Decomposition (SVD).
	
	We obtained $\chi^2=109$ for SI and $\chi^2=97$ for DSI, considering 97 independent degrees of freedom after SVD. The data exhibits good agreement with the odd parity level in the DSI simulations. The larger value ($\Delta \chi^2 \simeq 12$) for SI could in principle be interpreted as a $\sqrt{12} \sigma \simeq 3.5\sigma$ significant deviation. But as it happened for $C_\ell$, this result should be taken with some caution. The histogram of $\hat{\Delta}_i$ values deviates from a Gaussian distribution, especially noticeable for larger values of $Z$. This non-Gaussian behavior is explicitly depicted in the bottom left panel of Fig.~\ref{fig:ZhistSMICA} across various bins $i$ corresponding to different $Z$ values. Such deviations make it challenging to interpret the $\chi^2$ test for model evaluation accurately.
	
	A more reliable approach is directly assessing the likelihood using p-values, similar to what was done for $C_\ell$ (See Appendix.~\ref{sec:harmonicparity}), as elaborated in the next section.

	\subsubsection{p-values}
	
	For a direct combination with the parity measurements in harmonic space, we will focus next on the mean and skewness of the $P(Z)$ distribution, which we call  $Z^1 \equiv \sum_i P_i Z_i$ and $Z^3 \equiv \sum_i P_i Z_i^3$. These moments are not as constraining as the full parity $P(Z)$ distribution, but they are more easily cast in terms of p-values.
	We compute the p-value by determining the fraction of data simulations with $Z^1$ and/or $Z^3$ values more extreme than each model's predictions. It's important to note that because of the way the p-values are defined, they are constrained to be smaller than 50\%. 
	
\begin{figure}
		\centering
		\includegraphics[width=.49\linewidth]{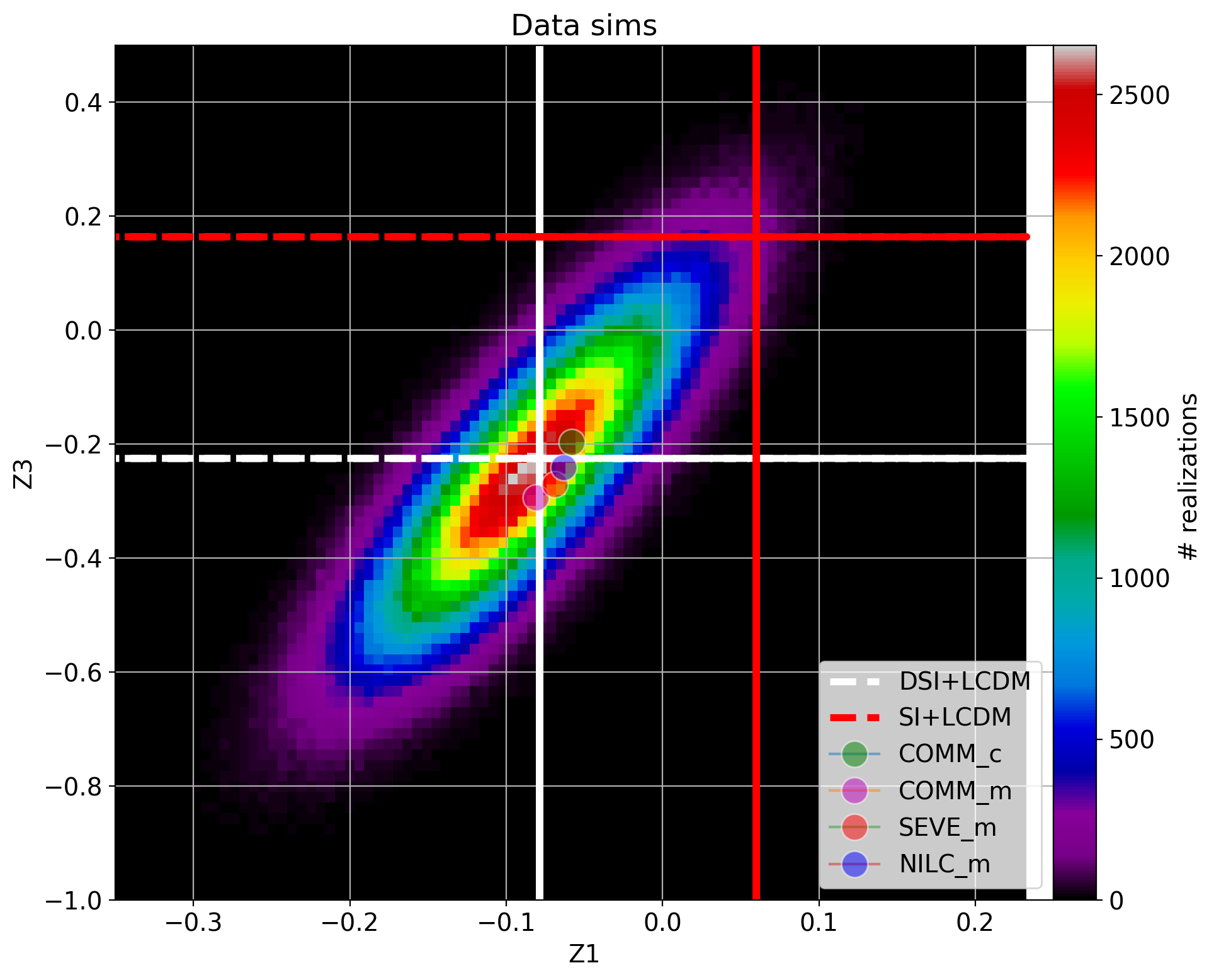}  
		\includegraphics[width=.49\linewidth]{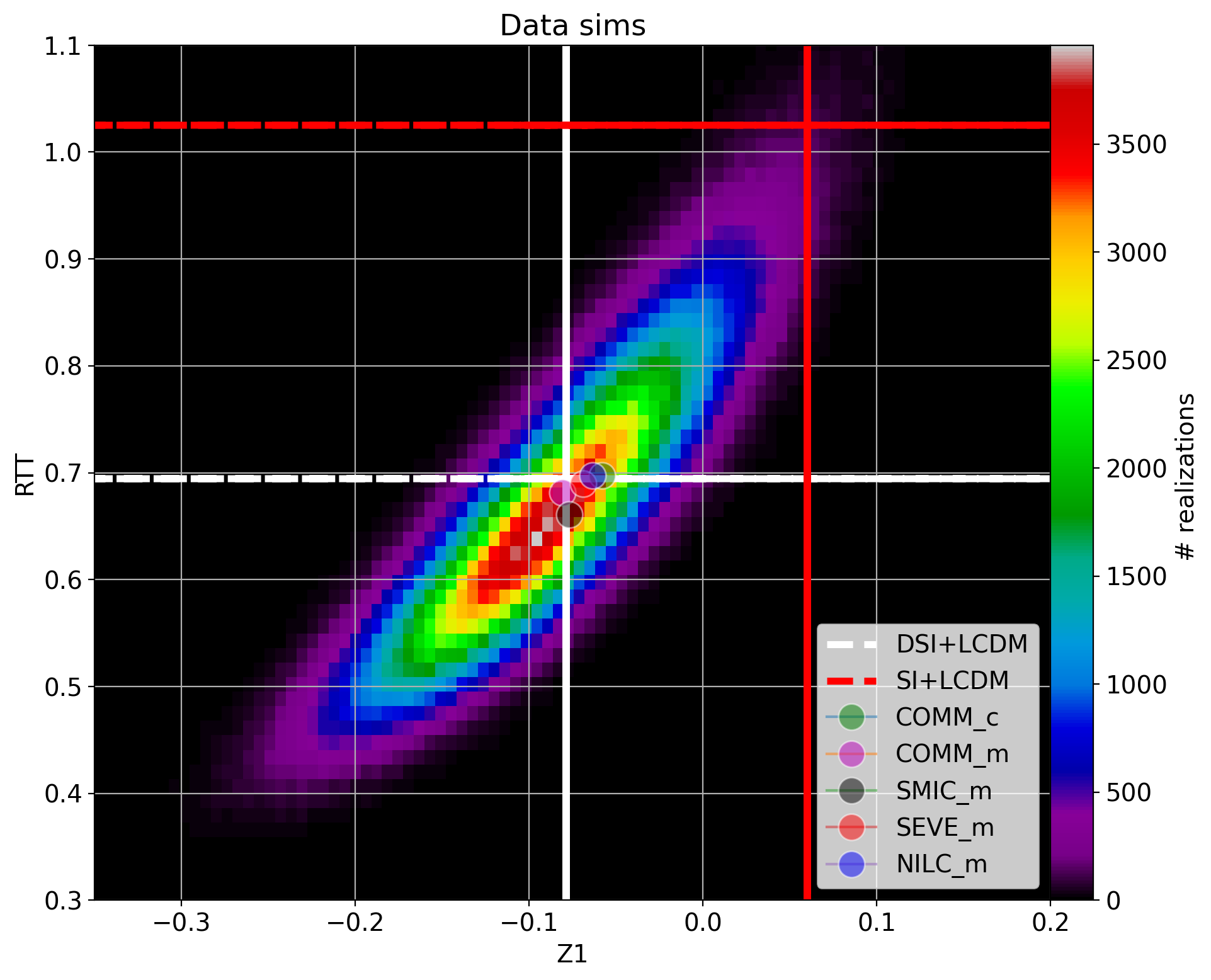}  
		\caption{
			Colored points show the density of parity measurements in data
			(with sampling variance errors from $10^6$ realizations) as compared to the corresponding expected values in 2 different models (horizontal and vertical lines): SI (red) and  DSI (white).
			The {\sc left panel} shows the skewness $Z^3$ and mean  $Z^1=w(\pi)$ of the $P(Z)$ parity distribution (i.e. parity in configuration space).
			The {\sc right panel} shows  $R^{TT}$ in Eq.~\ref{eq:RTT} (i.e. parity in harmonic space) versus the mean  $Z^1$.
			We also show the data measurements for different  CMB Planck component separation methods (circles). The external boundaries of the yellow, blue, and magenta points correspond to 1, 2, and 3$\sigma$ confidence levels (see also Fig.~\ref{fig:corner}).} 
		\label{fig:Z3RTT}
	\end{figure}

	The results are presented in Table.~\ref{tab1:annomalies}, where we also include the corresponding p-values (representing the probabilities of the model given the data: $p[M|D]$) for parity asymmetry in the quadrupole $C_2$ and the harmonic space even-odd ratio $R^{TT}$ in Eq.~\ref{eq:RTT}. Using the same $10^6$ data simulations for each case enables us to combine them, as illustrated in Fig.~\ref{fig:Z3RTT}. 

 \begin{table}
		\caption{Parity probability  $p[M|D]$ ($\times10^2$) of a model  given the data (based on $10^6$ sky realizations of the data). Each CMB parity estimate (lines) is compared with the two different models: Standard  Inflation quantum fluctuations with LCDM (SI) and direct-sum inflationary quantum fluctuations (DSI). 
			Data is estimated from the Planck 2018 masked SMICA component separation map. Very similar results are found for the other maps (e.g. see  Fig.~\ref{fig:Z3RTT}).
			In the last column 'ratio' refers to the ratio of probabilities between the DSI and SI predictions.}
		\begin{center}
			\label{tab1:annomalies}
			\begin{tabular}{r | c c | c}
				p-value: $p[M|D] \times 10^{2}$   & SI  & DSI  & ratio  \\ \hline \hline \quad {\sc  Harmonic} space:  & & &  DSI/SI \\ \hline
				$C_2$  & 0.09 & 3.3 & 37 \\         
				$R^{TT}$ ($\ell \le 20$) & 0.7 & 39.5 & 56  \\ 
				$C_2 ,R^{TT}(\ell \le 20)$  &  0.003 &  1.96 & 653 \\ \hline
				\quad {\sc  Configuration} space:  & & &  \\ \hline
				$Z^1 = w[\pi]$ & 1.12 & 45.3 & 40 \\
				$Z^3$  & 2.10 & 45.6 & 22\\
				$Z^1,Z^3$  &  0.67 &  36.3 & 54 \\ \hline
				\quad {\sc  Combined}:  & & &  \\ \hline       
				$Z^1 ,R^{TT}(\ell \le 20)$  &  0.45 &  34.6 & 77 \\
				$Z^1 , C_2$   &  0.016 &   2.65 & 166 \\ \hline
				\hline
				
			\end{tabular}
		\end{center}
	\end{table}

\begin{figure}
		\vskip -2.2cm
		\includegraphics[width=.95\linewidth]{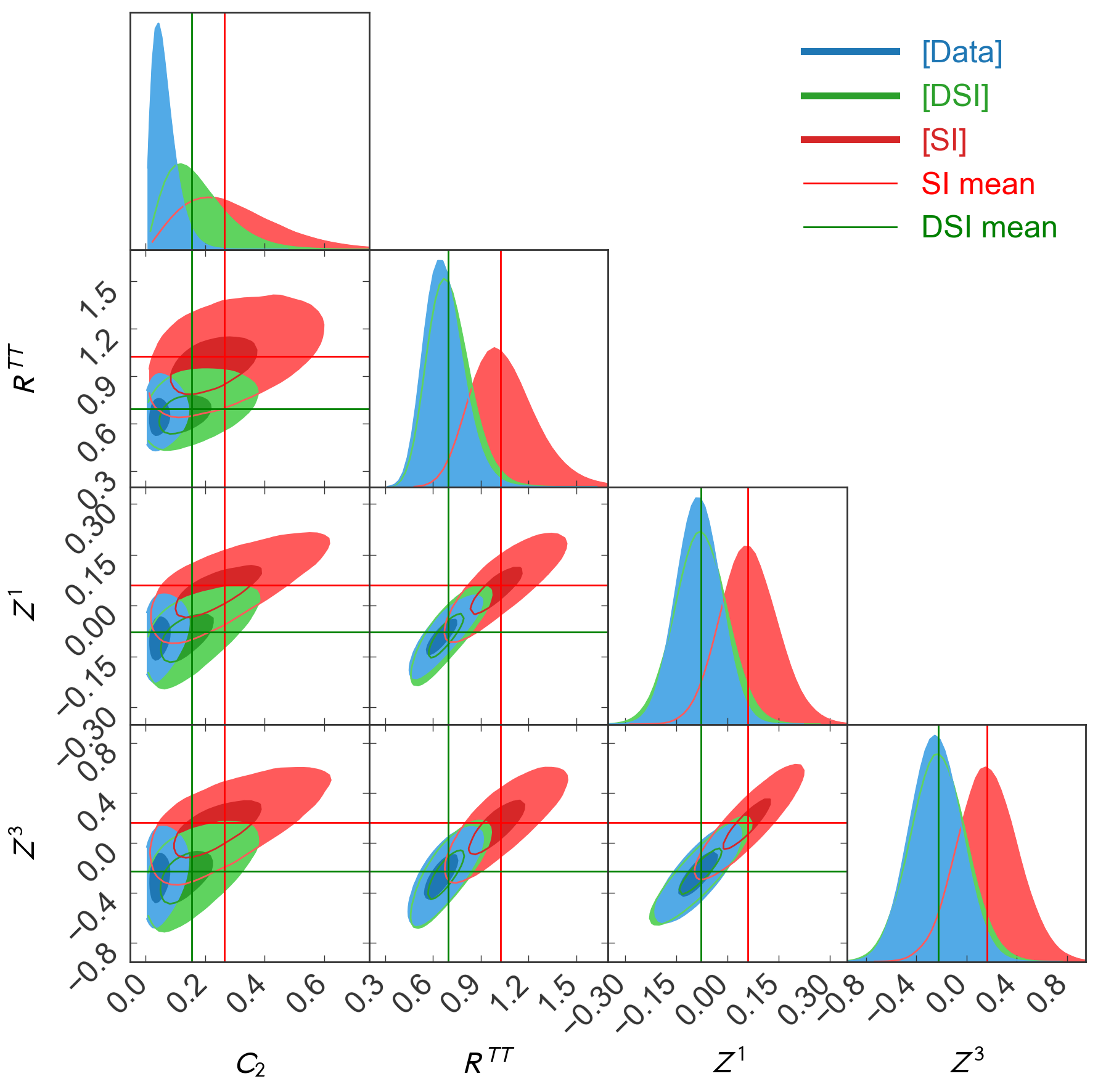}
		\caption{Similar to Fig.~\ref{fig:Z3RTT}, this shows parity measurements for contours of uncertainties (from $10^6$ realizations each) of the data (blue), SI (red), and DSI (green) models. The shaded regions correspond to 1 and 2-sigma confidence levels in the simulations. Note that the errors are dominated by cosmic variance, which is included in both the data and the models. We can either look at $P[D|M]$: i.e., where the mean data lies within the model realizations, or at $P[M|D]$: where the mean of the models lies within the data realizations. The latter is more constraining and well-defined. The mean DSI model is within the 1-2$\sigma$ data contours, whereas SI is ruled out with more than 3$\sigma$ confidence (see also Table.~\ref{tab1:annomalies}).}
		\label{fig:corner}
	\end{figure}

	Fig.~\ref{fig:corner} shows the other combinations presented in Table~\ref{tab1:annomalies}. 
	The model predictions for SI (red) and DSI (blue) are shown as vertical and horizontal lines in Fig.~\ref{fig:corner}. The data prefers DSI over SI with a significance of more than $3\sigma$, in good agreement with the direct $P(Z)$ test (shown in Fig.~\ref{fig:ZhistSMICA}) or the p-values in Table.~\ref{tab1:annomalies}. In particular, SI is ruled out at more than $3\sigma$ in any combination involving $C_2$, while DSI can explain the $C_2$ and all other parity measurements within $2\sigma$ confidence interval.

 \begin{table}
		\caption{Probability ($\times10^2$) $p[D|M]$ of the data being more extreme than a model realization (based on $10^6$ masked sky simulations of the model). Each CMB anomaly (lines) is compared with 2 different models (columns): Standard  Inflation quantum fluctuations with LCDM (SI+LCDM) and direct-sum inflationary quantum fluctuations (DSI+LCDM). 
			Data is estimated from the Planck 2018 masked SMICA component separation map. Very similar results are found for the other maps (e.g., see Fig.~\ref{fig:Z3RTT}).
			In the last column, 'ratio' refers to the ratio between the DSI and SI predictions. }
		
		\begin{center}
			\label{tab2:annomalies}
			\begin{tabular}{r | c c | c}
				p-value: $p[D|M] \times 10^{2}$  & SI & DSI & ratio \\ \hline \hline
				\quad {\sc Parity}:  &  & & DSI/SI  \\ \hline \hline
				$C_2$  & 2.62   &  8.88  &  3.4 \\ 
				$R^{TT}$ ($\ell \le 20$) & 1.00 & 39.7 & 40 \\ 
				$Z^1 = w[\pi]$ & 3.89 & 46.3 & 12  \\
				$Z^3$  & 1.78  & 37.1 & 21 \\ 
				$Z^3, R^{TT}$ & 0.58 & 32 & 55 \\     
				$C_2, R^{TT}$ & 0.12 & 5.59 & 47 \\
				\hline \hline
			\end{tabular}
		\end{center}
	\end{table}
	
	The other Planck component separation maps plotted as circles in Fig.~\ref{fig:Z3RTT}, are close to SMIC$_m$ and the small differences can mostly be attributed to the difference in the size of the masking areas.
	The DSI results exhibit close similarity to data. Given that both SI and DSI models share the same number of parameters (as dictated by the best LCDM fit used as $C_\ell$ input to the simulations), we can confidently assert that DSI is up to $\simeq 80-650$ times more probable than SI. This agrees well with the cumulative likelihood ratios based on the $C_\ell$ shown as a blue line in the bottom left panel of Fig.~\ref{fig:clfit}.

	Table.~\ref{tab2:annomalies} presents the corresponding probabilities based on model simulations (probabilities of the data given the model: $p[D|M]$), and the results agree qualitatively with Table \ref{tab1:annomalies}. Generally, the latter provides more stringent constraints, as the sampling variance errors are smaller for realizations based on the data (due to the smaller $C_2$ value).
	The p-values obtained for SI in Table.~\ref{tab2:annomalies} align with those found in previous literature (e.g., see Fig.~3 of \cite{Muir:2018hjv} and references therein).

	
	 	When $C_2$ is fixed in SI to match the data (model SI-$C_2$), the $P(Z)$ predictions get closer to observations. This makes sense because $C_2$ is the first and largest of the odd parity oscillations.
	However, additional evidence in favor of odd parity remains without $C_2$, as indicated by comparing p-values in the third columns of Table.~\ref{tab2:annomalies} and Table.~\ref{tab3:annomalies}.

 \begin{figure}
		\centering
		\includegraphics[width=0.49\linewidth]{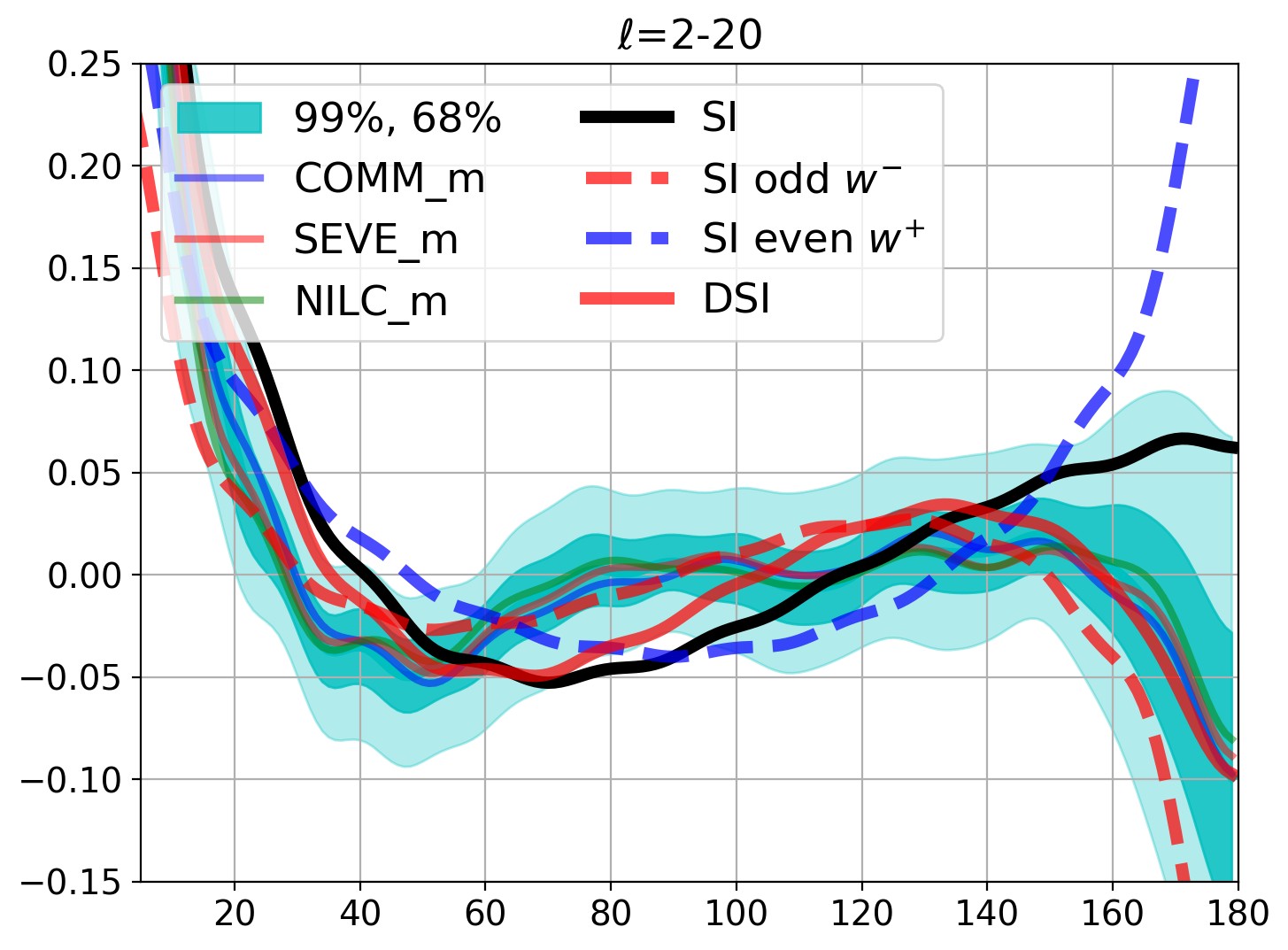}
		\includegraphics[width=0.49\linewidth]{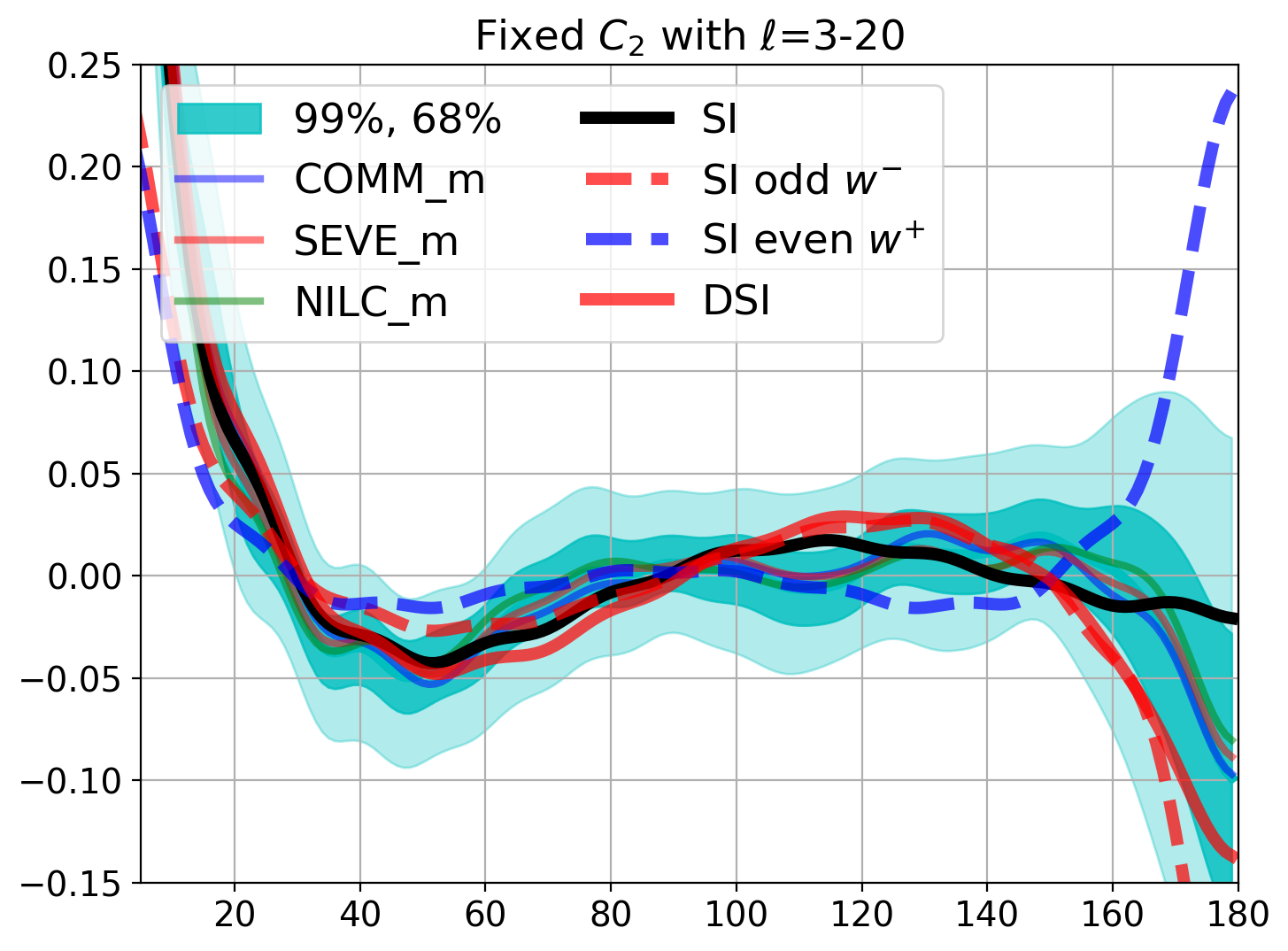}     
		\caption{Comparison of the 2-pt temperature correlation $w(\theta)$ estimated from $C_\ell$ in Eq.~\ref{eq:C_theta} with $\ell_{max}=20$.
			Results for larger $\ell_{max}$ are the same for separations $\theta>8^\circ$.
			The 99\% and 68\% regions (cyan) allowed by the SMIC$_m$ Planck map (with $10^6$ sampling realizations)  are compared with the SI best-fit model ($w$, black) split into odd ($w^{-}$, red dashed) and even ($w^{+}$, blue dashed) parity contributions. The data is closer to the odd component.
			The right panel shows the result of replacing the quadrupole $C_2$ in the SI and DSI models with the one measured in the SMIC$_m$ map (i.e. model SI-$C_2$ in Table.~\ref{tab3:annomalies}). This makes all the curves closer to the data and resolves the anomaly in $S_{1/2}$ (see Eq.~\ref{eq:S12} and Table.~\ref{tab3:annomalies}). But even with a low $C_2$ the data is more odd than the average SI model, with strong antisymmetry between antipodal angles.}
		\label{fig:Ctheta}
	\end{figure}

\subsection{Connection between lack of correlations at $\theta>60^\circ$ and lower quadrupole}

\label{sec:H60}

From Tables.~\ref{tab1:annomalies} and \ref{tab2:annomalies}, we can read that the DSI model predicts a lower quadrupole with a higher $p-$value than in comparison with the SI model. We can see this in Figs.~\ref{fig:clfit} and  Fig.~\ref{fig:corner} as well. Another CMB anomaly other than parity is called $S_{1/2}$ and is related to the lack of angular correlations Eq.~\ref{eq:C_theta} at $\theta>60^\circ$ (See Appendix.~\ref{sec:characterization}). 
 Fig.~\ref{fig:Ctheta} illustrates the decomposition of  $w[\theta]$ in Eq.~\ref{eq:C_theta} into even (red) and odd (blue) parity contributions using the same $C_\ell$ values displayed in Fig.~\ref{fig:clfit}. The Planck map and simulations align more closely with the odd component of the SI model, in agreement with the odd-even oscillations in $C_\ell$ that we showed in Fig.~\ref{fig:clfit}.
It's remarkable to observe that the antisymmetric part of Fig.~\ref{fig:Ctheta}—depicted by the red dashed line, representing the odd component of temperature fluctuation in the SI model—successfully connects antipodes at $\theta \simeq 180^\circ$ to $\theta \simeq 0^\circ$ with comparable amplitudes but opposite signs. This alignment provides compelling evidence that the odd component predominantly influences the CMB sky, confirming our visual observation in Fig.~\ref{fig:smicaSA}.
 Thus, observations not only imply homogeneity (due to the absence of anisotropies on the largest scales: $\theta>60^\circ$ or $\ell<4$) but also reveal an excess of odd parity (evidenced by the equal amount of negative and positive correlations at the largest and smallest scales). Despite the potential impact of the mask on the $C_\ell$ measurements, a direct estimation of $w(\theta)$ in configuration space yields results very similar to the straightforward sum of $C_\ell$ in harmonic space presented here (see, for instance, \cite{gazta2003, Camacho-Quevedo:2021bvt}). 
From the right panel of Fig.~\ref{fig:Ctheta}, where the SI+LCDM model fixes the value of the quadrupole $C_2$ to align with CMB data measurement, we can infer that this model better accounts for the absence of correlations at $\theta>60^\circ$. However, SI+LCDM still falls short in addressing the observed anticorrelation at $\theta=180^\circ$, a challenge that the DSI model meets. Considering that the likelihood of DSI explaining the quadrupole is 37 times higher than SI (see Table.~\ref{tab1:annomalies}), it becomes evident that DSI not only accounts for the observed lack of correlations at large angular scales but also captures the antipodal odd symmetry.

We have quantified the anomaly in $S_{1/2}$ in Eq.~\ref{eq:S12} and 
	presented the corresponding p-values in Table.~\ref{tab3:annomalies}. The obtained p-values are notably low, aligning with previous findings (see, for instance, \cite{Muir:2018hjv,Jones:2023ncn} and references therein, and also \cite{Camacho-Quevedo:2021bvt} for a related study).
	As demonstrated in the second column of Table.~\ref{tab3:annomalies}, the low p-value is primarily a consequence of the reduced quadrupole $C_2$ in the data. When we hold $C_2$ fixed at the observed value (in SMIC$_m$), the probability increases by a factor of 122. This shift is visually represented in Fig.~\ref{fig:Ctheta}.

 
	\subsection{Revising HPA with  de-biased simulations 
 and lower-quadrupole}
\label{sec:ruleoutHPA}

 \begin{table}
		\caption{Probability ($\times10^2$) $p[D|M]$ of the data being more extreme than a test model realization (based on $10^6$ masked sky simulations of the model). Each CMB anomaly (lines) is compared with 2 different test models (columns) against Standard  Inflation with LCDM (SI+LCDM): The second column is SI+LCDM but with quadrupole $C_2$ being fixed to data (SI-$C_2$),  anisotropic dipolar modulation (DPM with $A=0.07$ or $A=0.5$). 
			Data is estimated from the Planck 2018 masked SMICA component separation map. Very similar results are found for the other maps (e.g. see Fig.~\ref{fig:Z3RTT}-Fig.~\ref{fig:corner}).
			In the bottom of the Table we show other anomalies of the SI+LCDM model, which are in principle not directly related to Parity. These anomalies vanish for the low $C_2$ version of the model.}
		\begin{center}
			\label{tab3:annomalies}
			\begin{tabular}{r | c c c}
				p-value: $p[D|M] \times 10^{2}$  & SI  & SI-$C_2$ & DPM  \\ \hline \hline
				\quad {\sc Parity}:  & &  &   A=0.07 (0.5)  \\ \hline \hline
				$C_2$  & 2.62  & 47.8 &  2.6 (2.1)  \\ 
				$R^{TT}$ ($\ell \le 20$) & 1.00 & 4.07 & 1.0 (0.4)  \\ 
				$Z^1 = w[\pi]$ & 3.89 & 19.3 & 4.0 (3.9)  \\
				$Z^3$  & 1.78 & 9.0 & 1.8 (1.8)  \\ 
				$Z^3, R^{TT}$ & 0.58 & 3.01 & 0.7 (0.3)  \\     
				$C_2, R^{TT}$ & 0.12 &  1.31 & 0.1 (0.03)  \\
				\hline \hline
				{\sc Other Anomalies:} \\ 
				\hline \hline
				{\sc 2-pt $w(\theta)$}: $S_{1/2}$   &  0.08  & 9.8 \\ 
				{\sc HPA}: $\sigma^2_{16}$ &  0.49  &  4.6 \\
				{\sc HPA}: $\sigma^2_{16}[db]$ &  2.50 &  14.9 \\ 
			\end{tabular}
		\end{center}
	\end{table}

In this section, we revisit one of the most explored CMB anomaly called Hemispherical Power Asymmetry (HPA) which implies CMB in a preferred direction has a maximal power asymmetry.  Visually you can see this annomaly in the top right panel of Fig.\ref{fig:MirrorMask} which shows the Planck map with a low resolution in Ecliptic coordinates. In those coordinates, the amplitude of fluctuations are clearly larger in the South than in the North Ecliptic cap. As a result there is an alignment of the $C_2$ and $C_3$ multipoles with the Ecliptic plane. This can be modeled as a dipolar modulation (DPM) and the empirical fit to the CMB data has been found to be the following \cite{Akrami:2014eta}
	\begin{equation}
		T(\hat{n})= \LF1+ A\, \hat{n}.\hat{m} \RF \, T_{iso}(\hat{n}), 
		\label{eq:DPM}
	\end{equation}
	where $T_{iso}$ will be model as the 
	SI+LCDM model with an isotropic spectrum of fluctuations and   
	$\hat{m}$ is a fixed direction in the sky (the origin of the anisotropy). $A \simeq 0.07$ is the amplitude of anisotropies that is consistent with the maximum measured HPA asymmetry in Planck data \cite{Namjoo:2014nra}.
The HPA was also claimed to be the feature in WMAP analysis \cite{Hansen:2008ym} which has triggered numerous theoretical studies (See \cite{Erickcek:2008jp,Erickcek:2008sm} and the cited of them). The appearance of HPA in the particular direction of ecliptic coordinates (i.e., the direction of the orientation of the solar system) is coined to be the \say{axis of evil}, and even some late-time observations have supported it \cite{Land:2005ad,Secrest:2020has}. 
Recently, this has gained much more attention in the form of conclusive evidence for the violation of isotropy or cosmological principle \cite{Jones:2023ncn,Aluri:2022hzs}. 
First of all, this finding, if true, does change our understanding of cosmology, but it is important to carefully scrutinize the finding even further, which is what we aimed to do in this section. Notably, in \cite{Quartin:2014yaa}, a critical analysis of HPA was performed and concluded to be insignificant. We also reached a similar conclusion in our study with our direction-de-biased simulations of HPA, which will be explained shortly.  In all the previous sections, our focus was only on parity, now we can ask a much more generic question which is if HPA advocated by the relation Eq.~\ref{eq:DPM} has anything to do with Parity. The value of $A\approx 0.07$ has very little impact in our parity test and we will therefore also try an extreme case of $A=0.5$. Another question we can ask is whether DPM or HPA can be originated from low-quadrupole $C_2$?
	
	As shown in the 4th column of Table.~\ref{tab3:annomalies}, the anisotropic DPM model yields very similar parity results to the SI+LCDM model, even when we increase the asymmetry in Eq.~\ref{eq:DPM} to an extreme dipole with $A=0.5$. The case of $A=0.5$ is illustrated in the bottom right panel of Fig.~\ref{fig:MirrorMask} (in Ecliptic coordinates) and in Fig.~\ref{fig:simParityDPM} (in Galactic coordinates). We found that even such a substantial anisotropy proves insufficient to account for the observed parity asymmetry, particularly in the $R^{TT}$ or $P(Z)$ measurements.
	
	This occurs because the anisotropy is generated in a single fixed direction in each simulation to match the corresponding anisotropy found in the data. In that direction, the $Z$ values will have a negative background contribution between antipodal points, akin to the odd-parity case. However, because we average over all directions, this has a minimal impact on the overall $P(Z)$ statistics. The $P(Z)$ distribution for DPM is very similar to that of the SI simulations in Fig.~\ref{fig:ZhistSMICA}.
	In terms of $C_\ell$, regardless of how large $A$ is, there is no alteration to the odd-even oscillations ($R^{TT}$) of each isotropic (LCDM) realization. Consequently, parity remains unaffected. These findings are also evident from the examination of Table.~\ref{tab2:annomalies}-\ref{tab3:annomalies} and underscore the fundamental distinctions between parity and anisotropies, i.e., Parity cannot be altered by anisotropies. Isotropy constitutes a rotational symmetry, whereas parity is a separate discrete symmetry that is superimposed on top.
 
	Additionally, we've calculated the Hemispherical Power Asymmetry (HPA), characterized by the low variance $\sigma_{16}^2 = < \Delta T^2>$ in the North ecliptic sky, employing a resolution of $N_{nside}=16$ and $r_{beam}=10.5^\circ$, as defined in \cite{Muir:2018hjv} (refer to our Fig.~\ref{fig:MirrorMask}). The corresponding p-values for each test are shown in the bottom entries of Table.~\ref{tab3:annomalies}.
	
	The $\sigma_{16}^2$ value in the Planck 2018 data reflects the outcome of a directional search to find the hemisphere in which $\sigma_{16}^2$ is lower for a specific direction. Such a search found it to be near the North ecliptic hemisphere. However, in the simulations, the $\sigma_{16}^2$ value is derived from a single direction in each realization of SI+LCDM (which is random for isotropic Gaussian fluctuations). This discrepancy introduces a substantial bias in the p-values, as we are comparing different statistics in data and simulations. To rectify this bias, we conduct the test $\sigma_{16}^2[db]$ (de-bias value) by searching for the smallest $\sigma_{16}^2$ in $10^3$ directions in each of the $10^6$ realizations for each model, mirroring the approach done with the data. In both SI and DSI models, the p-value for $\sigma_{16}^2[db]$ is approximately 5 times larger than in $\sigma_{16}^2$.
	
	These anomalies are not inherently linked to parity asymmetry. However, their statistics are influenced by the parity in the data, as evidenced by the comparison between SI and SI-$C_2$ in the lower entries of Table.~\ref{tab2:annomalies}. When $C_2$ is low, the significance of the HPA or DPM-related anomalies largely disappear\footnote{The DSI model was initially proposed in the context of HPA \cite{Kumar:2022zff}. But in fact, the paper \cite{Kumar:2022zff} was only addressing the maximal difference between power spectra in the direction of $\hat{m}$ and $-\hat{m}$ when $\hat{n}\cdot \hat{m}=1$. This is nothing but looking at parity asymmetry in just one direction. DSI model by construction leads to parity-asymmetry as proposed in \cite{Kumar:2022zff} (which is explained in further detail in Sec.~\ref{sec:theory}), and this is precisely what we witnessed in the CMB data that is strongly confirmed by the Z-statistics of the previous section. Thus, the DSI theory proposed in \cite{Kumar:2022zff} is factually correct but not its interpretation towards HPA.}. We can, therefore, interpret all these anomalies as resulting from the odd parity in the data, which explains the low $C_2$ (with a p-value of approximately 9\% in Table.~\ref{tab2:annomalies} and a p-value about 40 times larger than SI in Table.~\ref{tab1:annomalies}), without the necessity of breaking isotropy or a deviation from the primordial scale-invariant spectrum. 
	
	It is intriguing how $P(Z)$ and the other CMB anomalies at large scales, while potentially explained by odd parity asymmetry, still leave room for more profound deviations. DSI provides an over 100 times better explanation of the CMB data than SI, yet it still has a low p-value for the combination of $C_2$ with several parity measures  (i.e., $R^{TT}$ or $Z^1$)  in Table.~\ref{tab1:annomalies}. The notion that the cosmological constant $\Lambda$ might play the role of a cosmic boundary condition opens up fascinating possibilities \cite{Gaztanaga2021, Gaztanaga2022, BHU1, BHU2, gaztanaga2023b}. It aligns well with this and other observations that indicate a bounded (or finite) universe \cite{Fosalba:2021, Gaztanaga:2022bdh,gaztanaga2023a}.

	\newpage
	\section{Inflationary quantum fluctuations with direct-sum QFT} 
	\label{sec:theory}
	
In this section, we discuss the formulation of DSI through the direct-sum quantum field theory (DQFT) in curved spacetime proposed in \cite{Kumar:2022zff,Kumar:2023ctp,Kumar:2023hbj}. 
In Appendix.~\ref{sec:reviewSI}, we first review the theory of quantum fluctuations in SI. We also discuss the fundamental concerns of SI.  In the following subsections, we lay out the theme of DSI emanating from the foundations of discrete spacetime transformations and their connection with the direct-sum construction of Hilbert/Fock space.

	\subsection{The $\Pc\Tc$ symmetry of de Sitter spacetime}
	
\begin{figure}
		\centering
		\includegraphics[width=0.7\linewidth]{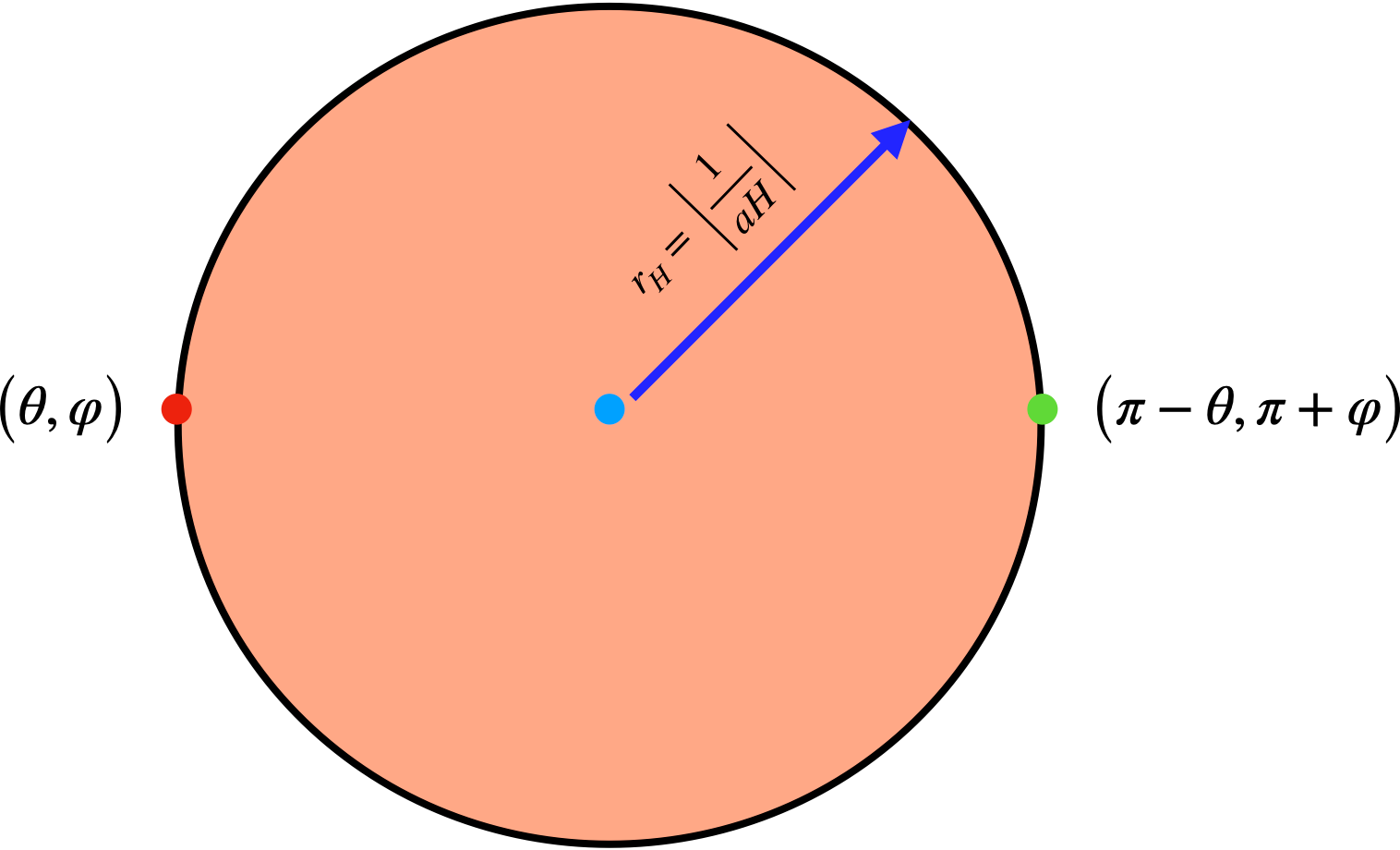}
		\caption{Depiction of the co-moving horizon in the de Sitter Universe in flat FLRW coordinates where the points $\LF \theta,\, \varphi  \RF$ and  $\LF \pi-\theta,\, \pi+\varphi  \RF$ are always causally separated.}
		\label{fig:horizon}
	\end{figure}

		To illustrate the \say{time} problem, let us look into the de-Sitter metric in flat FLRW coordinates 
	\begin{equation}
		ds^2 = -dt^2 + e^{2Ht}d\textbf{x}^2 \,.
		\label{dSmetric}
	\end{equation}
	The above metric illustrates an expanding Universe with a growing scale factor which acts as a \say{clock} that records the events within the particle horizon 
	\begin{equation}
		r_H = \Big\vert \frac{1}{aH}\Big\vert 
	\end{equation}
	In Fig.~\ref{fig:horizon} we can see that at a moment of a given size of particle horizon, we have infinitely many pairs of parity conjugate points in the spacetime that are space-like separated. 
	The description of a quantum field in such a curved spacetime is very non-trivial, especially when the quantum field is described by a vacuum that contains an infinite pair of parity conjugate points on the horizon that are space-like separated. In other words, it is inconceivable to decipher a single propagator that connects the points $\LF \theta,\,\varphi \RF$ and $\LF \pi-\theta,\, \pi+\varphi \RF$ on the particle horizon, otherwise we would violate causality. In the standard QFT in Minkowski spacetime, the Feynman propagator vanishes exponentially outside the light cone. 
	So conceptually, it is convincing to think that the propagator of a quantum field in any given spacetime must not involve space-like distances. However, within this principle, we run into trouble describing the quantum field in an inflationary Universe because, by construction, the pairs of parity opposite points on the horizon are always space-like separated at every moment of the expansion. 
	
	Let us turn now our attention to the metric Eq.~\ref{dSmetric} which has a symmetry\footnote{\begin{equation}
			{\rm Expanding\,Universe:} \implies \begin{cases}
				t: -\infty \to +\infty,\quad  H>0 & (\tau: -\infty \to 0)\\ 
				t: +\infty \to -\infty,\quad  H<0 & (\tau: \infty \to 0)
			\end{cases}
			\label{expcon}
	\end{equation}}
	\begin{equation}
		\Big\{	t\to -t,\, H\to -H \Big\} \implies \Big\{\tau\to -\tau\Big\},\quad \Big\{ \textbf{x}\to -\textbf{x} \Big\} 
		\label{dessym}
	\end{equation}
 where $\tau$ is conformal time.
	Time reversal and space reflection are the usual $\Pc\Tc$ operations in quantum field theory. Here, we see a new (quantum) parameter $H$ with reflection symmetry in de Sitter space, which does not affect the value of scalar curvature $R= 12H^2$. This can be understood in another way by rewriting the metric Eq.~\ref{dSmetric} in conformal time $\tau$ 
	\begin{equation}
		ds^2 = \frac{1}{\tau^2 H^2} \LF -d\tau^2+ d\textbf{x}^2 \RF
		\label{cds}
	\end{equation}
	This metric is conformal to Minkowski spacetime with the same discrete symmetry Eq.~\ref{dessym}, which is time ($\tau$) reversal and space-reflection. 
	This conveys that the nature of the expanding Universe is associated with the growth of scale factor (or shrinking co-moving particle horizon), consistent with both signs $\tau<0$ and $\tau>0$. Often in the literature, $\tau<0$ is fixed before quantization of a field in de Sitter space. This leads to conundrums since another choice of time $\tau>0$ is equally possible. A physical conundrum here is how can quantum fields in the time-symmetric ($\tau\to -\tau$) spacetime Eq.~\ref{cds} distinguish $\tau<0$ with $\tau>0$? It is important to note here that the singularity $\tau\to 0$ in the metric Eq.~\ref{cds} is a coordinate singularity as the de Sitter metric Eq.~\ref{dSmetric} is an entirely regular one. Therefore, $\tau\to 0$ singularity cannot be a reason to make a conscious choice $\tau>0$ or $\tau<0$. 
	
	\subsubsection{Observer complimentarity principle}
	
	A regular practice in theoretical cosmology is to assume $\tau<0$
	or $\tau>0$ which are often called Poincar\'e patches of de Sitter space and they have always been read as either two (entangled) Universes that are causally disconnected or only one universe is the real physical world while other being regarded as unphysical by hand \cite{Spradlin:2001pw,Hartman:2020khs}. It is remarkable that Schr\"{o}dinger in 1956 rejected the idea of two Universes and demanded there should be only one Universe \cite{Schrodinger1956,Parikh:2002py}. This is based on a basic principle that the physics of any observer should not depend on the physics happening outside his/her horizon. This is called the observer complementarity principle, which has been realized in the form of DQFT with recent developments \cite{Kumar:2022zff,Kumar:2023ctp,Kumar:2023hbj}.\footnote{Schr\"{o}dinger proposed a so-called elliptic de Sitter space to resolve the issue of discrete symmetries and two realizations of time for the expanding Universe. The details of his proposal are irrelevant to the subject of investigation in this work.}

	\subsection{DQFT for inflationary quantum fluctuations}
	
	It is good to recall again that the inflationary paradigm by definition is a regime of quasi-de Sitter phase in the early Universe \cite{Starobinsky:1980te,Guth:1980zm}. The de Sitter metric Eq.~\ref{dSmetric} has a discrete symmetry ($\Pc\Tc$) Eq.~\ref{dessym} that inflationary spacetime breaks by the background dynamics of the scalar field that rolls down the potential. The slow-roll parameters Eq.~\ref{slp} break the symmetry Eq.~\ref{dessym} classically. A legitimate question here is: What is the response of the quantum fluctuations to this symmetry breaking, and what is its imprint in the CMB? The standard formulation of inflationary quantum fluctuations is insensitive to this fundamental fact about classical inflationary background breaking the discrete symmetry of de Sitter spacetime. It is worth pointing out here that discrete spacetime transformations take a central spot in our understanding of quantum theory. We refer the reader to \cite{Kumar:2023ctp} for detailed discussions. 
	
	\subsubsection{Direct-sum QM and QFT}
	
	Direct-sum quantum mechanics (QM) is rewriting standard QM in a more symmetric form. Here \say{Direct-sum} means 
	\begin{center}
		{\it We define a quantum state as a direct-sum of two components: the states in parity conjugate regions of space corresponding to two Hilbert spaces whose direct-sum forms the total Hilbert space. Thus, a single quantum state in direct-sum QM evolves forward and backward in time in the parity conjugate regions of physical space. }
	\end{center}
	The above text can be translated to the equations as 
	\begin{equation}
		\vert \Psi \rangle = \frac{1}{\sqrt{2}} \LF \vert \Psi_{+}\rangle \oplus \vert \Psi_{-} \rangle \RF = \frac{1}{\sqrt{2}} \begin{pmatrix}
			\vert \Psi_{+}\rangle \\ \vert \Psi_{-}\rangle
		\end{pmatrix}
		\label{psid}
	\end{equation}
	which obeys the direct-sum Schr\"{o}dinger equation 
	\begin{equation}
		i\frac{\pd\vert \Psi\rangle}{\pd t_p} = \begin{pmatrix}
			\hat{\mathbb{H}} & 0 \\ 
			0 & -\hat{\mathbb{H}}
		\end{pmatrix} \vert \Psi\rangle 
	\end{equation}
	where $\mathbb{H}$ is the time-independent Hamiltonian (which is Hermitian) that is $\Pc\Tc$ symmetric. Note that 
 \begin{equation}
     \hat{\mathbb{H}}\vert \Psi_\pm\rangle = \LF   \hat{\mathbb{H}}_+ \oplus  \hat{\mathbb{H}}_-\RF \vert \Psi_\pm\rangle =\hat{\mathbb{H}}_\pm\LF  \hat{x}_\pm,\,\hat{p}_\pm \RF\vert \Psi_\pm\rangle 
 \end{equation}
where $\left(\hat{x}_\pm,\,\hat{p}_\pm\right)$ being the position and momentum operators of the parity conjugate regions. 
 One can immediately notice from the above equation that the components evolve as
	\begin{equation}
		\vert \Psi_+\rangle_{t_p}= e^{-i\mathbb{H}t_p}\vert \Psi_+\rangle_{0},\quad \vert \Psi_{-} \rangle_{t_p} =  e^{+i\mathbb{H}t_p}\vert \Psi_-\rangle_{0}
	\end{equation}
	which are positive energy components evolving forward and backward in time. To be more precise, the component $\vert \Psi\rangle_+$ is defined only in a spatial region $x\gtrsim 0$ whose time evolution is governed by
	\begin{equation}
		i\frac{\pd}{\pd t_p}\vert\Psi_+\rangle = \mathbb{H} \vert\Psi_+\rangle
	\end{equation}
	which is a positive energy state with a parametric convention on the arrow of time $t_p: -\infty\to \infty$ whereas the component $\vert \Psi\rangle_-$ is defined only in a spatial position $x\lesssim 0$ whose time evolution is governed by
	\begin{equation}
		-i\frac{\pd}{\pd t_p}\vert\Psi_-\rangle = \mathbb{H} \vert\Psi_-\rangle
	\end{equation}
	which is a positive energy state with a parametric convention for the arrow of time $t_p: \infty \to -\infty$.  Here $\Psi_+$ and $\Psi_-$ are the states corresponding to two direct-sum Hilbert spaces associated with the parity conjugate spatial regions $x>0$ and $x<0$, respectively.
	\begin{equation}
		\mathtt{H} = \mathtt{H}_+ \oplus \mathtt{H}_- \implies \Big[ \hat{\Oc}_+,\, \hat{\Oc}_-\Big]=0\,.
	\end{equation}
	where $\Oc_+,\,\Oc_-$ are the operators that act on the states of Hilbert spaces $\mathtt{H}_+,\, \mathtt{H}_-$ respectively. 
	
	From Eq.~\ref{psid} we can deduce that 
	\begin{equation}
		\int_{-\infty}^\infty dx\langle \Psi\vert \Psi\rangle = \frac{1}{2} \int_{-\infty}^0 dx_+ \langle \Psi_+ \vert \Psi_+\rangle + \frac{1}{2} \int_{0}^\infty dx_-  \langle \Psi_- \vert \Psi_-\rangle = 1, 
	\end{equation}
	Since probabilities $\int_{-\infty}^0 dx_+\langle \Psi_+  \vert \Psi_+ \rangle = \int_{0}^\infty dx_-\langle \Psi_- \vert \Psi_- \rangle =1$ in each Hilbert space are unity. 
	The wavefunction is obtained by
	\begin{equation}
		\begin{aligned}
			\Psi\LF x,\,t_p \RF & =  \langle x\vert \Psi\rangle_{t_p} = \frac{1}{2}\begin{pmatrix}
				\langle x_+\vert & \langle x_- \vert
			\end{pmatrix}\begin{pmatrix}
				\vert \Psi_+\rangle_{t_p} \\ \vert \Psi_- \rangle_{t_p}
			\end{pmatrix} = \frac{1}{2} \langle x_+\vert \Psi_+\rangle_{t_p} + \frac{1}{2} \langle x_-\vert \Psi_-\rangle_{t_p} \\
			& = \frac{1}{2} \Psi_+\LF x_+ \RF+ \frac{1}{2} \Psi_-\LF x_- \RF
		\end{aligned}\,, 
	\end{equation}
	which conveys the information that the total wave function of the physical system is the sum of the wave functions of the parity conjugate regions defined by the position vectors $\vert x_+\rangle$ and $\vert x_-\rangle$, respectively. We can witness here that the wave-function of direct-sum QM is $\Pc\Tc$ symmetric for a $\Pc\Tc$ symmetric physical system described by $\mathbb{H}$. We can easily apply the direct-sum QM for the harmonic oscillator case, which does not change any results in practice, but rather, it brings a new understanding of having the $\Pc\Tc$ symmetry in the physical system. The quantum harmonic oscillator in direct-sum QM is described by
	\begin{equation}
		\begin{aligned}
			\hat{x}_+ & = \sqrt{\frac{1}{2}}\LF a+a^\dagger \RF,\quad \hat{p}_+ = -i\frac{d}{dx_+}= \frac{i}{\sqrt{2}} \LF a^\dagger-a \RF \\ 
			\hat{x}_- & = \sqrt{\frac{1}{2}}\LF b+b^\dagger \RF,\quad \hat{p}_- = i\frac{d}{dx_-}= -\frac{i}{\sqrt{2}} \LF b^\dagger-b \RF 
		\end{aligned}
	\end{equation}
	with 
	\begin{equation}
		\begin{aligned}
			\Big[\hat{x}_+,\,\hat{p}_+\Big] & = i,\quad \Big[\hat{x}_-,\,\hat{p}_-\Big]= -i\\
			\Big[a,\,a^\dagger\Big] & =     \Big[b,\,b^\dagger\Big] =1,\quad  \Big[a,\,b^\dagger\Big]=  \Big[a,\,b\Big]=0\,. 
		\end{aligned}
	\end{equation}
	The wavefunction of the Harmonic oscillator is 
	\begin{equation}
		\begin{aligned}
			\Psi\LF x,\,t_p \RF & = \frac{1}{\sqrt{2}}\Psi_+\LF t_p,\,x_+ \RF + \frac{1}{\sqrt{2}}\Psi_-\LF -t_p,\,x_- \RF \\ & = 
			\frac{1}{\sqrt{2^{n+1}n!}}\LF \frac{1}{\pi} \RF^{1/4}e^{-x_+^2} H_n\LF x_+ \RF e^{-iE_n t_p}+ \frac{1}{\sqrt{2^{n+1}n!}}\LF \frac{1}{\pi} \RF^{1/4}e^{-x_-^2} H_n\LF x_- \RF e^{iE_n t_p}
		\end{aligned}
	\end{equation}
	where $H_n(z)$ is the Hermite Polynomial. 
	and the probability density is
	\begin{equation}
		\begin{aligned}
			\Big\vert \Psi \LF x,\,t_p \RF\Big\vert^2 & =    {}_{t_p}\langle \Psi\vert \Psi \rangle_{t_p}= \frac{1}{2} \Big\vert \Psi_+ \LF x_+,\,t_p \RF\Big\vert^2+ \frac{1}{2}\Big\vert \Psi_- \LF x_-,\,-t_p \RF\Big\vert^2 \\
		\end{aligned}
	\end{equation}
	Now we turn our attention to the DQFT in Minkowski spacetime 
	\begin{equation}
		ds^2 = -dt_m^2+d\textbf{x}^2\,. 
	\end{equation}
	According to DQFT, the Klein-Gordon field in Minkowski spacetime is quantized as
	\begin{equation}
		\begin{aligned}
			\hat{\phi}\LF x \RF & = \frac{1}{\sqrt{2}}  \hat{\phi}_{+}  \LF t_m,\, \textbf{x} \RF \oplus \frac{1}{\sqrt{2}} \hat{\phi}_{-} \LF -t_m,\,-\textbf{x} \RF \\ 
			& = \frac{1}{\sqrt{2}} \begin{pmatrix}
				\hat{\phi}_{+} & 0 \\ 
				0 & 	\hat{\phi}_{-}
			\end{pmatrix}
		\end{aligned}
		\label{disum}
	\end{equation}
	where
	\begin{equation}
		\begin{aligned}
			\hat{\phi}_{+}  \LF x \RF &  = 	\int \frac{d^3k}{\LF 2\pi\RF ^{3/2}}	\frac{1}{\sqrt{2\vert k_0 \vert}} \Bigg[\hat a_{(+)\,\textbf{k}}  e^{ik\cdot x}+\hat a^\dagger_{(+)\,\textbf{k}} e^{-ik\cdot x} \Bigg]   \\ 
			\hat{\phi}_{-}  \LF -x\RF &  = \int \frac{d^3k}{\LF 2\pi\RF ^{3/2}}	\frac{1}{\sqrt{2\vert k_0 \vert}} \Bigg[\hat a_{(-)\,\textbf{k}}  e^{-ik\cdot x}+\hat a^\dagger_{(-)\,\textbf{k}} e^{ik\cdot x} \Bigg] 
		\end{aligned}
		\label{fiedDQFTMin}
	\end{equation}
	The creation and annihilation operators satisfy
	\begin{equation}
		\begin{aligned}
			[\hat{a}_{(+)\,\textbf{k}},\,\hat{a}_{(+)\,\textbf{k}^\prime}^\dagger] & = 	[\hat{a}_{(-)\,\textbf{k}},\,\hat{a}_{(-)\,\textbf{k}^\prime}^\dagger] = \delta^{(3)}\LF \textbf{k}-\textbf{k}^\prime \RF\\
			[\hat{a}_{(+)\,\textbf{k}},\,\hat{a}_{(-)\,\textbf{k}^\prime}] &=	[\hat{a}_{(+)\,\textbf{k}},\,\hat{a}_{(-)\,\textbf{k}^\prime}^\dagger]   =0\,.
		\end{aligned}
		\label{comcan}
	\end{equation}
	Furthermore, the commutation relations of the field and the corresponding conjugate momenta become 
	\begin{equation}
		[\hat{	\phi}_{(+)}\LF t_m, \textbf{x} \RF,\,\pi_{(+)}\LF t_m,\,\textbf{x}^\prime \RF] = i \delta\LF \textbf{x}-\textbf{x}^\prime \RF,\quad [\hat{	\phi}_{(-)}\LF- t_m, -\textbf{x} \RF,\,\pi_{(-)}\LF -t_m,\,-\textbf{x}^\prime \RF] = -i \delta\LF \textbf{x}-\textbf{x}^\prime \RF,
		\label{canDQFT}
	\end{equation}
	where 
	\begin{equation}
		\pi_{(+)} \LF t_m,\,\textbf{x} \RF = \frac{\pd\Lc_{\rm KG}}{\pd\LF \pd_{t_m } \phi_{(+)}\RF},\quad \pi_{(-)} \LF t_m,\,\textbf{x} \RF = - \frac{\pd\Lc_{\rm KG}}{\pd\LF \pd_{t_m}  \phi_{(-)}\RF}
	\end{equation}
	where $\Lc$ is the Lagrangian of the Klein-Gordan field.
	We now define Fock space vacuums as 
	\begin{equation}
		\hat{a}_{+\,\textbf{k}}\vert 0_+\rangle = 0,\quad 	\hat{a}_{-\,\textbf{k}}\vert 0_-\rangle = 0 \, ,
	\end{equation}
	The total Fock space vacuum state is given by 
	\begin{equation}
		\vert 0_T\rangle = \vert 0_+\rangle \oplus \vert 0_-\rangle =  \begin{pmatrix}
			\vert 0_+\rangle \\ \vert 0_-\rangle\
		\end{pmatrix}\,.
		\label{tFs}
	\end{equation}
	In the DQFT, the scalar field propagator, which is by definition a time-ordered product of a two-point function, is split into two parts as
	\begin{equation}
		\langle 0_T \vert \Tc\hat{\phi}\LF x \RF\hat{\phi}\LF y \RF\vert 0_T\rangle = \langle 0_+ \vert \Tc\hat{\phi}_+\LF x \RF\hat{\phi}_+\LF y \RF\vert 0_+\rangle+\langle 0_- \vert \Tc\hat{\phi}_-\LF x \RF\hat{\phi}_-\LF y \RF\vert 0_-\rangle
	\end{equation}
	According to DQFT in Minkowski spacetime, $\hat{\phi}\vert 0\rangle_{T} = \frac{1}{\sqrt{2}} \begin{pmatrix}
		\hat{\phi}_+\vert 0_+\rangle \\ \hat{\phi}_-\vert 0_-\rangle\ 	
		\end{pmatrix}$ leads to a quantum state that evolve forward in time at $\textbf{x}$ and backward in time at $-\textbf{x}$.  
	Further details of DQFT in Minkowski spacetime can be found in \cite{Kumar:2023ctp}.
	
	\subsection{DQFT inflationary fluctuations}
	
	When we do quantization of inflationary fluctuations we promote the canonical variable Eq.~\ref{canvar} as an operator $\hat{v}$ which according to the rules of DQFT transformed as 
	\begin{equation}
		\begin{aligned}
			\hat{v} & = \frac{1}{\sqrt{2}} \hat{v}_{(+)}\LF \tau,\, \textbf{x} \RF \oplus  \frac{1}{\sqrt{2}} \hat{v}_{(-)}\LF -\tau,\, -\textbf{x} \RF \\ & = \frac{1}{\sqrt{2}} \begin{pmatrix}
				\hat{v}_{(+)} \LF \tau,\,\textbf{x} \RF & 0 \\ 
				0 & \hat{v}_- \LF -\tau,\, -\textbf{x} \RF
			\end{pmatrix}
		\end{aligned}
		\label{fieldmat}
	\end{equation}
	where
	\begin{equation}
		\begin{aligned}
			& \hat{v}_{\pm}  \LF \pm\tau,\, \pm\textbf{x} \RF= 
			\int \frac{ d^3k}{\LF 2\pi \RF^{3/2}} \Bigg[ c_{\LF\pm \RF\textbf{k}} {v}_{\pm,\,k} e^{\pm i\textbf{k}\cdot \textbf{x}} + c_{\LF\pm \RF\textbf{k}}^\dagger {v}_{\pm,\,k}^\ast e^{\mp i\textbf{k}\cdot \textbf{x}} \Bigg]
		\end{aligned}
		\label{vid}
	\end{equation}
	with $c_{\LF\pm \RF\textbf{k}},\, c_{\LF\pm \RF\textbf{k}}^\dagger$ being the creation and annihilation operators
	of quasi de Sitter (qdS) vacua defined by
	\begin{equation}
		\begin{aligned}
			c_{\LF + \RF\textbf{k}}\vert 0\rangle_{\rm qdS_I} & = 0\quad c_{\LF - \RF\textbf{k}}\vert 0\rangle_{\rm qdS_{II}} = 0 \\ 
			\big[c_{\LF \pm \RF\textbf{k}},\, c_{\LF \pm \RF\textbf{k}^\prime}^\dagger\big] & = \delta\LF \textbf{k}-\textbf{k}^\prime \RF,\quad \big[c^\dagger_{\LF \pm \RF\textbf{k}},\, c^\dagger_{\LF \mp \RF\textbf{k}^\prime}\big]  = 0 
			\label{qds1}
		\end{aligned}
	\end{equation}
	The total inflationary vacuum is a direct-sum 
	\begin{equation}
		\vert 0\rangle_{\rm qdS} = \vert 0 \rangle_{\rm qdS_{I}} \oplus \vert 0\rangle_{\rm qdS_{II}} = \begin{pmatrix}
			\vert 0\rangle_{\rm qdS_{I}} \\ 
			\vert 0\rangle_{\rm qdS_{II}}
		\end{pmatrix}
		\label{qdSmat}
	\end{equation}
	and $ {v}_{\pm,\,k}$ are the mode functions obtained by solving Mukhanov-Sasaki (MS) equation for $v_{\pm}$ 
	\begin{equation}
		v_{\pm,\,k}^{\prime\prime}+ \LF k^2-\frac{{\nu}_s^{\LF \pm\RF 2}-\frac{1}{4}}{\tau^2} \RF v_{\pm,\,k}^2 =0\,.
		\label{MS-equation}
	\end{equation}
	where
	\begin{equation}
		\nu_s^{\pm} \approx \frac{3}{2}\pm\epsilon\pm\frac{\eta}{2}
	\end{equation}
	Here where $v_{(+)}\vert 0\rangle_{\rm qdS_I}$ gives a field component evolving forward in time\footnote{Note that in \cite{Kumar:2022zff,Kumar:2023ctp} conformal time is defined with a minus sign as $\tau = -\int \frac{dt}{a} =\frac{1}{aH}$.  } ($\tau: -\infty \to 0$) whereas $v_{(-)}\vert 0\rangle_{\rm qdS_{II}}$ ($\tau: \infty \to 0$) in the parity conjugate regions $\textbf{x}$ and $-\textbf{x}$ respectively. Note that time reversal operation (in a completely quantum mechanical sense) is given by 
	\begin{equation}
		\tau\to -\tau \implies t\to -t,\, H\to -H,\, \epsilon\to -\epsilon\,, \eta\to -\eta\,. 
  \label{dsisym}
	\end{equation}
	which conveys that in the context of gravity time reversal operation is much more than what we know in the context of quantum mechanics. Here, the slow-roll parameters $\epsilon,\,\eta$ classically break the time-reversal symmetry of de Sitter space Eq.~\ref{dessym}. Therefore, the quantum field in the inflationary space-time breaks the $\Pc\Tc$ symmetry of de Sitter space due to Eq.~\ref{dsisym} in our framework of DQFT. 
	
\begin{figure}
		\centering
		\includegraphics[width=0.7\linewidth]{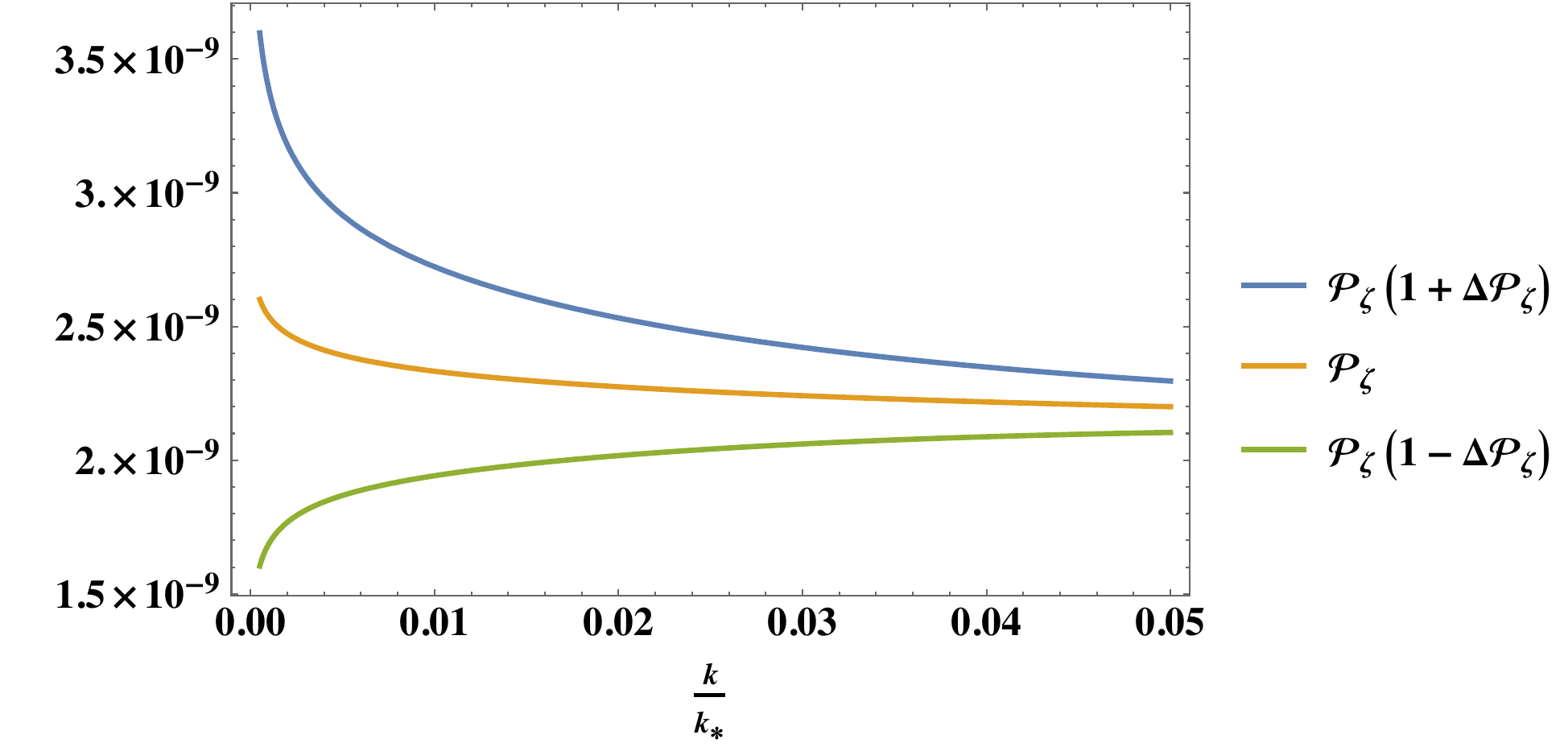}
		\caption{In this plot, we depict the power spectrum of curvature perturbation $\Pc_\zeta$ Eq.~\ref{totpw} with its antisymmetric components. The corresponding values are displayed in Table \ref{tab:DSI}.}
		\label{fig:pws}
	\end{figure}

	The solutions for the mode functions are  
	\begin{equation}
		\begin{aligned}
			&	{v}_{\pm ,\,k}   = \frac{\sqrt{\mp \pi \tau}}{2} e^{\LF i\nu_s^{\pm}+1\RF} \Bigg[C_k^{\pm} H^{(1)}_{\nu_s^{\pm}}\LF \mp k \tau \RF
			+ D_k^{\pm} H^{(2)}_{\nu_s^{\pm}}\LF \mp k  \tau \RF\Bigg].
			\label{new-vac1}
		\end{aligned}
	\end{equation}
	The canonical commutation relations 
	\begin{equation}
		\Big[\hat{v}_{\pm},\,\hat{v}_{\pm}^\prime\Big] =  \pm i 
	\end{equation}
	gives us 
	\begin{equation}
		\vert C_k^{\pm}\vert^2 -\vert D_k^{\pm}\vert^2 = 1\,. 
	\end{equation}
	We choose the adiabatic vacuum that corresponds to 
	\begin{equation}
		\LF C_k^{\pm},\, D_k^{\pm} \RF = \LF 1,\,0 \RF\,. 
	\end{equation}
	The two-point function of the canonical variable is 
	\begin{equation}
		\begin{aligned}
			{}_{\rm qdS}\langle 0 \vert \hat{v}\LF \tau,\,\textbf{x} \RF  \hat{v}\LF \tau,\,\textbf{y} \RF \vert 0 \rangle_{\rm qdS} = &  \,	\frac{1}{2}{}_{\rm qdS_{I}}\langle 0 \vert \hat{v}_{+}\LF \tau,\,\textbf{x} \RF  \hat{v}_{(+)}\LF \tau,\,\textbf{y} \RF \vert 0 \rangle_{\rm qdS_{I}} \\ & + \frac{1}{2} {}_{\rm qdS_{II}}\langle 0 \vert \hat{v}_{-}\LF -\tau,\,-\textbf{x} \RF  \hat{v}_{-}\LF -\tau,\,-\textbf{y} \RF \vert 0 \rangle_{\rm qdS_{II}}\\ 
			=&\, \frac{1}{2}\int \frac{dk}{k}\frac{k^3}{2\pi^2}\LF\vert v_{+\,k}  \vert^2+\vert v_{-\,k}  \vert^2 \RF \frac{\sin{kL}}{kL}
		\end{aligned}
	\end{equation}
	where $L=\vert \textbf{x}-\textbf{y}\vert$. The power spectrum related to the canonical variable is given by
	\begin{equation}
		\begin{aligned}
			{}_{\rm qdS}\langle 0\vert \hat{v}_{\textbf{k}}\hat{v}_{\textbf{k}^\prime}\vert 0\rangle_{\rm qdS} & =    \frac{1}{2}{}_{\rm qdS_{I}}\langle 0\vert \hat{v}_{(+),\,\textbf{k}}\hat{v}_{(+),\,\textbf{k}^\prime}\vert 0\rangle_{\rm qdS_{I}}+\frac{1}{2}{}_{\rm qdS_{II}}\langle 0\vert \hat{v}_{(-),\,\textbf{k}}\hat{v}_{(-),\,\textbf{k}^\prime}\vert 0\rangle_{\rm qdS_{II}} \\ 
			&  = \delta\LF \textbf{k}+\textbf{k}^\prime \RF \frac{2\pi^2}{k^3} \Pc_v
		\end{aligned}
	\end{equation}
	where 
	\begin{equation}
		\hat{v}_{(\pm)\,\textbf{k}}   =  \int \frac{d^3x}{\LF 2\pi \RF^{3/2}} \hat{v}_{(\pm)}\LF \pm \tau,\,\pm\textbf{x} \RF e^{-i\textbf{k}\cdot \textbf{x}}
	\end{equation}
	and
	\begin{equation}
		\begin{aligned}
			\Pc_v & = \frac{1}{2}\vert v_{(+)\,k}  \vert^2+\frac{1}{2}\vert v_{(-)\,k}  \vert^2 
		\end{aligned}
	\end{equation} 
	where 
	\begin{equation}
		\begin{aligned}
			\Pc_v^{\pm}= \vert v_{(\pm)\,k}\vert^2    \approx\frac{1}{2} \frac{\pi \vert \tau\vert }{2} \Bigg\vert H_{3/2}^{(1)}\LF \mp k\tau \RF\Bigg\vert^2   \LF 1 \pm \Delta \Pc_v  \RF
		\end{aligned}
	\end{equation}
	which conveys the power spectrum of quantum fluctuation $v$ has a parity antisymmetric component 
	\begin{equation}
		\Delta \Pc_v= \LF 2\epsilon+\eta\RF \operatorname{Re}\LT \frac{2}{H_{3/2}^{(1)} \LF \mp k\tau \RF} \frac{\pd H^{(1)}_{\nu_s}\LF \mp k\tau \RF}{\pd\nu_s} \Bigg\vert_{\nu_s=\frac{3}{2}} \RT
		\label{delpP}
	\end{equation}
 Note that an additional 2 factor in the above expression is attributed to the rescaling factor of $\sqrt{2}$ for every mode crossing the Horizon, as discussed in the previous section. 
	The power spectrum of the curvature perturbation can be obtained by (classical) rescaling  from Eq.~\ref{canvar}, which we evaluate at a moment of a mode $k_\ast = a_\ast H_\ast$ which exits the horizon approximately 55-60 e-folds before the end of inflation\footnote{Note that the power spectrum is by definition \cite{Lochan:2018pzs}  is a Fourier transform of two-point function evaluated at a particular time $\tau_0$ 
 \begin{equation}
     \Pc_{\tilde{\phi}} \LF k,\,\tau_0 \RF = \int \frac{d^3x}{\LF 2\pi \RF^{3/2}} e^{-i\textbf{k}\cdot \textbf{x}} G(\textbf{x},\,\tau_0)
 \end{equation}
 where $\textbf{x}= \vert \textbf{x}_1-\textbf{x}_2\vert $
 \begin{equation}
     G(\textbf{x},\,\tau_0) = \langle 0 \vert \tilde{\phi}(\textbf{x}_1,\,\tau_0  )\tilde{\phi}(\textbf{x}_2,\,\tau_0  )\vert 0\rangle 
 \end{equation}
 In our case, the two-point function is split into two parts by $\Pc\Tc$ due to the direct-sum split of the quantum field $\tilde{\phi}$
 \begin{equation}
     G(\textbf{x},\,\tau_0) =  \Theta(\tau_0)\theta(\textbf{x}) G_+(\textbf{x},\,\tau_0)+ \Theta(-\tau_0)\theta(-\textbf{x}) G_-(\textbf{x},\,\tau_0)
 \end{equation}
 which results in a power spectrum 
 \begin{equation}
      \Pc_{\tilde{\phi}} \LF k,\,\tau_0 \RF= \Theta(\tau_0)\theta(\textbf{x}) \Pc_{\tilde{\phi}_+} +\Theta(-\tau_0)\theta(-\textbf{x}) \Pc_{\tilde{\phi}_-} 
 \end{equation}
 }
	\begin{eqnarray}
			\Pc_\zeta & = & \frac{k^3}{2\pi^2} \frac{1}{2a^2\epsilon}\Bigg\vert_{\rm classical} \Pc_v \Bigg\vert_{\tau = \mp \frac{1}{a_\ast H_\ast}} 		\label{totpw} \\ \nonumber
   			& \approx &  \frac{H_\ast^2}{8\pi\epsilon_\ast}\LF \frac{k}{k_\ast} \RF^{n_s-1} \frac{1}{2} \LT 2+ \Theta(\tau)\Theta(\textbf{x}) \Delta\Pc_v\LF \frac{k}{k_\ast} \RF - \Theta(-\tau)\Theta(-\textbf{x}) \Delta\Pc_v\LF \frac{k}{k_\ast} \RF \RT 
	\end{eqnarray}
	where $A_s= 2.2\times 10^{-9}$ from the Planck data with $k_\ast = 0.05 {\rm Mpc}^{-1}$. 
	In our DSI approach developed above, the antisymmetric component of the primordial fluctuations $\Delta \Pc_v$ will result in an antisymmetric $\Delta T$ CMB component under parity transformation: $\Delta  \Tc \LF  \hat{n} \RF =- \Delta  \Tc \LF  -\hat{n} \RF$ which translates into  $\Delta a_{\ell m}$ of $\Delta \Pc_v$, as shown in Eq.~\ref{eq:Tparity}. With our method of calculation, the scale-dependent correction $\Delta \Pc_v\LF \frac{k}{k_\ast} \RF$ is accurate enough for the modes $k<k_c$ that must have already exited the horizon much before $k_\ast$ which corresponds to $55-$efoldings in our consideration. From the latest Planck data \cite{Planck:2018jri} the value of spectral index $n_s$ corresponding to the pivot scale $k_\ast$ is 
	\begin{equation}
		n_s = 1-2\epsilon_\ast-\eta_\ast \approx  0.9634 \pm 0.0048\quad \LF \rm{Planck TT+TE+EE} \RF
	\end{equation}
	which we use it to estimate $\Delta \Pc_v$ as shown in Fig.~\ref{fig:pws}. 

	\begin{table}
		\centering
		\begin{tabular}{c|c|c|c|c|c}
			\hline
			$k<k_c$ & DSI25   & DSI  & DSI50 & 15.8 percentile & 84.1 percentile \\ \hline \hline
			$\ell$ & $k_c=k_\ast/25$ & $k_c=k_\ast/40$ & $k_c=k_\ast/50$ &  \multicolumn{2}{c}{$10^6$ SMIC$_m$ realizations} \\ \hline \hline
			1 & 1.475 & 1.471 & 1.468 \\ 
			2 & 0.582 & 0.589 & 0.594 & 0.473 & 1.824\\ \hline
			3 & 1.385 & 1.373 & 1.363 & 0.549 & 1.637\\ 
			4 & 0.641 & 0.657 & 0.667 & 0.593 & 1.545 \\ \hline
			5 & 1.339 & 1.320 & 1.305 & 0.634 & 1.470\\
			6 & 0.678 & 0.705 & 0.725 & 0.649 & 1.455\\ \hline
			7 & 1.305 & 1.278 & 1.258 & 0.674 & 1.412 \\ 
			8 & 0.708 & 0.743 & 0.770 & 0.687 & 1.395\\ \hline
			9 & 1.277 & 1.235 & 1.199 & 0.705 & 1.365\\ 
			10 & 0.734 & 0.776 & 0.814 & 0.714 & 1.355 \\ \hline
			11 & 1.254 & 1.203 & 1.168 & 0.727 & 1.331\\
			12 & 0.759 & 0.824 & 0.875 & 0.734 & 1.324\\ \hline
			13 & 1.233 & 1.165 & 1.073 & 0.745 & 1.309\\ 
			14 & 0.781 & 0.845 & 0.967 & 0.750 & 1.301\\ \hline
			15 & 1.208 & 1.126 & 1.012 & 0.759 & 1.288\\ 
			16 & 0.799 & 0.916 & 0.996 & 0.759 & 1.288 \\ \hline
			17 & 1.188 & 1.045 & 1.001 & 0.759 & 1.288\\ 
			18 & 0.825 & 0.980 & 1.000 & 0.759 & 1.288\\ \hline
			19 & 1.169 & 1.007 & 1.000 & 0.780 & 1.259\\ 
			20 & 0.840 & 0.998 & 1.000 & 0.783 & 1.256\\ \hline
			21 & 1.143 & 1.001 & 1.000 & 0.783 & 1.256 \\ 
			22 & 0.871 & 1.000 & 1.000 & 0.792 & 1.244\\ \hline
			23 & 1.124 & 1.000 & 1.000 & 0.797 & 1.237\\ 
			24 & 0.882 & 1.000 & 1.000 & 0.799 & 1.234\\ \hline
			25 & 1.102 & 1.000 & 1.000 & 0.799 & 1.234\\
			26 & 0.925 & 1.000 & 1.000 & 0.799 & 1.234\\ \hline
			27 & 1.047 & 1.000 & 1.000 & 0.799 & 1.234\\
			28 & 0.974 & 1.000 & 1.000 & 0.811 & 1.219\\ \hline
			29 & 1.012 & 1.000 & 1.000 & 0.814 & 1.214\\ 
			30 & 0.995 & 1.000 & 1.000 & 0.828 & 1.229\\\hline
			31 & 1.002 & 1.000 & 1.000 & 0.841 & 1.239\\
			32 & 0.999 & 1.000 & 1.000 & 0.779 & 1.144\\ \hline
			33 & 1.000 & 1.000 & 1.000 & 0.809 & 1.180\\
		\end{tabular}
		\caption{Values for $1\pm \Delta \Cc_\ell$ in Eq.~\ref{eq:Cl_DQFT} for the DSI model with different approximations for the cut-off $k_c$. The last two columns give the percentiles 15.8 and 84.1 (corresponding to $\pm 1\sigma$ range for a Gaussian distribution) in the relative $C_\ell$ measurements over $10^6$ sampling variance realizations of the SMIC$_m$ Planck CMB maps. The corresponding simulated outputs are displayed in Fig.~\ref{fig:clLCDM} and Fig.~\ref{fig:clfit}. The significance of odd parity in the first individual multipoles is only about $1\sigma$, but each multipole is approximately independent, which means that such odd parity can potentially be measured with a large significance within $\ell<30$.}
		\label{tab:DSI}
	\end{table}

	Using Eq.~\ref{Cllstan} we obtain the temperature angular power spectrum for the even and odd multipoles as     
	\begin{equation}
		\begin{aligned}
			&   C_{\ell}^{odd}  =  \frac{2}{9\pi} \int_{0}^{k_c} \frac{dk}{k}j_{\ell}^2\LF \frac{k}{k_s} \RF
			\frac{H^2}{8\pi\epsilon}\LF 1+\Delta \Pc_v \RF 
			\\
			& C_{\ell}^{even}  = \frac{2}{9\pi}\int_{0}^{k_c} \frac{dk}{k} j_{\ell}^2\LF \frac{k}{k_s} \RF\frac{H^2}{8\pi\epsilon}\LF 1-\Delta \Pc_v \RF  
		\end{aligned}
	\end{equation}
	
	In summary, according to DQFT, quantum fluctuations evolve forward and backward in time in the two parity conjugate regions. Although the total power spectrum of the fluctuation matches with the standard (near) scale-invariant power spectrum, the antisymmetric component Eq.~\ref{delpP} leads to an asymmetry in power in the parity conjugate regions. This is pictorially depicted in Fig.~\ref{fig:mirror}.
	
	\subsection{The power spectrum of DSI and the first modes to exit  horizon}

\label{subsec:StoacIn}
 
	At first glance, the power spectrum of DSI Eq.~\ref{totpw} may appear to be very simple but the physics of it is rather profound. Recalling once again the concept that in DSI a quantum fluctuation is generated as a direct-sum of two opposite time-evolving components at the parity pair of points in physical space. The total power spectrum  Eq.~\ref{totpw} is the average of power at the parity conjugate pair of points in the homogeneous and isotropic spacetime. In other words, in DSI the primordial power is distributed unequally into the parity conjugate points in physical space, and when we take the average we recover exactly the power-law form of the power spectrum which has been the best-fit description of the CMB excluding the low-$\ell$ anomalies. 
	
	The antisymmetric component of the power spectrum Eq.~\ref{delpP} is evaluated for the modes of interest $k\lesssim 0.001\, \rm{Mpc}^{-1}$ which are the ones to exit the horizon much before the mode $k= 0.05 \,\rm{Mpc}^{-1}$. The reason we impose $k\lesssim 0.001\, \rm{Mpc}^{-1}$ for the antisymmetric component of the power spectrum is that the modes do not immediately become frozen after the horizon exit. From now on we refer to the results of SI in the literature whose conclusions are generically valid for DSI as well. The time evolution of curvature perturbation is determined by the following equation 
	\begin{equation}
		\zeta_k^{\prime\prime}+2\frac{z^\prime}{z}\zeta_k+k^2\zeta_k =0\,. 
		\label{curpeq}
	\end{equation}
	where $z= a\frac{\dot{\phi}}{H}$ which is the classical rescaling factor that relates the canonical variable and curvature perturbation Eq.~\ref{canvar}. At the horizon exit i.e., when $k\sim aH$ the decaying component of the curvature perturbation given below takes a few e-folds (about 3 to 4) after which the constant component dominates \cite{Polarski:1995jg,Julien} 
	\begin{equation}
		\zeta_k = G_k + F_k \int \frac{dt}{a^3\epsilon}
		\label{cons-decay}
	\end{equation}
	where $G_k,\, F_k$ are the $k$-dependent constants. The solution for $\zeta_k$ Eq.~\ref{cons-decay} can be obtained from solving Eq.~\ref{curpeq} in the limit $k\to 0$. Following the numerical evaluations of Eq.~\ref{MSeq} and the corresponding curvature perturbation in SI, one can reach sufficient accuracy in evaluating the primordial power spectrum by evolving each mode at least a factor of 50 times before the mode exits the horizon \cite{Julien}. Alternatively, we determine the power spectrum of $k$ at a moment of $aH$ at least 50 times larger. This is exactly the upper cut-off we used in our calculation of the angular power spectrum and the antisymmetric component in DSI. Moreover, the scales $10^{-4} \, \rm{Mpc}^{-1} < k \lesssim 0.001\, \rm{Mpc}^{-1}$ corresponds to approximately $\ell = 2-30 $ where CMB anomalies are centered in. These are the scales $k< 0.02k_\ast$ where the pivot scale $k_\ast$ corresponds to $\ell=200$. 
	Therefore, for the investigation carried out in this paper, we sufficiently computed the primordial power spectrum of DSI to the particular modes of interest (that have exited the horizon much before the onset of inflation corresponding to the pivot scale $k_\ast$) to confront against the latest CMB data. From the point of view of DSI, the parity asymmetry of the power spectrum for the high-energy modes is expected to become small because the effects of curved spacetime on quantum fields should cease to exist as we go to high-energy modes. This is exactly what we can observe in Fig.~\ref{fig:pws} which is accurate up to an order of magnitude smaller than the pivot scale as the remaining modes would be in the process of quantum-to-classical transition impacted by the modes that have already left the horizon. 
	
	The quantum-to-classical transition of inflationary fluctuations is the whole subject of separate investigation \cite{Kiefer:1998qe,Sudarsky:2009za,Landau:2011ljv}. Inflationary fluctuations once they become classical would leave their imprint as in-homogeneities and anisotropies which we can probe now using CMB and primordial gravitational waves. The first scalar modes that leave the horizon at the beginning of inflation would leave large-scale in-homogeneities which affect the modes that leave at a later stage. This is because inflationary quantum fluctuations leaving the horizon and evolving into classical ones is a non-Markovian process \cite{Cruces:2022imf}. 
	To accurately capture the correlations of all modes, it is more appropriate to employ the scheme of Stochastic inflation \cite{Starobinsky:1983zz} where the scalar field fluctuations are split into the long and short wavelength modes as
	\begin{equation}
		\phi\LF N,\,\textbf{x} \RF =\phi_{cg}\LF N,\,\textbf{x} \RF+ \delta\hat{\phi}\LF N,\, \textbf{x} \RF
		\label{ssplit}
	\end{equation}
	where $\phi_{cg}\LF \tau,\,\textbf{x} \RF$ is the coarse-grained field (non-expanding) at a fixed scale that contains the classical modes that have already left the horizon long before the later modes. This contribution already leaves in-homogeneities within the Hubble patch which determine the short-wavelength modes which are called the Stochastic noise or white noise described as
	\begin{equation}
		\delta\hat{\phi} = \frac{1}{\LF 2\pi \RF^{3/2}}\int d^3k \,\Theta(k-\sigma aH)  \LT\hat{s} \phi_k\LF \tau \RF e^{-i\textbf{k}\cdot\textbf{x}}+ \hat{s}^\dagger \phi_k^\ast\LF \tau \RF e^{i\textbf{k}\cdot \textbf{x}}\RT\,,
		\label{stocp}
	\end{equation} 
	where $\sigma\ll 1$ determine coarse-graining modes. The annihilation and creation operators characterize the modes that are shorter in wavelength than the coarse-graining modes. The splitting of the field into classical and quantum/stochastic parts Eq.~\ref{ssplit} in Stochastic formalism gives the effective quantum dynamics of long-wavelength modes. In Eq.~\ref{stocp} is the most popular way to introduce a cut-off between infrared (IR) long wavelength and UV short wavelength modes, but instead of a sharp cut-off more realistic smooth functions have been studied recently \cite{Mahbub:2022osb}.
	As inflation proceeds contribution to the coarse-grained field increases due to the continuous \say{quantum} kicks from the short-wavelength modes that later evolve to be classical. This is called the quantum backreaction which has been explored in the context of primordial black holes during inflation \cite{Pattison:2017mbe,Pattison:2019hef}.
	The stochastic framework is to study inflationary fluctuations in a non-perturbative manner and the technique in several instances has been proven to be powerful and match the predictions of standard QFT methods \cite{Finelli:2001bn,Tsamis:2005hd,Finelli:2011gd,Finelli:2008zg}. The free parameter $\sigma\ll 1$ is used for gradient expansion of the perturbed metric of the long wavelength modes in the Stochastic framework. 
	
	Having the long wavelength modes becoming classical, the Universe will only be locally homogeneous and isotropic with the following local scale factor and the Hubble parameter which are estimated to the first order in gradient expansion ($\sigma$) of the perturbed quantities \cite{Cruces:2022imf}
	\begin{equation}
		a_\ell \approx a(t)e^{-\int \pd_i \Nc_{cg}^i dt },\quad H_\ell \approx \frac{H}{\Nc_{cg}}-\frac{1}{3\Nc_{cg}}\pd_i\Nc_{cg}^i\,.  
	\end{equation}
	Once we split the perturbed quantities into the IR and UV parts by the characteristic scale $k_\sigma = \sigma aH$
	\begin{equation}
		\zeta_k = \zeta_{k<k_\sigma}+\zeta_{k>k_\sigma}\,, 
	\end{equation}
	the Lapse and Shift functions $\Nc,\, \Nc_i$ are also split into IR and UV parts. In the spatially flat gauge, the curvature perturbation is proportional to the inflaton field fluctuation as 
	\begin{equation}
		\zeta = -\frac{H}{\dot\phi} \delta\phi\,, 
	\end{equation}
	and the Lapse and Shift are given by 
	\begin{equation}
		\Nc \approx 1+\sqrt{\frac{\epsilon}{2}}\delta\phi,\quad \pd_i\Nc^i \approx -aH^2\LF \frac{\pd\delta\phi}{\pd N}-\frac{\eta}{2}\delta\phi \RF 
	\end{equation}
	The non-Markovian nature of inflationary quantum fluctuations can be seen by the Stochastic differential equations (in the units of $M_p=1$) given by 
	\begin{equation}
		\begin{aligned}
			& \pi_{cg} = \frac{\pd \phi_{cg}}{\pd N}+\xi_1 \\ 
			& \frac{\pd \pi_{cg}}{\pd N}+\LF 3-\frac{\pi_{cg}^2}{2} \RF \LF \pi_{cg} + \frac{V_{,\,\phi_{cg}}\LF \phi_{cg}\RF}{V} \RF = -\xi_2\,.
		\end{aligned}
		\label{non-markoeq}
	\end{equation}
	where 
	\begin{equation}
		\begin{aligned}
			\xi_1 & = -\frac{\pd}{\pd N}\LF \frac{\sigma aH}{\Nc_{cg}} \RF \int \frac{d^3k}{\LF 2\pi \RF^{3/2}} \delta\LF k-\frac{\sigma aH}{\Nc_{cg}} \delta\hat{\phi}_k \RF  \\ 
			\xi_2 & = -\frac{\pd}{\pd N}\LF \frac{\sigma aH}{\Nc_{cg}} \RF \int \frac{d^3k}{\LF 2\pi \RF^{3/2}} \delta\LF k-\frac{\sigma aH}{\Nc_{cg}} \frac{\pd \delta\hat{\phi}_k}{\pd N} \RF
		\end{aligned}
	\end{equation}
 	
	\begin{figure}
		\centering
		\includegraphics[width=0.6\linewidth]{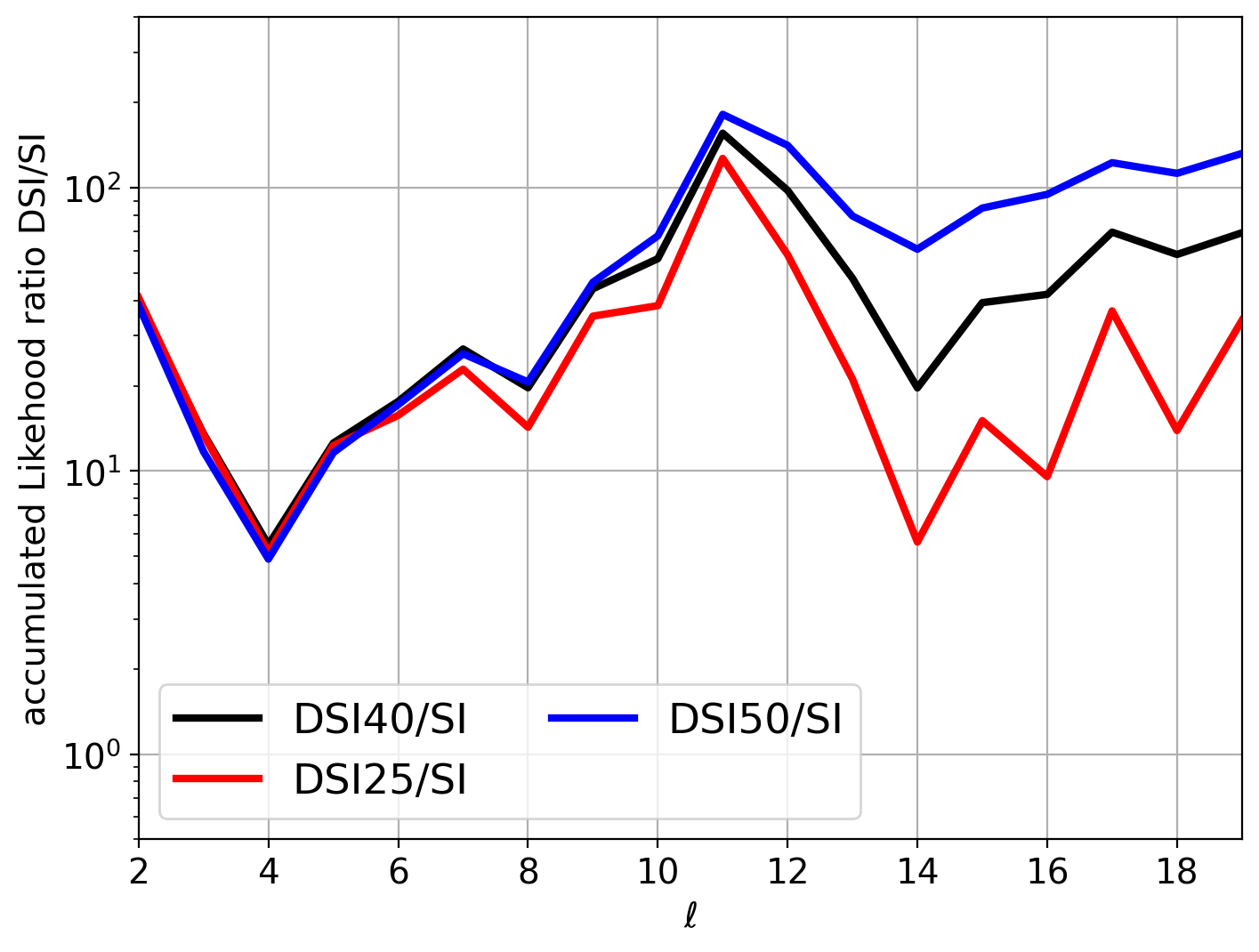}
		\caption{Cumulative likelihood ratio between DSI and SI predictions compared to the CMB data (as a function of the multipole $\ell$). The differnt colors correspond to different values of the cutoff:
			$\sigma=0.02$ (DSI50, blue) , $0.025$  (DSI40, black) and $0.04$  (DSI25, red).}
		\label{fig:LikeHood}
	\end{figure}
	In the Starobinsky's formulation \cite{Starobinsky:1986fx,Starobinsky:1994bd} the 
	the dynamics of the field fluctuations in Stochastic formalism become Markovian and they are determined by the Langevin equation \cite{Starobinsky:1994bd}
	\begin{equation}
		\frac{d\phi}{d N} = -\frac{1}{3H^2} \frac{dV}{d\phi} +\sigma aH\hat{\phi}\LF N,\,\textbf{x}\RF\,. 
	\end{equation}
	Solving the above Langevin equation even numerically is a very non-trivial tast and this was recently achieved within and also beyond the slow-roll approaximations using the so-called stochastic $\delta N$ formalism (See \cite{Fujita:2014tja,Vennin:2015hra,Pattison:2019hef,Mishra:2023lhe,Jackson:2022unc} and references therein). However, 
it is the non-Markovian system of equations Eq.~\ref{non-markoeq} one needs to adopt and solve for DSI to determine the parity asymmetries in the small angular scales or large multipoles. This exercise is entirely non-trivial and is beyond the scope of the present paper. 
Furthermore, the Parity asymmetries are only significant at large angular scales or small multipoles, therefore, we only look at the infrared modes or course graining modes $k\ll k_\ast$ with the value of $\sigma = 0.02  $. 
	In Table \ref{tab:DSI} and 
	Fig.~\ref{fig:LikeHood} we show how the value of $\sigma$ changes the DSI estimates for different values $\sigma= k_c/k_\ast=0.02, 0.025, 0.04$. The results only change slightly in this range, but data seems to prefer
	$\sigma=0.02$, as expected from the above discussion. Therefore, for the first time, we exactly specify what should be the value of $\sigma$ which is most often treated as a free parameter in the stochastic inflation literature \cite{Mishra:2023lhe,Fujita:2014tja}.  Furthermore, 
	this is an important input to further develop the framework of the stochastic formalism of inflationary fluctuations together with QFT in the inflationary spacetime.

	\section{Conclusions and outlook}

 \label{sec:concl}

 		CMB measurements starting from COBE to the latest Planck data have been consistently telling us about the anomalies \cite{Schwarz:2015cma} at lower multi-poles ($\ell< 30$) or large angular scales $\theta>6^\circ$. The well-celebrated fit of near scale-invariant power spectrum Eq.~\ref{HZpower} together with LCDM is perfect only for the physics of angular scales $\theta<1^\circ$ or for $\ell>200$. Numerous investigations have been carried out in the last decades with WMAP and Planck data sets to characterize and understand the low-$\ell$ angular power spectrum.  The notable challenges have been the low-quadrupole ($C_2$), there is more power observed in the odd-multipoles rather than in even-multipoles, the anti-correlation between two-point temperature correlation at the anti-podes of the CMB (i.e., at $\theta\simeq 180^\circ$) and the apparent lack of power at $\theta>60^\circ$. The common origin of these anomalies has been the major concern for theoretical and observational cosmologists. All these anomalies point to one aspect, which is parity, which is a discrete symmetry by nature and has nothing to do with the statistically isotropic feature of CMB. Therefore, parity does not seem to conflict with the cosmological principle. By meticulously analyzing the CMB map in the configuration space, we have unveiled a remarkable fact that CMB is parity asymmetric by a significant preference for odd parity. This can be seen by the naked eye in Fig.~\ref{fig:smicaSA}, where the total CMB map (A+S) is very similar to its antisymmetric counterpart (A). This tells us there are large-scale temperature fluctuations in the CMB (which also means there are large-scale structures in the Universe) that not only look flipped in shape by parity but also asymmetric in amplitude. This picture conveys a clear parity asymmetric feature of CMB and is calling our attention to better understand the (quantum) physics associated with the early Universe. This is because discrete (a)symmetries most likely must be related to quantum physics rather than classical physics. History is teaching us about it; it is the understanding and evidence for discrete (a)symmetries such as charge conjugation ($\Cc$), Parity ($\Pc$), and time reversal ($\Tc$) and their combinations $\Cc\Pc$ and $\Cc\Pc\Tc$ have helped us to build quantum field theory and the standard model of particle physics. 
	
Analyzing CMB data is the most tricky aspect, especially since it requires a guiding cue, which is the most plausible theory with the least number of free parameters. In the context of early Universe cosmology, the most plausible theory is advocated to be the near-scale invariant primordial power spectrum Eq.~\ref{HZpower} together with the LCDM. The power spectrum Eq.~\ref{HZpower} is widely advocated to be the robust prediction from quantum fluctuations in standard inflation (SI). Since SI+LCDM is the only model with the least parameters, low-$\ell$ CMB data is studied with simulations and sampling variance calculated from SI+LCDM. Naturally, if SI+LCDM is not the right model to address the low-$\ell$ anomalies of CMB, it is not correct to consider sampling variance from the model. Studies in the past have strongly suggested the inconsistency of SI+LCDM at large angular scales, and in this paper, we show that the rather remarkable parity asymmetric feature of CMB is quite at odds with SI.  
Here in this paper, we study a new theory of inflationary quantum fluctuations called Direct-Sum Inflation (DSI) (See Sec.~\ref{sec:theory}) where quantum fluctuations have parity asymmetric time evolution (different time evolution at parity conjugate points in physical space) during inflation. This theory is developed based on an observation that time reversal symmetry would be broken during inflation, and corresponding parity should also be broken spontaneously. This framework strongly resonates with what we see in the CMB data, i.e., a strong preference for odd parity (See Sec.~\ref{sec:priseed} and Fig.~\ref{fig:mirror}). 

Given that we have two theories of inflationary quantum fluctuations, SI and DSI, with an equal number of parameters, we can directly compare them. We carefully look if they are or not consistent with
the low-$\ell$ parity asymmetric CMB observables such as: the mean and skewness of the Z-parity (Eqs. ~\ref{eq:z1}-\ref{eq:skewness}), the even-odd asymmetry of power in harmonic space $R^{TT}$ (Eq.~\ref{eq:RTT}) and the low-quadrupole $C_2$. We perform our analysis in two ways, one is we do data simulations (up to $10^6$ realizations of CMB sky), and ask the question given the data how probable DSI is in comparison with SI to explain the data. We find robust answers to this question, which are summarized in Table.~\ref{tab1:annomalies} and Figs.~\ref{fig:clfit}-\ref{fig:corner}. We also analyze the data by performing the simulations of SI and DSI models (where we consider the cosmic variance from each model rather than the data). The results of this second approach are depicted in Table.~\ref{tab2:annomalies}. In the first approach, we found the DSI is 653 times more probable than SI in explaining the low-quadrupole and $R^{TT}$ given the data. With the Z-parity statistics in configuration space, DSI becomes 54 times more probable than SI, and with both Z-parity and $C_2$, DSI is 166 times more favored than SI. In the second approach of finding how probable the models match with data, we still find DSI is 47-55 times more probable than SI. In the whole analysis, the $p-$ values of SI happen to be only $\Oc\LF 10^{-4}-10^{-3} \RF$ whereas the DSI stands out to be consistent with data with p-values of $\Oc\LF 2-50\% \RF$. Another notable point here is that the cosmic variance of DSI theory is much closer to the data in comparison with SI, which again strongly indicates the power of parity asymmetric inflationary quantum fluctuations. We have found that the lower quadrupole can explain another anomaly associated $S_{1/2}$ (which indicates the lack of correlations in the very large scales $\theta>60^\circ$) by a factor of 122 times the p-value found in the SI model. Interestingly the DSI does predict a lower value for quadrupole and it does naturally explain the $S_{1/2}$ anomaly as we can see in Fig.~\ref{fig:Ctheta}.
All of this makes the DSI+LCDM the Planck data's best-fit theory as we can witness in our Fig.~\ref{fig:corner}. 

We revisited one of the most discussed anomalies called HPA and its related observables, such as the amplitude of DPM (defined in Eq.~\ref{eq:DPM}) and $\sigma_{16}^2$, which is the variance at $N_{side}=16$. This anomaly has been advocated to challenge the cosmological principle by showcasing evidence for statistical anisotropy in the direction of the ecliptic plane (the direction our solar system is oriented) that suggests our place in the Universe as a special one. HPA has been shown as a significant anomaly in the literature by doing isotropic Gaussian simulations of SI+LCDM in a specific direction in each realization. As this introduces a bias, we instead ran a million SI+LCDM simulations and looked for HPA in a thousand directions in each realization. Thus, we found evidence that HPA has reduced by the p-value of 2.5\%, which turned out to be close to the p-value of quadrupolar power in SI (See Table.~\ref{tab3:annomalies}). This establishes a direct connection between HPA and lower quadrupole. To test this, we studied the SI+LCDM model by fixing the quadrupole as low as the data, and this has dramatically made the HPA anomaly statistically very insignificant with a p-value of 15\%. Furthermore, we have also shown that any dipolar modulation in CMB does not affect the parity measurements such as Z-parity and $R^{TT}$, which proves parity and isotropy are totally different.  

Our study only concentrated on parity asymmetric CMB at low-$\ell$ because this is where the major anomalies have been reported from WMAP to Planck. From the mechanism of inflationary quantum fluctuations in DSI, we should expect parity asymmetry at all scales, but there are two limitations to explore in the present paper. The CMB data is not that accurate to tell us about parity asymmetry at smaller scales (large $\ell$). Since the inflationary quantum fluctuations becoming classical is a non-Markovian process, computing parity asymmetry for the smaller scales is completely non-trivial, and it requires employing the stochastic formalism and high numerical analysis as we discussed in Sec.~\ref{sec:theory}. We apply a cut-off in wavenumber of inflationary quantum fluctuations $k_c=0.02 k_\ast$, which is the scale that affects the angular power spectrum at low-$\ell$. This determines the so-called coarse gaining modes in stochastic formalism. We greatly expect this will help further develop the framework and, correspondingly, the quantum field theory in curved spacetime. 

In the coming years, there will be Stage 4 CMB observations, which are expected to observe more precisely the polarization of the CMB photons \cite{Abazajian:2016yjj}. This will add complementary parity information to the $C_\ell$ measurements at low-$\ell$, providing a further test to DSI. Furthermore, DSI also predicts parity asymmetry for primordial gravitational wave background \cite{Kumar:2022zff}, which can be extracted from future possible detection of CMB B-modes. 
The DSI model also predicts that the primordial dipole $C_1$ at about 50\% larger than SI (see Table \ref{tab:DSI}). Such predicted primordial dipole cannot be easily separated from the kinematic dipole. This translates into a $\simeq 3$Km/s systematic uncertainty in establishing the local motion of comoving observers.  This corresponds to $\Delta H_0 \simeq 3$Km/s/Mpc for the distance ladder calibration \cite{2021ApJ...908L...6R} of $H_0$ at 1 Mpc distance. Further work is needed to address the possible implications of such uncertainty. Given that the Hubble tension of late-time cosmology has been gaining significant attention in recent years \cite{DiValentino:2021izs}, it indeed be worth investigating the effect of DSI's primordial dipole in the local expansion measurements.  

Finding a signature for the quantum mechanical nature of gravity (i.e., quantum gravity) is vital to understanding early Universe and Black Hole physics and building a consistent ultraviolet complete theory of quantum gravity towards the Planck scales. What else would be the most important signature for quantum gravity than a parity asymmetric CMB emerging from primordial quantum fluctuations in the early Universe? The Direct-Sum-Inflation resolves the conundrum between gravity and quantum mechanics by separating the concept of time in GR with the concept of time for quantum fluctuations in curved spacetime. In the DSI, the clock in the expanding FLRW Universe would be the scale factor, and the quantum-mechanical (parametric) time allows the quantum fields to respect the (a)symmetries of background spacetime. The evidence we found for DSI in this paper is only the first step in further exploring the role of gravity and quantum mechanics at the spacetime horizons.\footnote{In this regard, it would be worth exploring what would be the observational implications of direct-sum QFT that is recently proposed in the context of Black-Hole physics \cite{Kumar:2023hbj}. One can ask interesting questions such as: Is Hawking radiation parity asymmetric similar to what we witness for CMB in this paper? Is the Hawking radiation non-Markovian in nature? An answer to which allows us to extract information from Black Holes.}

\acknowledgments

EG acknowledges grants from Spain Plan Nacional (PGC2018-102021-B-100) and Maria de Maeztu (CEX2020-001058-M). KSK would like to thank the Royal Society
for the Newton International Fellowship.
We thank David Bacon, Benjamin Camacho-Quevedo, Andrew Gow, Mathew Hull,  Jo\~ao Marto, Alexei A. Starobinsky, David Wands, Eiichiro Komatsu and Kazuya Koyama for useful discussions and encouragement.

	\newpage
	
	\appendix
	
	\section{Isotropy and parity}
	\label{sec:isotropy}

\begin{figure}
		\centering
		\includegraphics[width=0.7\linewidth]{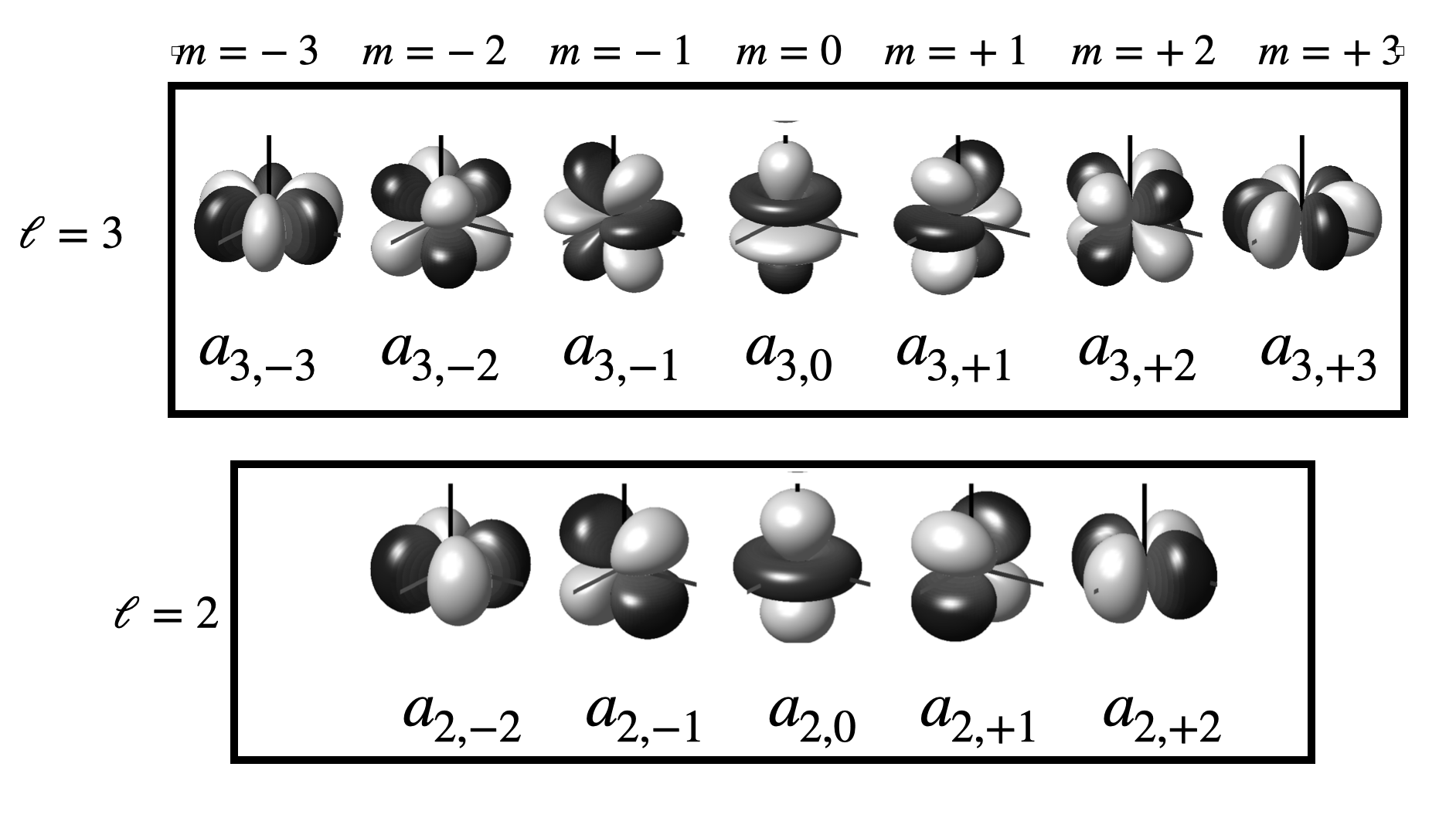}
		\caption{Representation of Spherical harmonic functions $Y_{\ell m}$ with the corresponding $a_{\ell m}$ coefficients for the quadrupole ($\ell=2$) and octopole ($\ell=3$). Note the even (odd) spherical $\Pc$ conjugate parity of $\ell=2$ ($\ell=3$).}
		\label{fig:alm}
	\end{figure}

CMB temperature sky fluctuations $\Delta T$ within an isotropic  background of mean temperature $T_0$ can be characterized by the local (configuration space) scalar field  $\Tc(\hat{n})$ or its Fourier counterpart, which in spherical coordinates is the harmonic $a_{\ell m}$ decomposition:
	\cite{Durrer:2008eom}
	\begin{equation}
		\Tc\LF \hat{n} \RF	\equiv \frac{\Delta T\LF \hat{n} \RF}{T_0} = \sum a_{\ell m} Y_{\ell m}\LF \hat{n} \RF \quad ; \quad
		a_{\ell m} = \int d\Omega \, Y_{\ell m}^\ast\LF \hat{n} \RF \Tc\LF \hat{n} \RF\,.
		\label{eq:alm}
	\end{equation}
	where the multipole $\ell=0,1,\dots,\ell_{max}$  corresponds to the inverse angular separation $\theta$ of each mode: $\ell \simeq \pi/\theta$ (e.g. $\ell_{max} \sim \pi/\theta_{min}$ corresponds to the map resolution $\theta_{min}$) and $m=-\ell \dots \ell$ is the direction 
	$\hat{\theta}$ of each mode (see Fig.~\ref{fig:alm} for an illustration). 
	In the flat sky limit (i.e. small angles) we have that the 2D Fourier modes are $\hat{k} = \ell e^{i \pi m/\ell}$. The CMB angular TT power spectrum is usually defined as:
	\begin{equation}
		C_{\ell} = \frac{1}{2\ell+1} \sum_m \vert a_{\ell m}\vert^2\,.
		\label{eq:cl}
	\end{equation}
	but often displayed as $D_\ell \equiv \ell(\ell+1) \, C_\ell/2\pi$, corresponding to the variance in each mode $\ell$.
	
	A key property of Eq.~\ref{eq:alm} is rotation invariance: the values of   $C_{\ell}$ are the same for any rotation of the coordinate system. This is not the case for parity transformations.  This vital distinction is the key element of this paper, as well as understanding the differences between a possible break of parity and a violation of isotropy (or the so-called Cosmological Principle). Both symmetries are independent. We, therefore, need to check if they are broken or conserved separately.
 	
	\section{Making  parity-conjugate CMB maps with Healpix}
	\label{sec:mirror}

\begin{figure}
		\centering
		\includegraphics[width=0.49\linewidth]{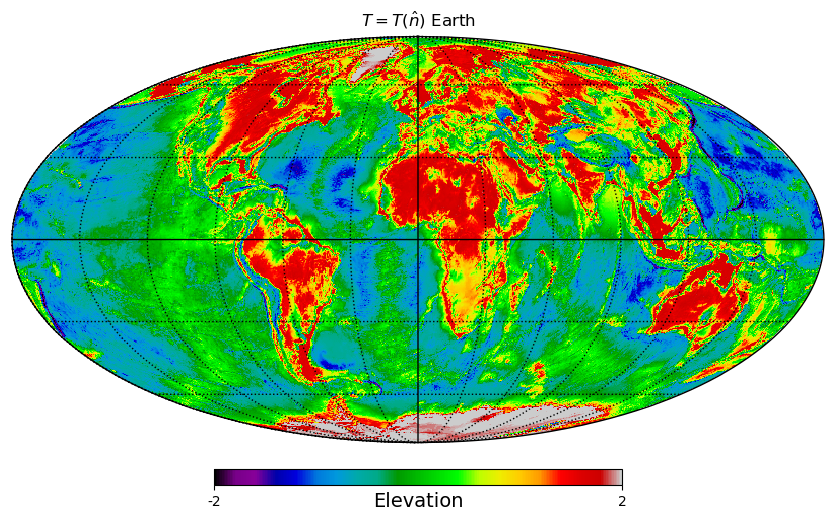}
		\includegraphics[width=0.49\linewidth]{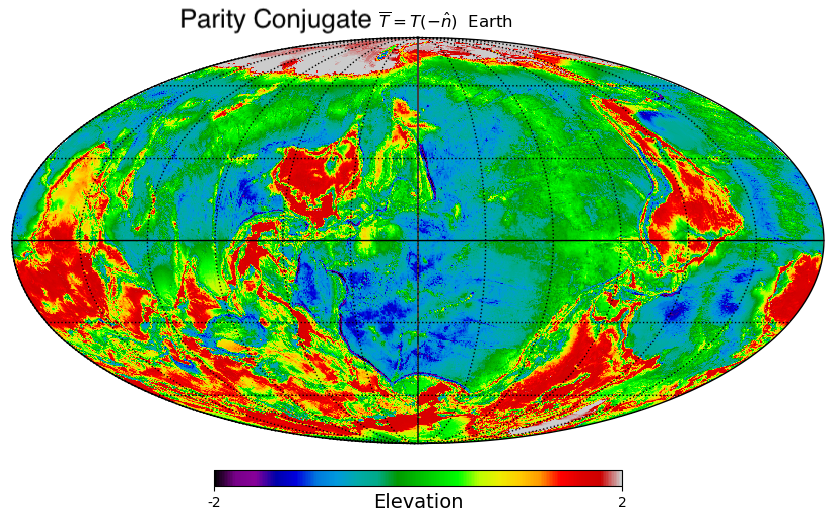} 
		\vskip -0.1cm
		\includegraphics[width=0.49\linewidth]{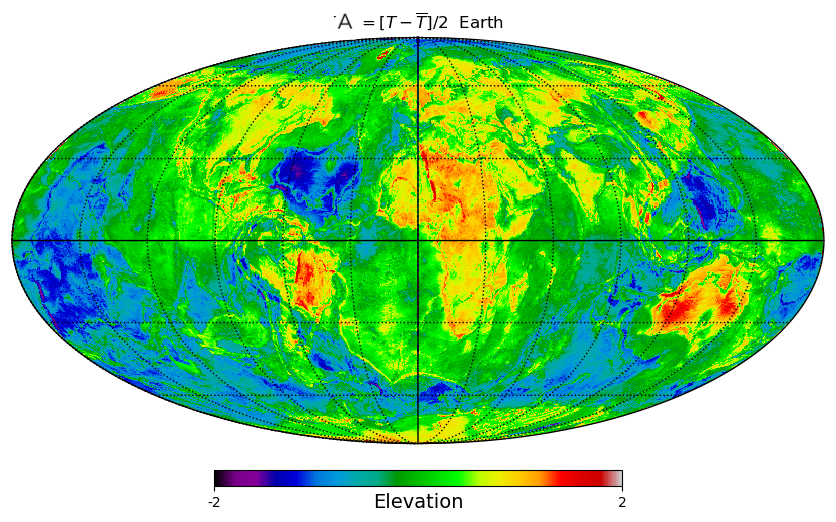}
		\includegraphics[width=0.49\linewidth]{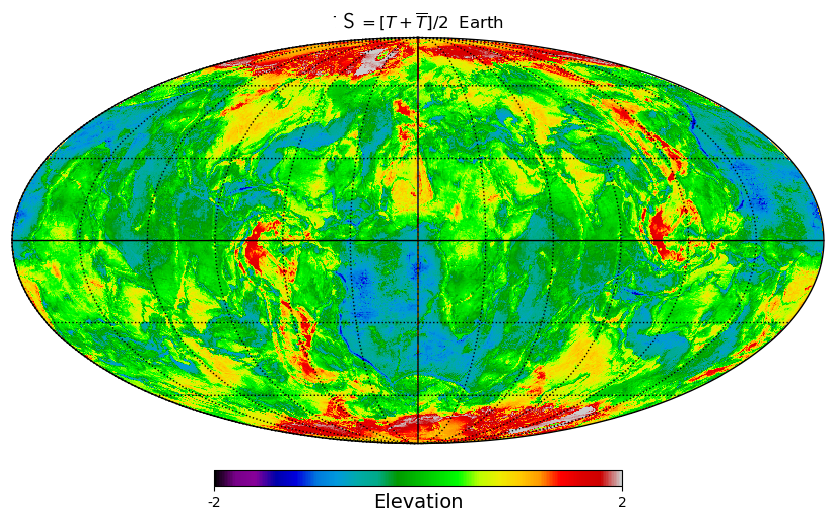}
		\vskip -0.1cm
		\includegraphics[width=0.49\linewidth]{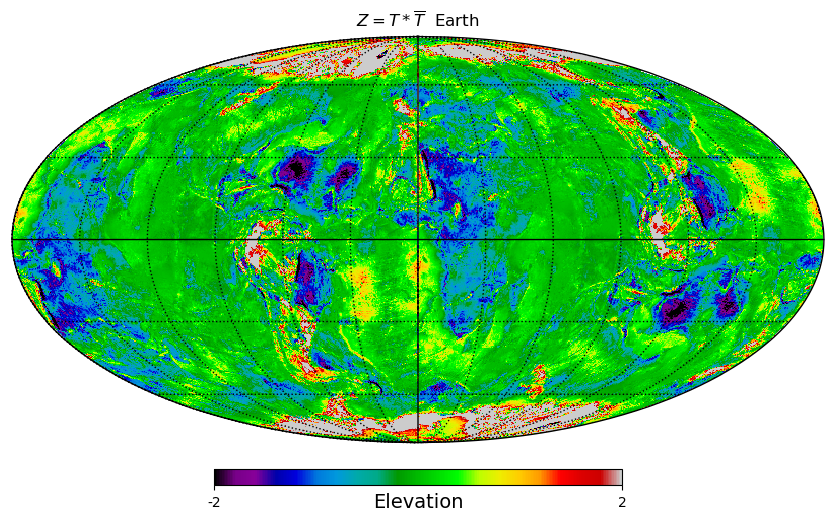}
		\includegraphics[width=0.49\linewidth]{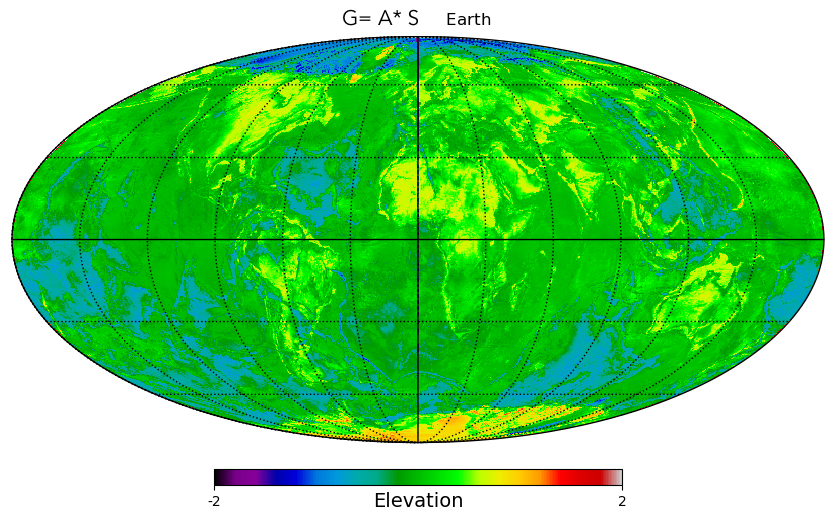}
		\caption{
			In these Mollview projections, we compare the antisymmetric ($A$ odd parity, middle left) and symmetric ($S$ even parity, middle right) decomposition of the Earth elevation map (top left). The Earth $\Pc$ conjugate image $\overline{T}(\hat{n})=T(-\hat{n})$ (top right) indicates where to find the antipodes of the original map. The odd parity map shows the locations where the antipodes of a given elevated continent match a deep ocean depression (like Australia), while the even parity matches regions with similar elevation in the antipodes (like the polar caps or Europe-New Zeland). This can also be seen in the $Z=T *\overline{T}$ map which shows blue for odd and red for even continents.
			The cross-map $G=A*S$ is predominantly noise close to zero, except in the polar caps.}
		\label{fig:smicaParityEarth}
	\end{figure}

	In the  {\sc HEALPX}  ({\sc hp}) format a map is just an ordered 1D array of ($\Delta T$) numbers {\sc map[pix]} where {\sc pix} is an index array that 
	runs from $0$ to {\sc Npix-1}. The number of pixels {\sc Npix} depends only on the map resolution {\sc NSIDE}: 
	{\sc  Npix = 12 NSIDE*NSIDE}. 
	The angular spherical coordinates {\sc (t,p)}$=(\theta,\varphi)$ of each pixel is just given by the order in the index array {\sc pix} with a transformation (i.e. the  {\sc HEALPX} pixelization scheme) that only depends on {\sc NSIDE}:  {\sc t, p = hp.pix2ang(nside,pix)}. 
	To apply a parity transformation on a sphere with {\sc HEALPX} we follow 2 simple steps:
	
	\begin{enumerate}

		\item Apply the parity transformation $(\theta,\varphi) \implies (\pi-\theta,\varphi+\pi)$ and get the corresponding pixel $\Pc$ conjugate array {\sc pix\_mirror}: \\
   {\sc t, p = hp.pix2ang(nside,pix)} \\
		{\sc pix\_mirror = hp.ang2pix(nside, $\pi$- t, p + $\pi$})
		
		\item The spherical $\Pc$ conjugate map, {\sc map\_mirror}, is then just given by: \\
		
		{\sc map\_mirror[pix\_mirror] = $\pm$ map[pix]}
		
	\end{enumerate}
	where the $\pm$ signs correspond to the odd and even parities. To illustrate this transformation with a familiar image we show in  
	Fig.~\ref{fig:smicaParityEarth} the different parity transformations of the Earth elevation map\footnote{\url{https://www.ncei.noaa.gov/products/etopo-global-relief-model}.}. The maps show that Earth's elevation shows a preferred odd parity, at least at the level of large continents, except the polar caps, which are even.
	
	\section{Sampling variance and Non-Gaussian distribution of errors}
	\label{sec:harmonicparity}
	
\begin{figure}
		\centering
		\includegraphics[width=0.49\linewidth]{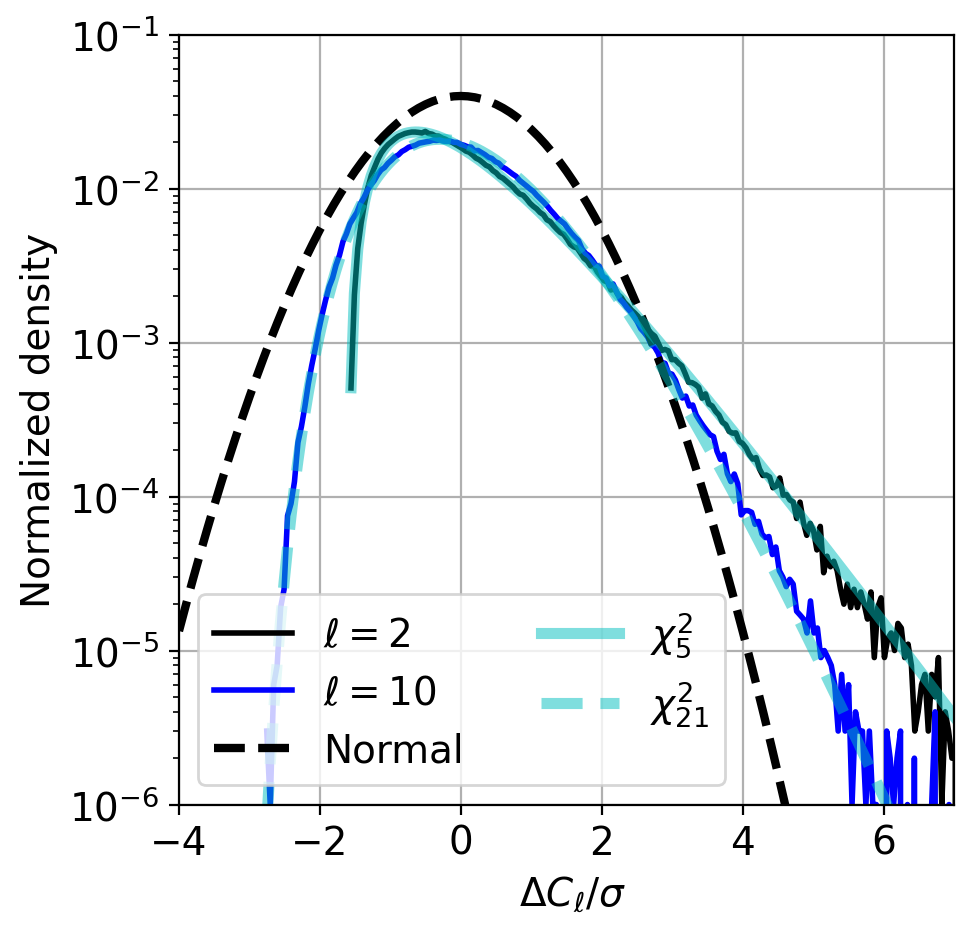} 
		\includegraphics[width=0.49\linewidth]{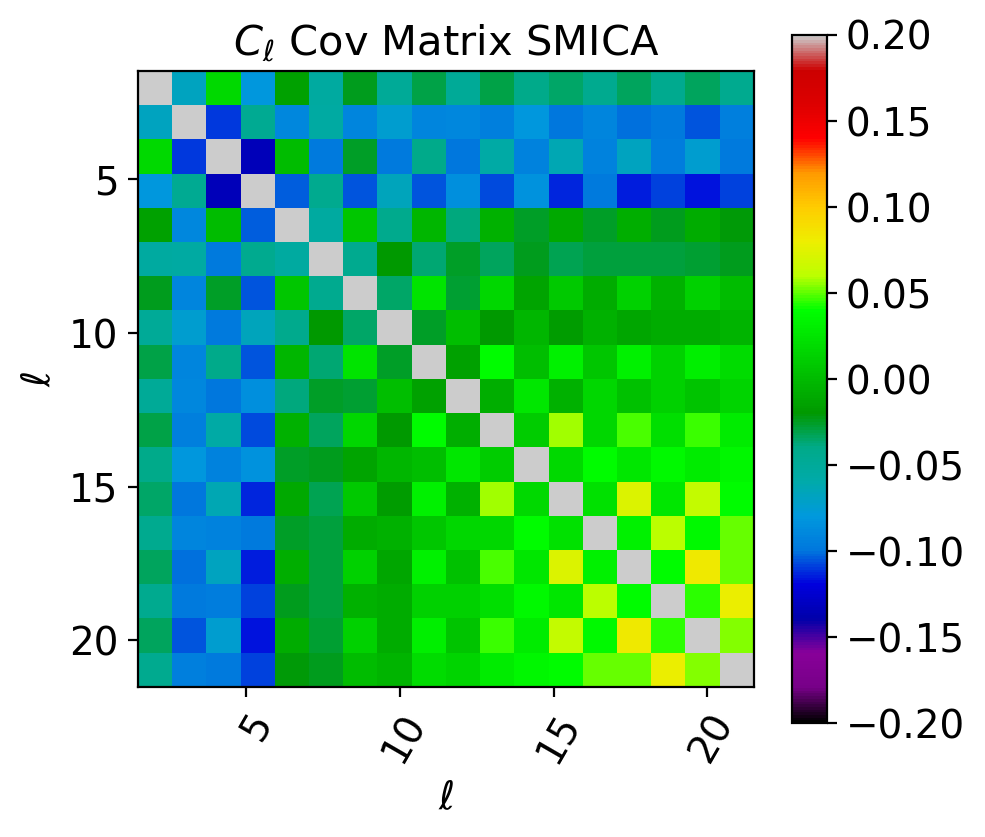} 
		\caption{{\sc Left:} Distribution of normalized variations $\delta_\ell \equiv \Delta C_\ell/\sigma$ for the lower order multipoles ($\ell=2-10$) in $10^6$ masked map realizations of the measured $C_\ell$ in the Planck SMICA$_m$ map, compared to the Normal distribution (dashed black). The distribution is non-Gaussian which makes a simple $\chi^2$ test hard to interpret. For example, a $\simeq +4\sigma$ fluctuation in $C_2$ is about $20$ more likely in the actual data than in the Gaussian prediction, while fluctuations approaching $\simeq -2\sigma$ is over $20$ times less likely. A $\chi^2$-distribution  with $2\ell+1$ degrees of freedom: $\chi^2_{2\ell+1}$  (cyan lines) gives a much better approximation. {\sc Right:} normalized covariance matrix $Cov[\ell,\ell']=<\delta_\ell \delta_{\ell'}>$  between $\delta_\ell$ in two different multipoles $[\ell,\ell']$. The cross-correlation Pearson coefficients are always below $15\%$.}
		\label{fig:CLpdf}
	\end{figure}

	Fig.~\ref{fig:clfit} shows a comparison of the measured $C_\ell$ for the lower multipoles (i.e. larger scales)  in the CMB temperature maps against the corresponding estimates in simulated maps that will be described below. 
	We also show the diagonal $\chi^2$ values of the data and models, given the simulations. The  $\chi^2$ values seem to prefer the odd-parity model (DSI+LCDM) over the standard inflation (SI+LCDM) model. But such a simple interpretation should be taken with caution for several reasons. First of all, the errorbars (or covariance matrix) 
	are dominated by sampling variance and therefore depend on the model assumption.
	This can be easily seen in the figure as the shaded region from the simulations is quite different for the SI model (top left), for the data (top right), or the odd-parity models (bottom panels). Second, the covariance matrix is not quite diagonal (because of the map masking). Third, for SI the large $\chi^2$ values are dominated by the low quadrupole $C_2$ in the data. This is a prediction of the odd parity model, but could also originate from a break in the primordial power spectrum, unrelated to parity symmetry. Fourth, the distribution of $C_\ell$ values is not Gaussian (e.g. see \cite{Bond2000} and references therein). This is illustrated in Fig.~\ref{fig:CLpdf}.
	More generally, when doing such $\chi^2$ we are testing different aspects of the model (like gaussianity, scale-invariant, or global isotropy) and not just directly parity.
	
	Fig.~\ref{fig:clfit} suggests that the octopole $C_3$ aligns more closely with SI than with DSI. While this observation holds true, the difference is less pronounced compared to $C_2$. The p-value of $C_3$ in DSI is approximately $9-10\%$, which is 3 times smaller than the $28\%$ for SI. This could mean that there is something else going on the largest scales: $\ell=2-3$ (i.e. lack of power on scales $\theta>60^\circ$).  However, a low $C_3$ remains a plausible random fluctuation given the range of multipole draws explored within $\ell<20$. The reason for the contrast in significance between $C_2$ and $C_3$ lies in the fact that sampling variance errors in $C_\ell$ are directly proportional to $C_\ell$:
	
	\begin{equation}
		\Delta C_\ell = \frac{C_\ell}{\sqrt{(2\ell+1)f_{sky}}}. 
	\end{equation}
	An increase in $C_\ell$ (occurring for odd $\ell$ under odd-parity) entails larger sampling errors compared to a decrease, as observed in the case of $C_2$. Consequently, deviations in even multipoles are more statistically significant than those in even ones. In addition, the lower $\ell$ multipoles have a more pronounced non-Gaussian distribution of errors.
	
	We can avoid some of these complications by just focusing on the alternating even-odd oscillation in the $C_\ell$  multipoles, which is the distinct feature of parity (e.g. see Eq.~\ref{eq:Tparity}).  Such an approach has already been reported in several analyses (see \cite{2005PhRvD..72j1302L, Kim:2012hf,Muir:2018hjv,Jones:2023ncn} and references therein). We will  use $R^{TT}$ defined in Eq.~\ref{eq:RTT} (e.g. see \cite{Muir:2018hjv} and references):
	Here, we will set $\ell_{max}=20$, motivated by our predictions. We have checked that our results remain similar for $\ell_{max}= 22, 27$, and $30$ which are the other values that have been used in the literature. 
	For SI we expect $R^{TT} \simeq 1$ because the fluctuations have no defined parity.  
	For the Planck 2018 masked SMICA map, \cite{Jones:2023ncn} finds $R^{TT}=0.79$  for $\ell_{max}=27$ with a p-value for SI of $\simeq 10^{-2}$ (this is the probability that a randomly generated SI realization produces a value of $R^{TT}$ as extreme as—in this case, lower than—the observed data).
	This result indicates a pronounced antisymmetric component (20\% larger odd parity) in the data. \footnote{Even-odd coupling from
		map masking and $C_\ell$ estimation have a much lower amplitude.}

	\section{CMB simulations}
	\label{sec:simulations}
	
\begin{figure}
		\centering
		\includegraphics[width=.49\linewidth]{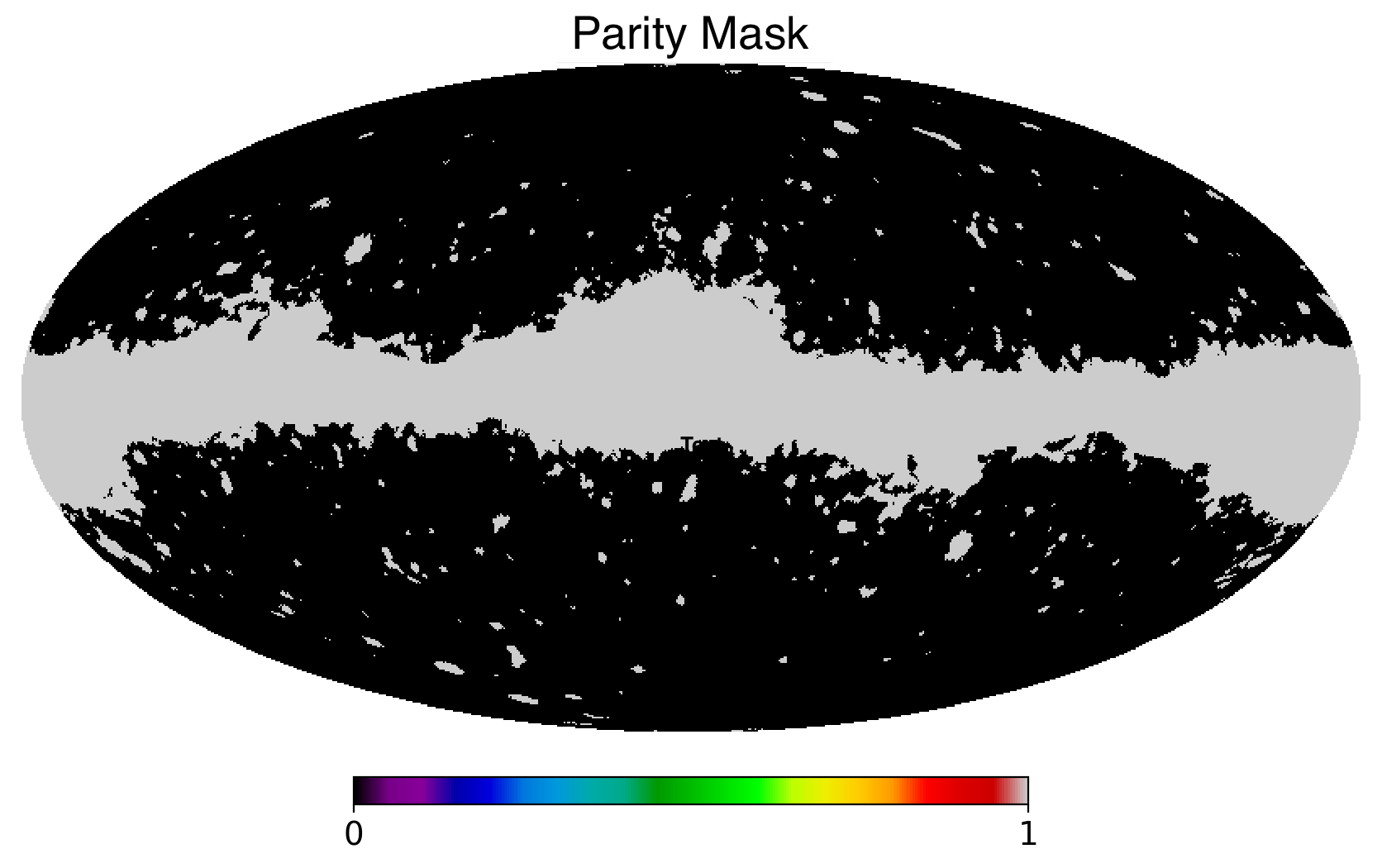}
		\includegraphics[width=.49\linewidth]{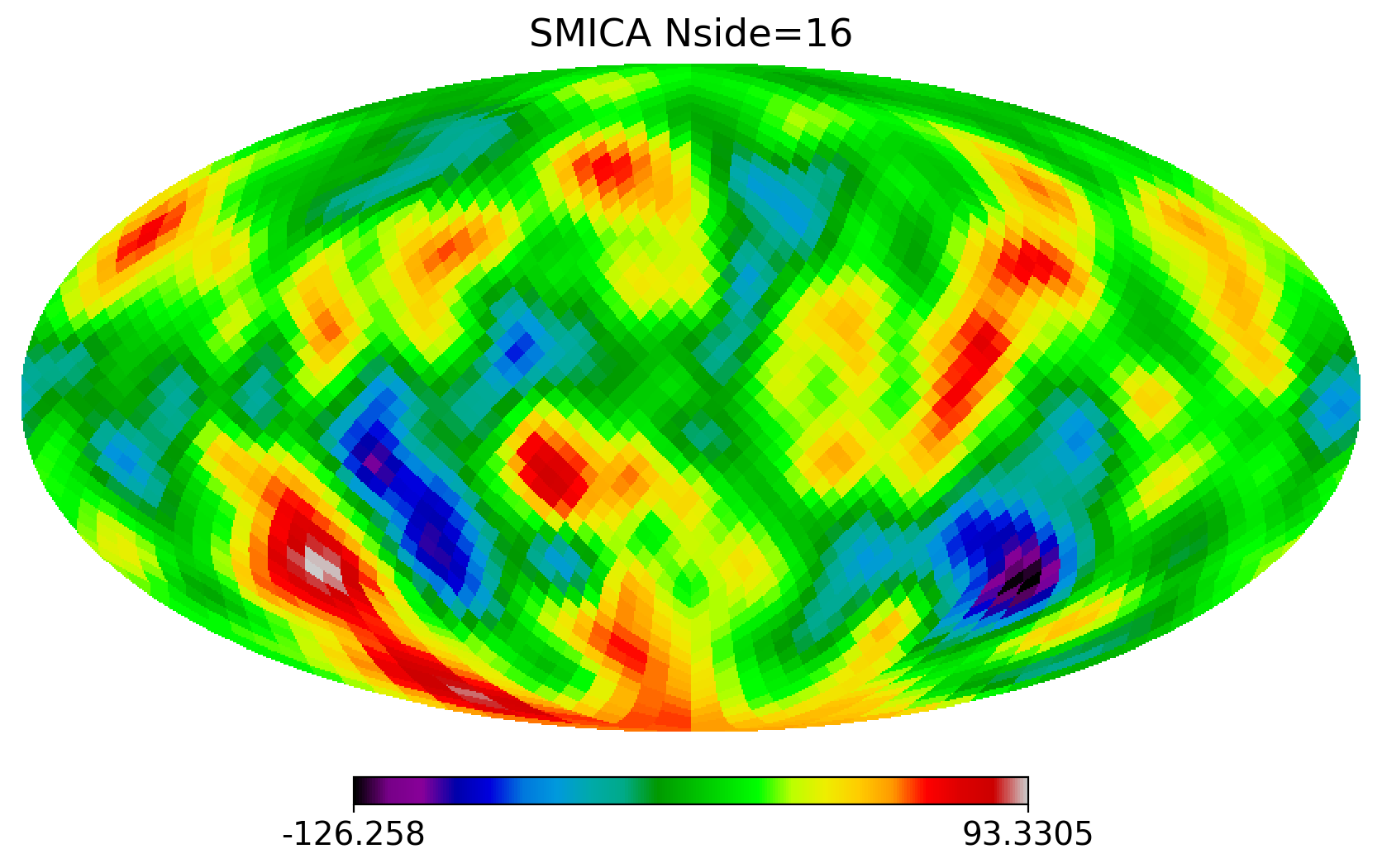}   
		\includegraphics[width=.49\linewidth]{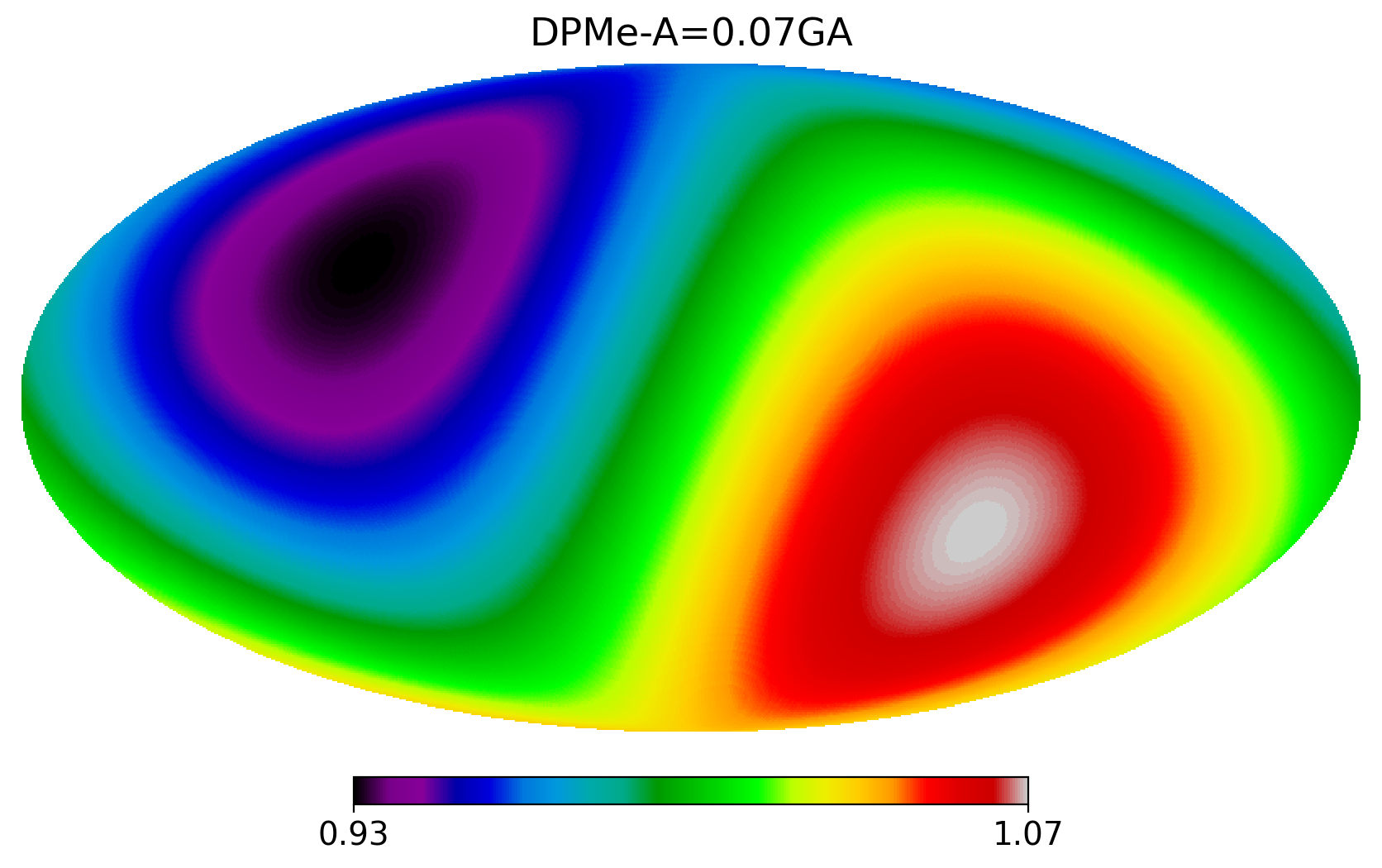}  
		\includegraphics[width=.49\linewidth]{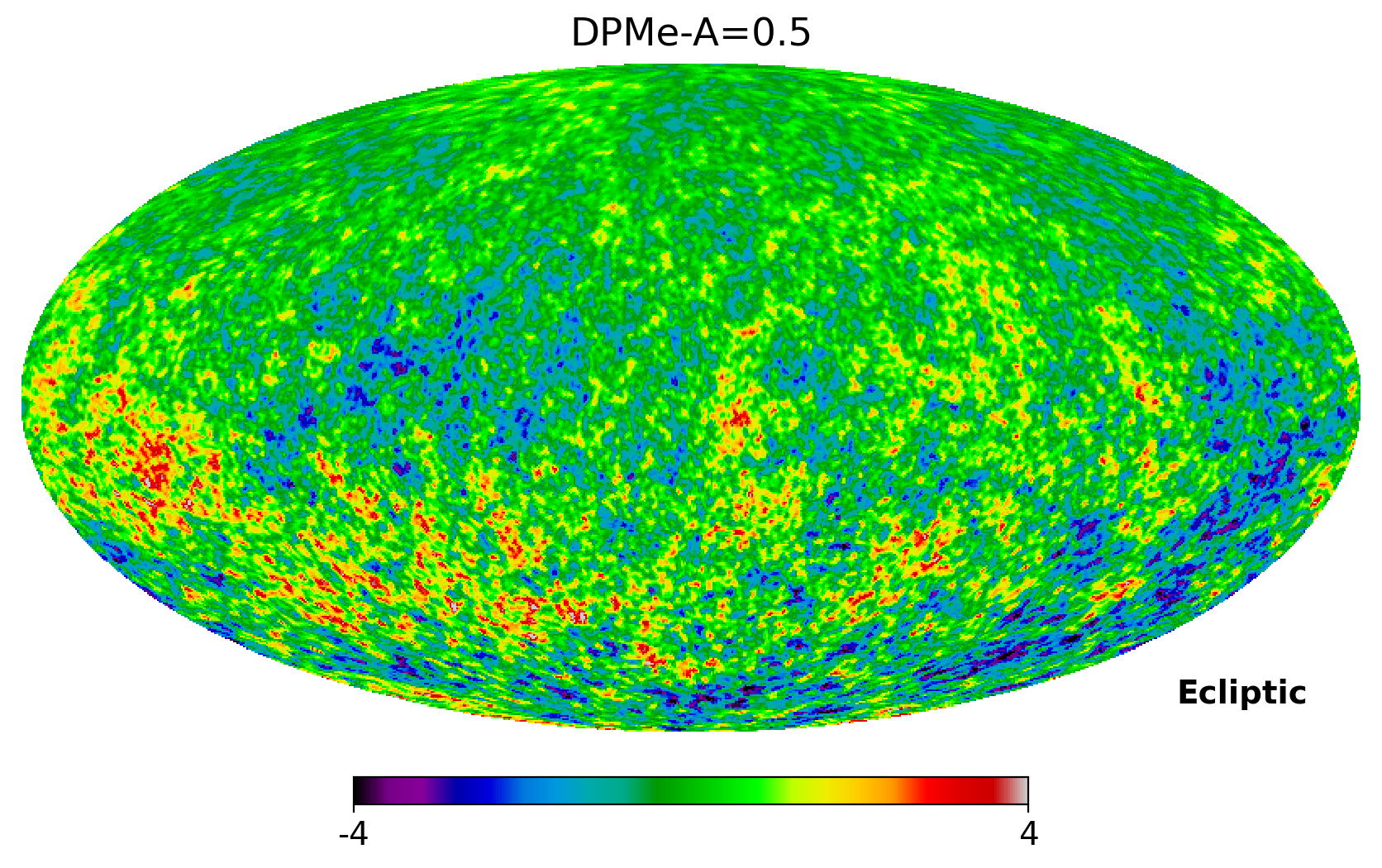}      
		\caption{{\sc Top Left:}
			$\Pc$ conjugate product image of the Planck common mask ($77.9\%$ of the sky), which is needed for the $Z$ map masking. This $Z$ masked map covers only 74\% of the sky. {\sc Top Right:} Planck 2018 (SMICA) maps in Ecliptic coordinates with $N_{side}=16$ and smoothing of $10.5^\circ$. This clearly shows less structure in the North Eclictic pole and is used to estimate the $\sigma_{16}^2$ HPA anomaly.
			{\sc Bottom Left:} A dipole modulation in Eq.~\ref{eq:DPM} with $A=0.07$ (shown in Galactic coordinates) pointing in the direction of the South Ecliptic pole. {\sc Bottom Right:} A LCDM Gaussian realization of the dipolar anisotropy modulation (DPM) model in Eq.~\ref{eq:DPM} with $A=0.5$ in Ecliptic coordinates (dipole is negative on the North Ecliptic Cap). Note that $A=0.5$ is much larger than the value $A=0.07$ allowed by the Planck data and is only used here to state the separation between isotropy and parity symmetries.
		} 	\label{fig:MirrorMask}
	\end{figure}

	Fig.~\ref{fig:MirrorMask} displays the version of the common mask employed for the parity $Z$ maps. This $\Pc$ conjugate mask results from combining the common mask with its $\Pc$ conjugate image (see Appendix \S\ref{sec:mirror}). Generally, our conclusions remain consistent across different component separations or masks, though the specific numerical values may differ. Unless specified otherwise, results will be presented for SMIC$_m$. Additionally, this is the default model simulated for estimating data error bars.
	
\begin{figure}
		\centering
		\includegraphics[width=.49\linewidth]{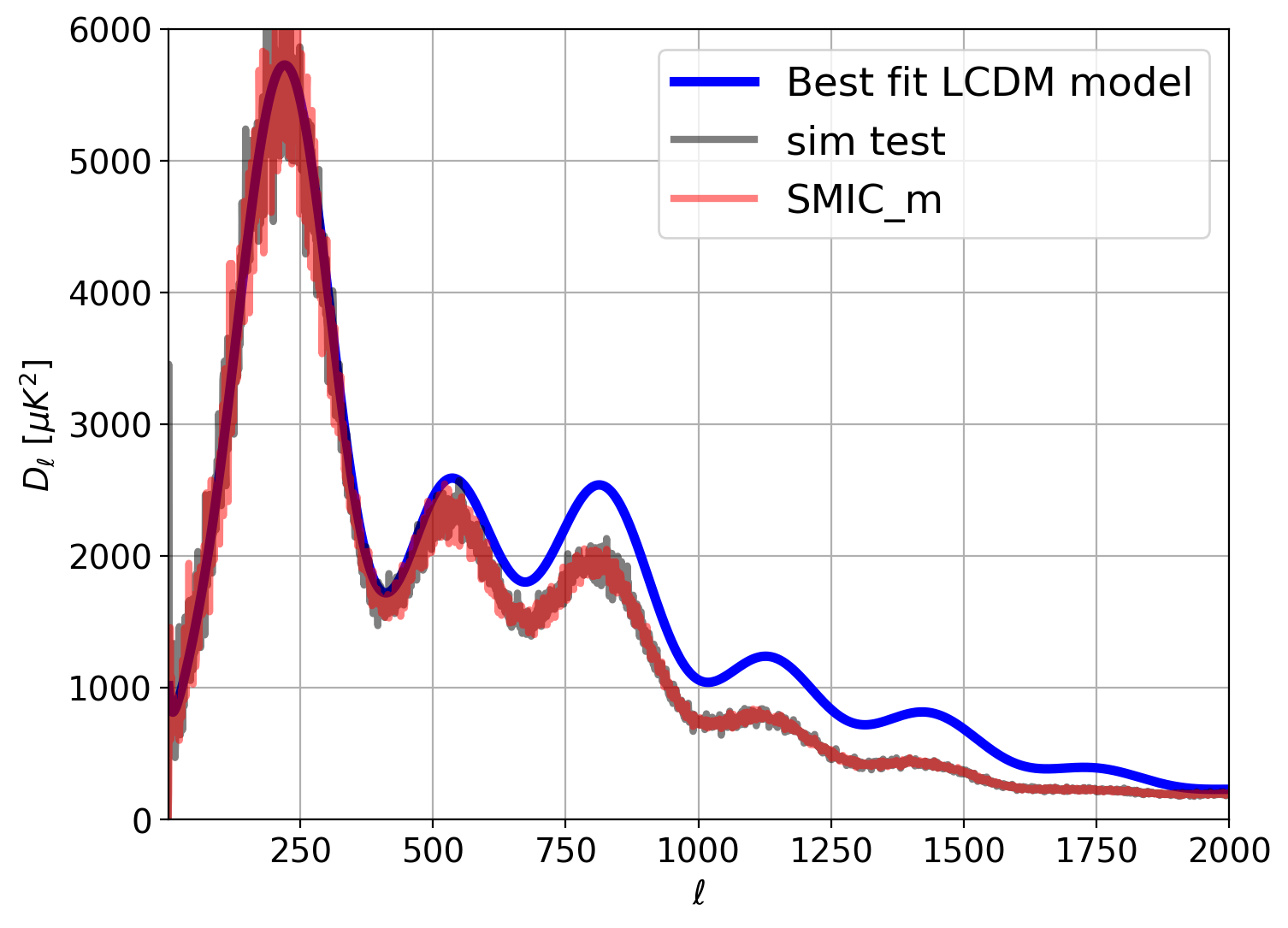}  
		\includegraphics[width=.49\linewidth]{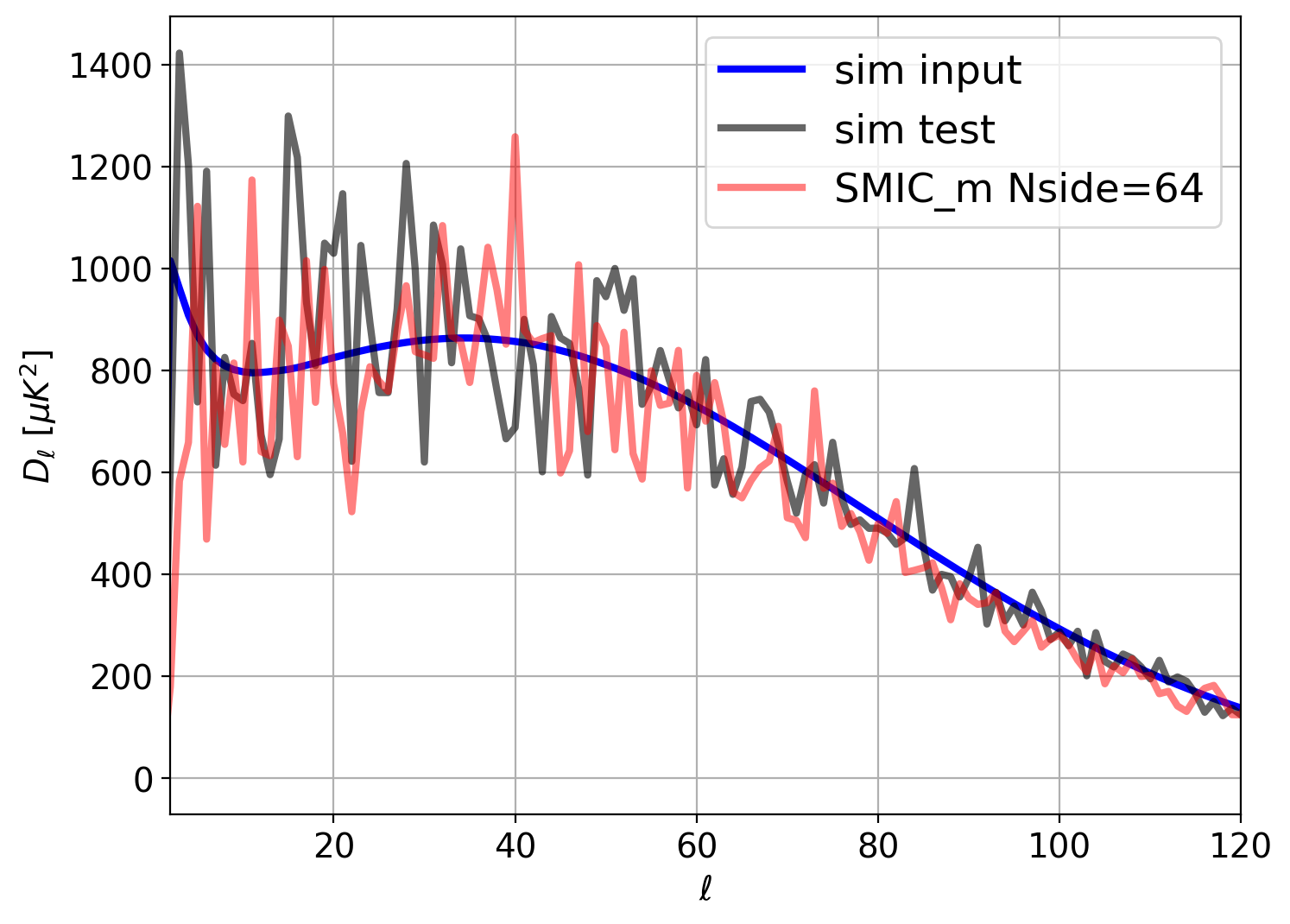}  
		\includegraphics[width=.49\linewidth]{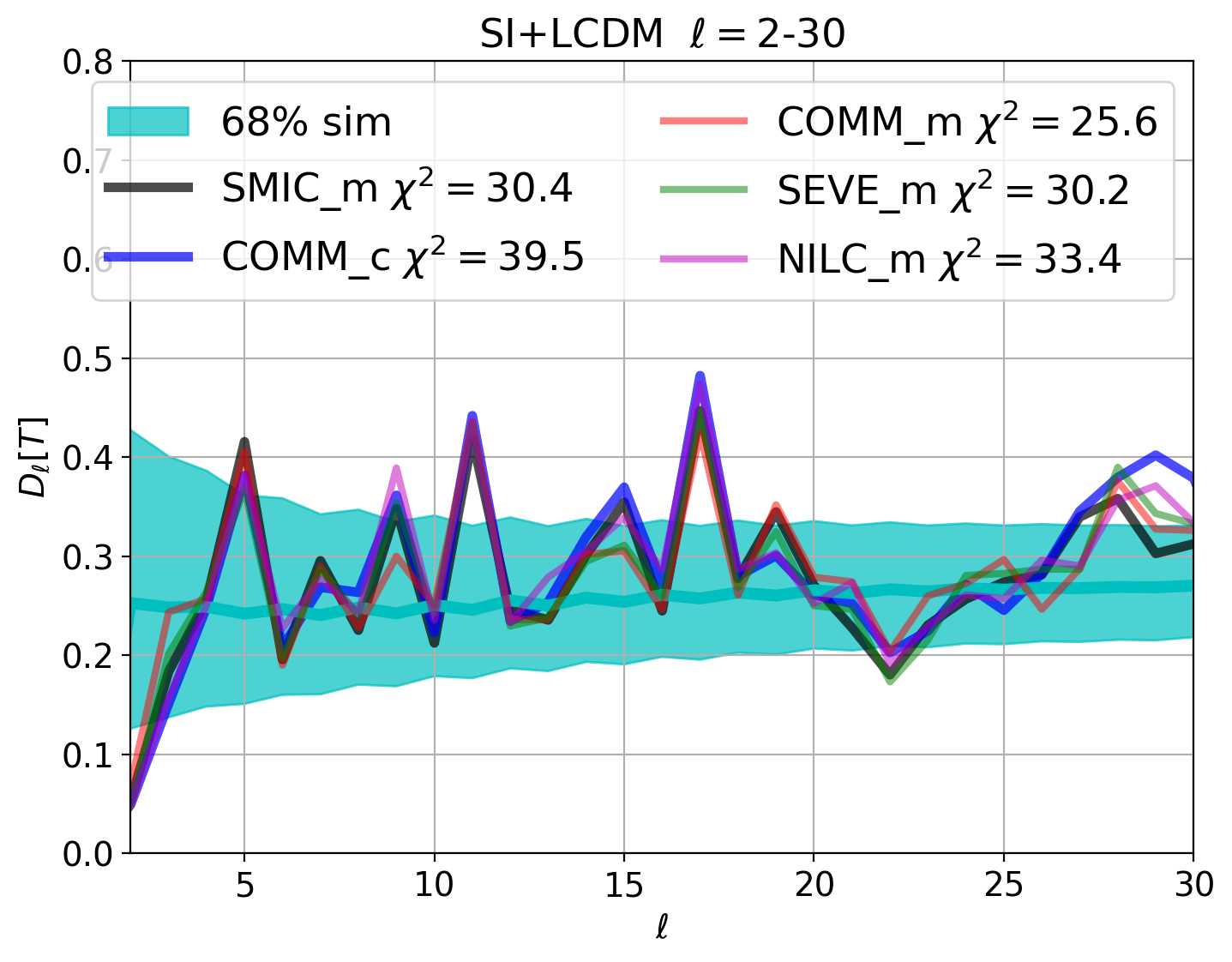}
		\includegraphics[width=.49\linewidth]{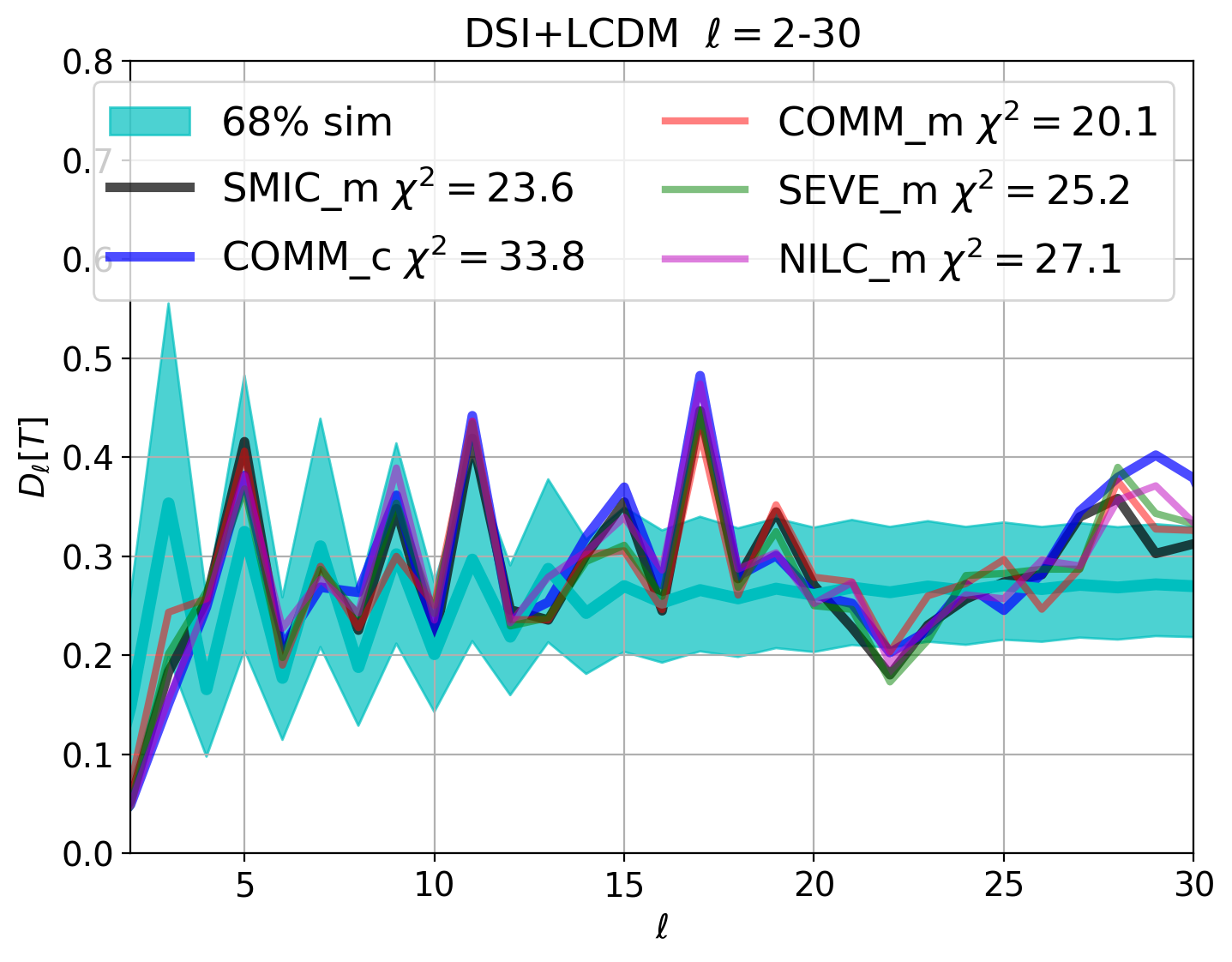}  
		\caption{{\sc Top left :} Planck 2018  best-fit LCDM model power spectrum $C_\ell$ (blue), which is the input for our simulations. In gray we show the measured $C_\ell$ from the simulation test with $N_{side}=2048$ and $r_{beam}=5'$, as in the Planck maps. It overlaps very well with the measurements in the Planck SMIC$_m$ map (shown in red). The differences for $\ell>500$ between the raw input and the measurements are due to the instrumental beam smoothing, noise, and masking. 
			{\sc Top Right:} Zoom in on the lower multipoles measured after degrading the maps to $N_{side}=65$ (with $r_{beam}=2^\circ$ Gaussian beam). This is the default resolution in our parity analysis. {\sc Bottom Panels:} Shaded (cyan) region corresponds to 68\% range of $C_\ell$ values around the mean from $10^6$ simulations in SI+LCDM (left) and DSI+LCDM (right).
			These models are compared to $C_\ell$ measured in the different Planck 2018 maps. We have degraded and smoothed the original maps and simulations from $N_{side}=2048$ to  $N_{side}=64$ and normalized them to unit variance.} 
		\label{fig:clLCDM}
	\end{figure}

	We generate Gaussian simulations based on the Planck 2018 best-fit LCDM model\footnote{COM-PowerSpect-CMB-base-plikHM-TTTEEE-lowl-lowE-lensing-minimum-theory-R3.01} using the {\sc Helapix} routine {\sc symfast} with $N_{side}=2048$. We introduce instrumental noise and a $\theta_B = 5$ arcmin Full Width at Half Maximum (FWHM) Gaussian beam. The Planck 2018  best fit is predominantly influenced by $\ell>40$, where the signal-to-noise is more significant, and the Baryon Acoustic Oscillation (BAO) peaks are prominent. The top left panel of Fig.~\ref{fig:clLCDM} illustrates this fit in blue.
	
	To assess the accuracy of our simulation, we simulate the blue line with {\sc symfast} (smoothed with a $\theta_B = 5$ arcmin beam) and measure the $C_{\ell}$ back from the simulated map using {\sc anafast}. The result is represented by the gray line labeled 'sim test' in the figure. The difference between gray and blue lines is due to the instrumental noise and instrumental beam smoothing. The gray line is further compared to the same direct measurement of $C_{\ell}$ in the Planck SMIC$_m$ maps (in red), post-application of the mask, and division by $f{sky}$. The two outputs exhibit a remarkable overlap for all multipoles, essentially lying on top of each other. This test emphasizes that our simulations faithfully reproduce the intended models.

	Since our focus is on studying the largest scales, there's no necessity to work with such a relatively large resolution of $N_{side}=2048$. Consequently, we downsampled the maps to $N_{side}=256,128,$ or $64$ and applied smoothing with $\theta_S = 0.5^\circ,1^\circ,$ or $2^\circ$.
	These processes were also applied to the Planck maps for consistency.
	
	
	The instrumental noise exhibits minimal impact on the smoothed $N_{side}<256$ maps, particularly at the low multipoles. This was previously demonstrated by \cite{Muir:2018hjv}, who conducted a comparison between simulations incorporating realistic noise (derived from FFP8.1 Planck public simulations) and simpler {\sc symfast} simulations. The findings indicated that both approaches yield very similar statistics for CMB anomalies at $\ell<30$.

	To focus on the parity symmetry, we normalize $\Delta T$ in the low-resolution maps by $\sigma_T$, defined as the root mean square (RMS) of the temperature map: $\sigma_T = \sqrt{\langle \Delta T^2 \rangle}$. In practice, we estimate $\sigma_T$ by determining the 68\% percentile width of the 1-point $\Delta T$ probability distribution function. This approach is more robust in the presence of outliers or artifacts.
	
	When estimating $\sigma_T$, it is crucial to utilize only the data outside the mask. This is because values within the mask possess different noise properties, and we want to avoid introducing such noise into the CMB signal. By doing so, the results become less sensitive to resolution or noise at the pixel level. The resulting pipeline statistics prove to be quite robust, yielding very similar results and conclusions across different resolutions. For simplicity, we default to presenting results for $\theta_S=2^\circ$ and $N_{side}=64$.
	
	The parity measurements yield very similar results whether one degrades higher resolution simulations from $N_{side}=2048$ to lower resolutions (e.g., $N_{side}=64$) or directly simulate CMB maps at the lower resolution without any instrumental noise. The efficiency gained from the latter approach adds practicality to the methodology.

	We have checked that the smoothed normalized low-resolution maps (with $N_{side}=64$)  also have the same average power spectrum $C_\ell$ as the corresponding data. This is shown in the right panel of Fig.~\ref{fig:clLCDM}.
	The bottom panels of  Fig.~\ref{fig:clLCDM} show a zoom-in around the lower multipoles.  

\begin{figure}
		\centering
		\includegraphics[width=.49\linewidth]{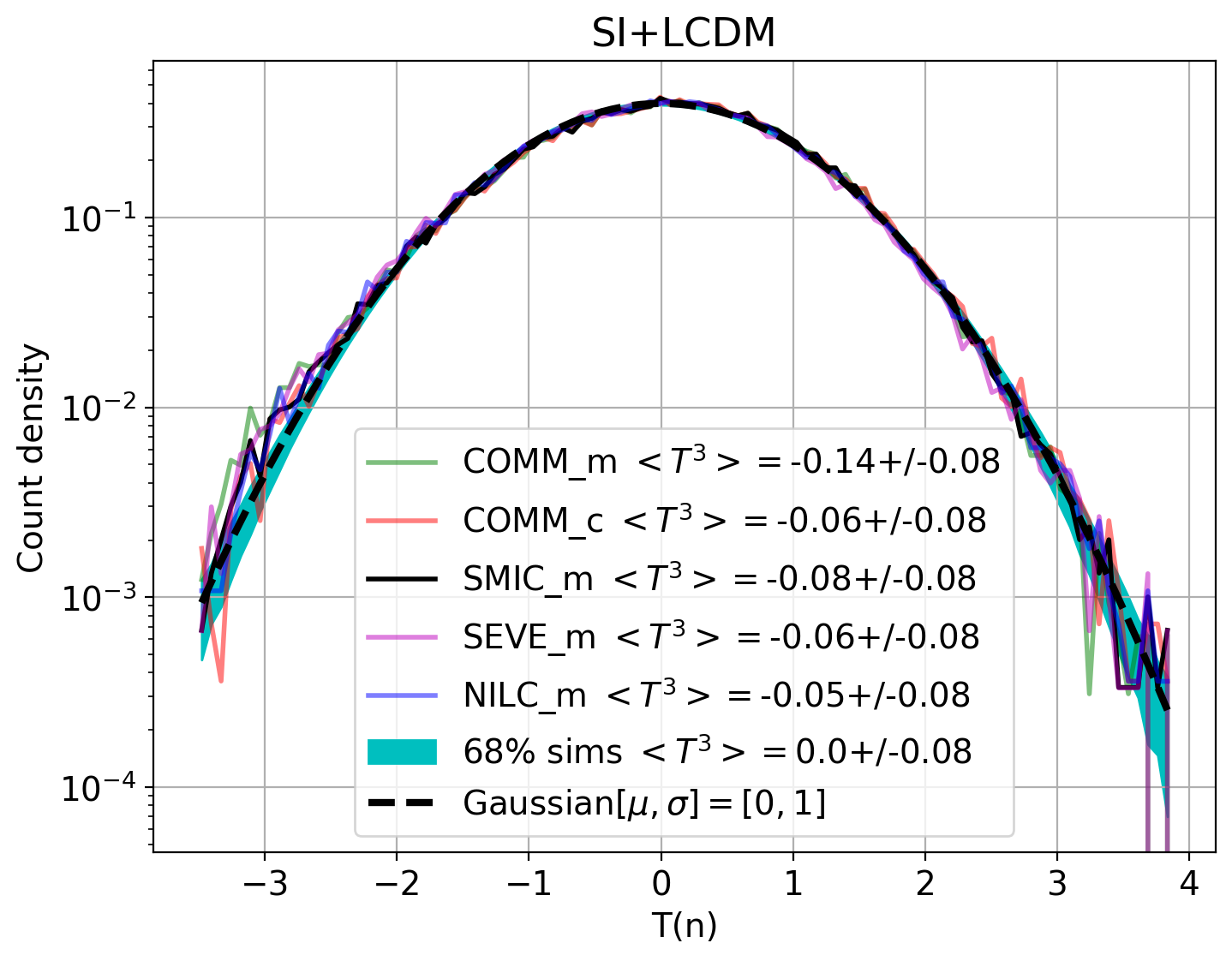}  
		\includegraphics[width=.49\linewidth]{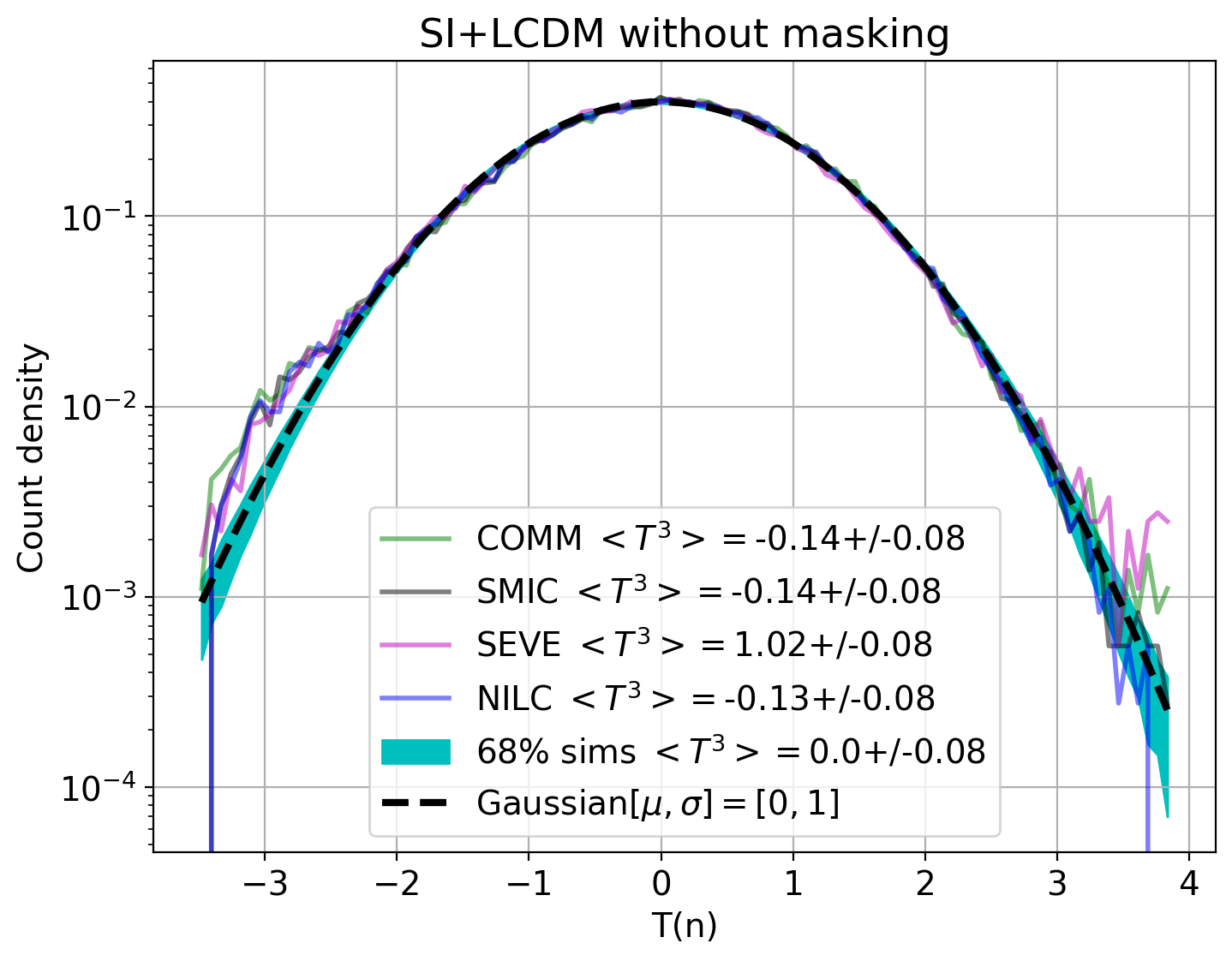}  
		\caption{Test of Gaussianity in the CMB maps and simulations, with (left) and without  (right) mask.
		} 
		\label{fig:dTcouns}
	\end{figure}

	In Fig.~\ref{fig:dTcouns}, we further test our simulations by demonstrating that both Planck data and simulations exhibit Gaussian 1-point statistics with unit variance (as the temperature is normalized to $\Delta T/\sigma_T$). While there is a small negative skewness in the Planck maps, it remains consistent with zero at the 1$\sigma$ level, contingent on the component separation map used. The most noticeable discrepancy (still within $2\sigma$) arises from the COMM$_m$ map, which has the smallest mask. When utilizing the common mask (labeled COMM$_c$ in the figure), the skewness in COMM$_m$ reduces to half its value. The skewness becomes more pronounced (deviating from zero) when the mask is not applied (right panel). This is an artifact. Moving forward, we'll exclusively present results outside the masked region.
	
	Simulating the SI+LCDM-C2 model involves repeating the same SI simulation, with the only modification being the fixation of the quadrupole $C_2$ to the value measured in the Planck SMICA masked map (SMIC$_m$). Comparable results are observed for the other component separations.

	To simulate the DPM model described in Eq.~\ref{eq:DPM}, the process begins by generating a fixed dipole using {\sc Healpix hpt.dipole\_pdf} with an amplitude $A=0.07$ and oriented toward the South Ecliptic pole (to emphasize power in the South). An extreme case with $A=0.5$ was also explored. This fixed dipole is then multiplied by each of the LCDM realizations. An illustrative example with $A=0.5$ is presented in the left panel of the bottom right of Fig.~\ref{fig:MirrorMask}, showcasing a visible dipolar anisotropy.

	To simulate the DSI we add a new component to the SI best-fit (+LCDM) model: 
	\begin{equation}
		C^{DSI}_\ell = C^{SI}_\ell \LF 1 \pm \Delta \Cc_\ell  \RF
		\label{eq:Cl_DQFT}
	\end{equation}
	where $+$ is for odd-$\ell$ and $-$ stands for even-$\ell$. The $\Delta \Cc_\ell$ is given by
\begin{equation}
    \Delta \Cc_\ell = \frac{1}{C_\ell^{SI}} \int_0^{k_c} \frac{dk}{k} A_s\LF \frac{k}{k_s} \RF^{n_s-1} j_{\ell}^2\LF \frac{k}{k_s} \RF\LF \Delta \Pc_v \RF
    \label{eq:RDcl}
\end{equation}
where $\Delta \Pc_v$ can be read from Eq.~\ref{eq:delpPR} and
\begin{equation}
   C_\ell^{SI} =  \int_0^{\infty} \frac{dk}{k} A_s\LF \frac{k}{k_s} \RF^{n_s-1} j_{\ell}^2\LF \frac{k}{k_s} \RF
   \label{eq:SICl}
\end{equation}
Since Eq.~\ref{eq:RDcl} is the ratio, the LCDM transfer function's contribution cancels out in the numerator and the denominator. This modification Eq.~\ref{eq:Cl_DQFT}  oscillates as shown in bottom panels of Fig.~\ref{fig:clLCDM} and Table~\ref{tab:DSI}. 
 	

	\begin{figure}
		\centering
		\includegraphics[width=0.49\linewidth]{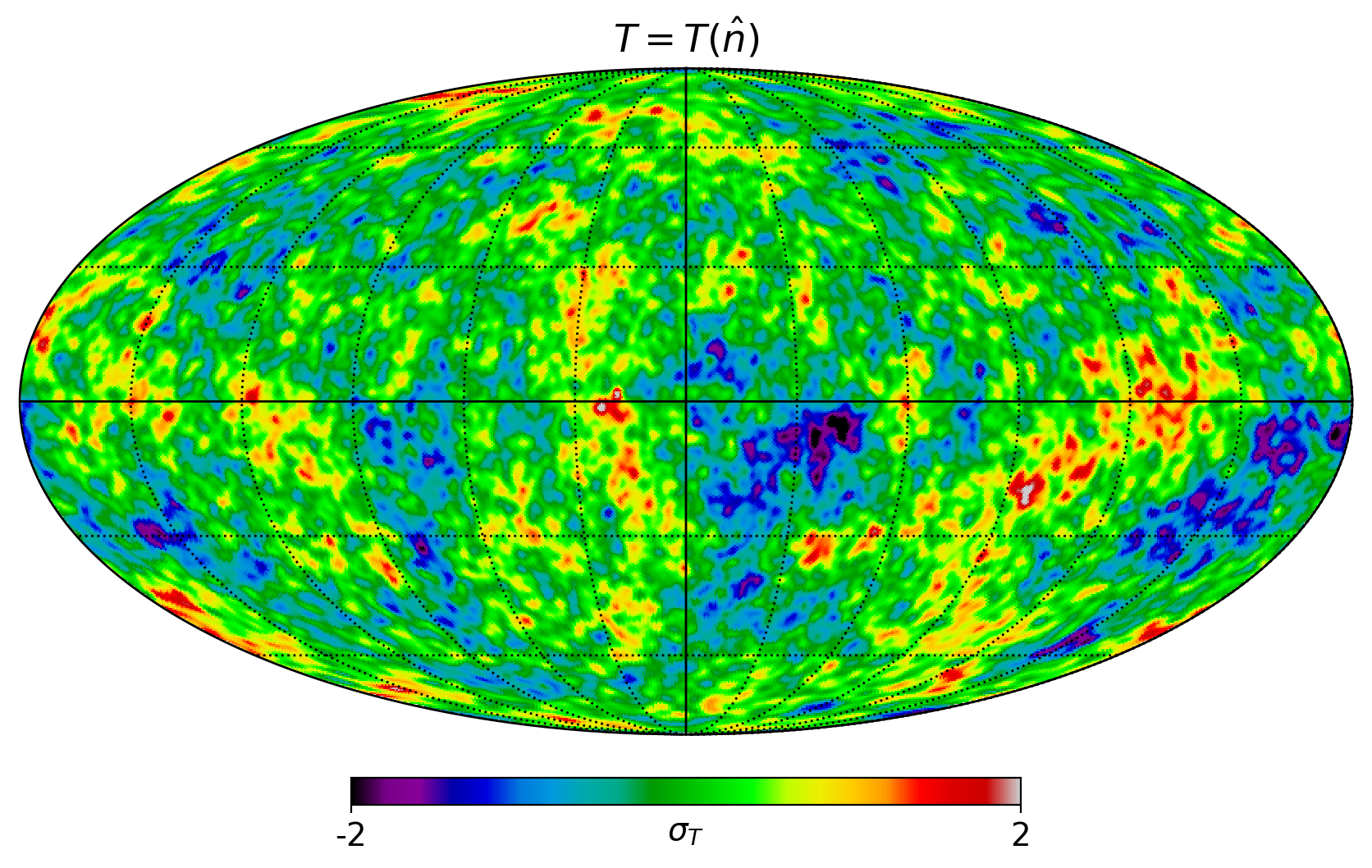}
		\includegraphics[width=0.49\linewidth]{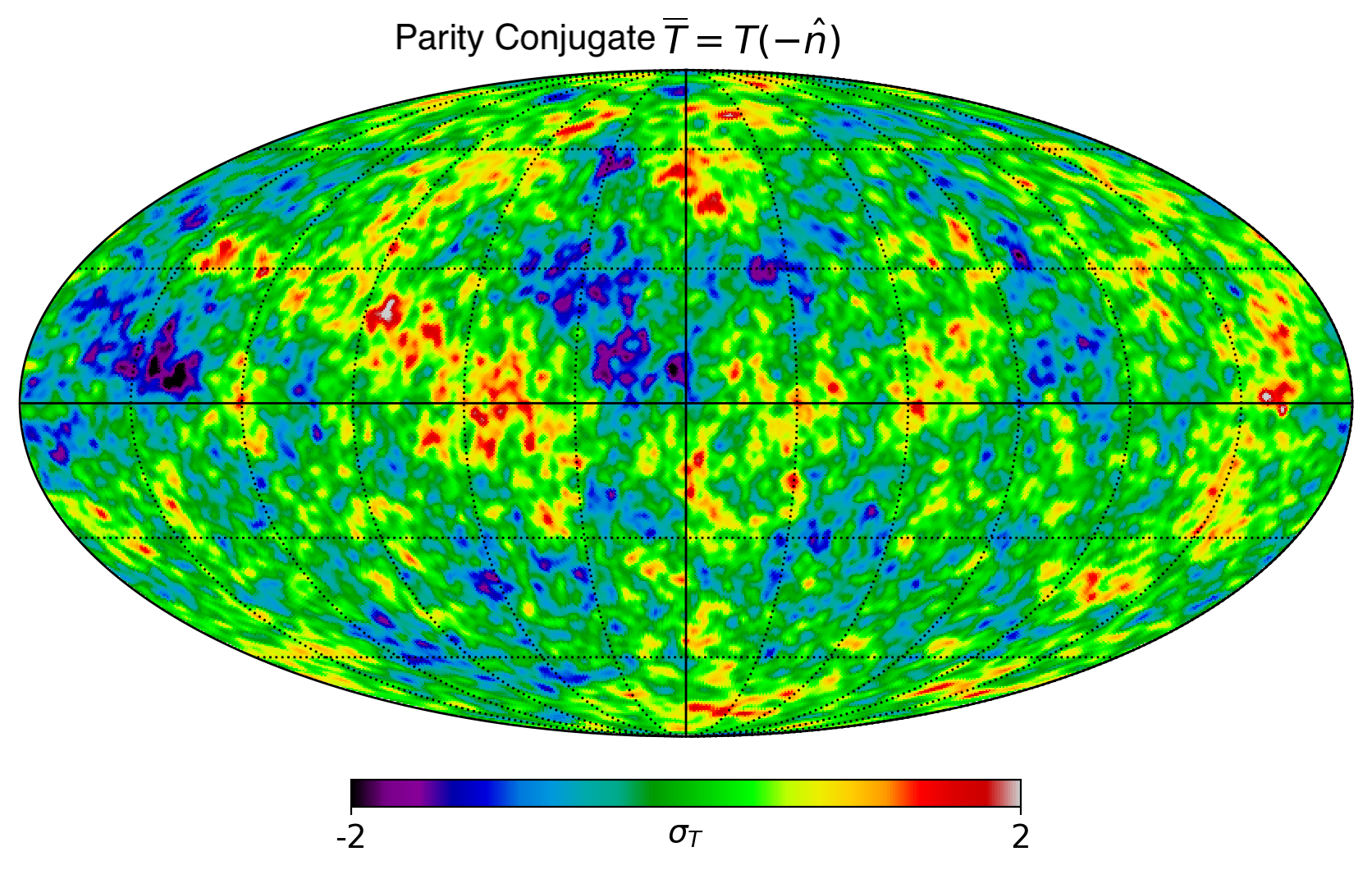} 
		\vskip -0.1cm
		\includegraphics[width=0.49\linewidth]{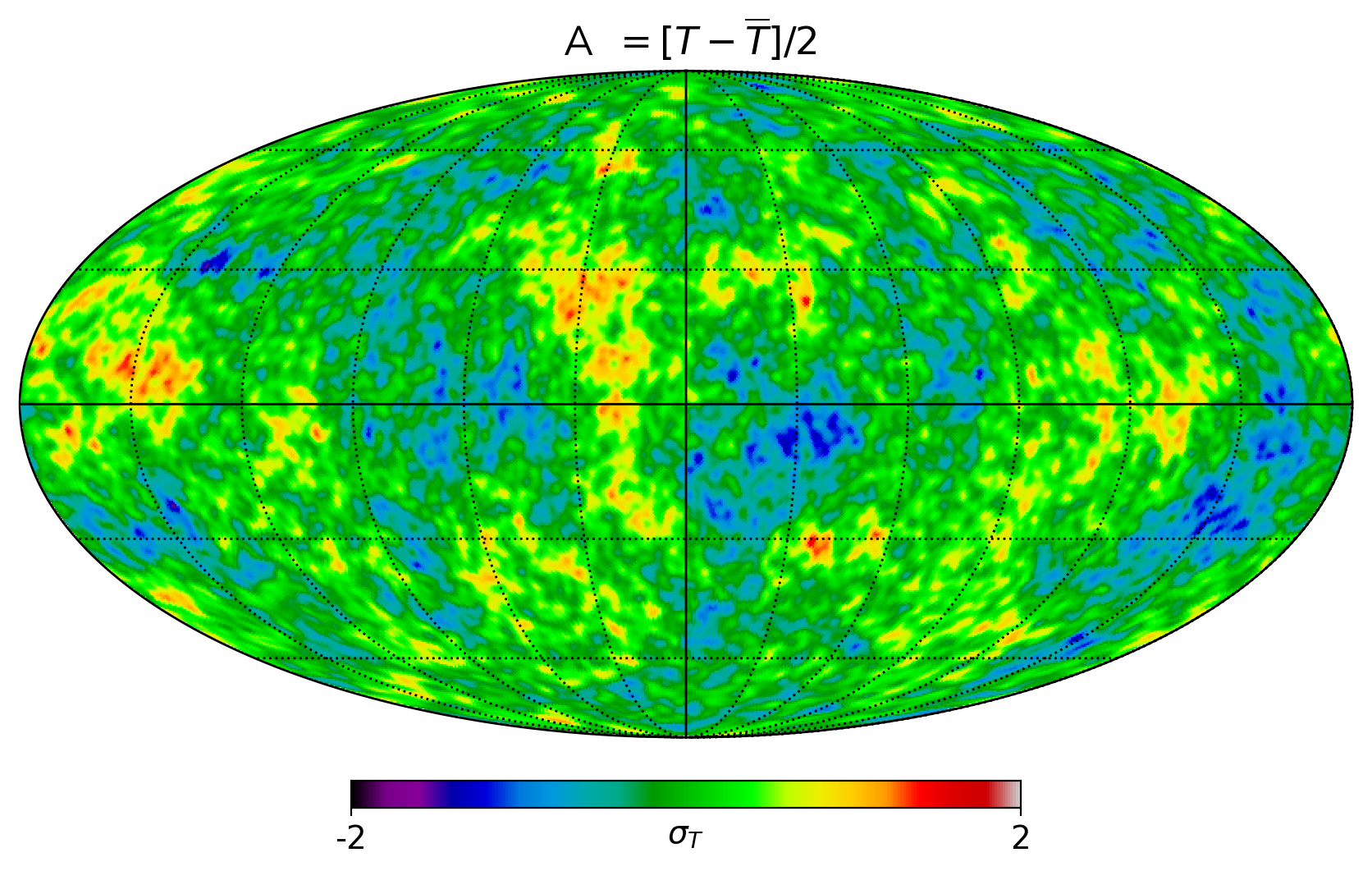}
		\includegraphics[width=0.49\linewidth]{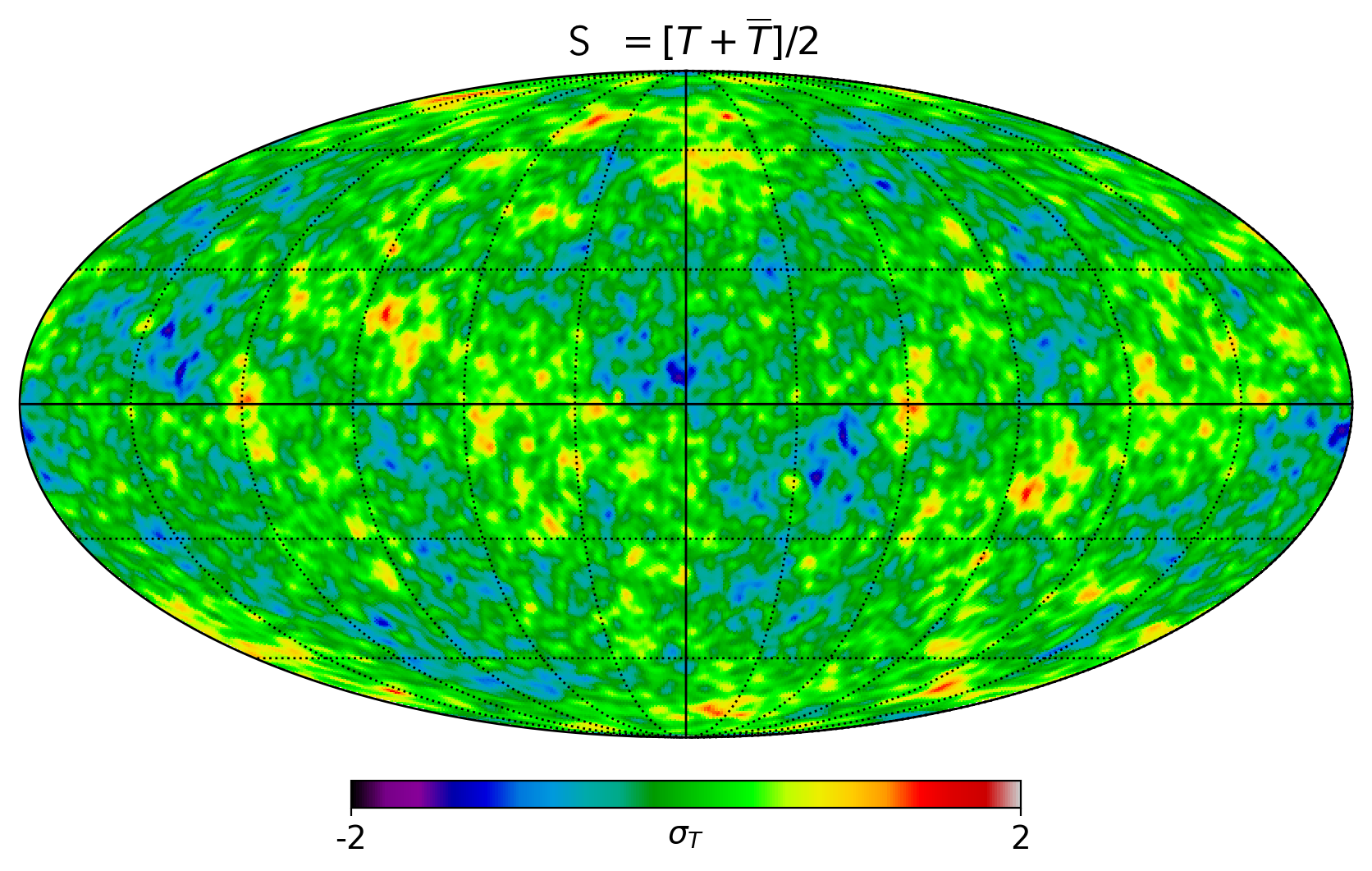}
		\vskip -0.1cm
		\includegraphics[width=0.49\linewidth]{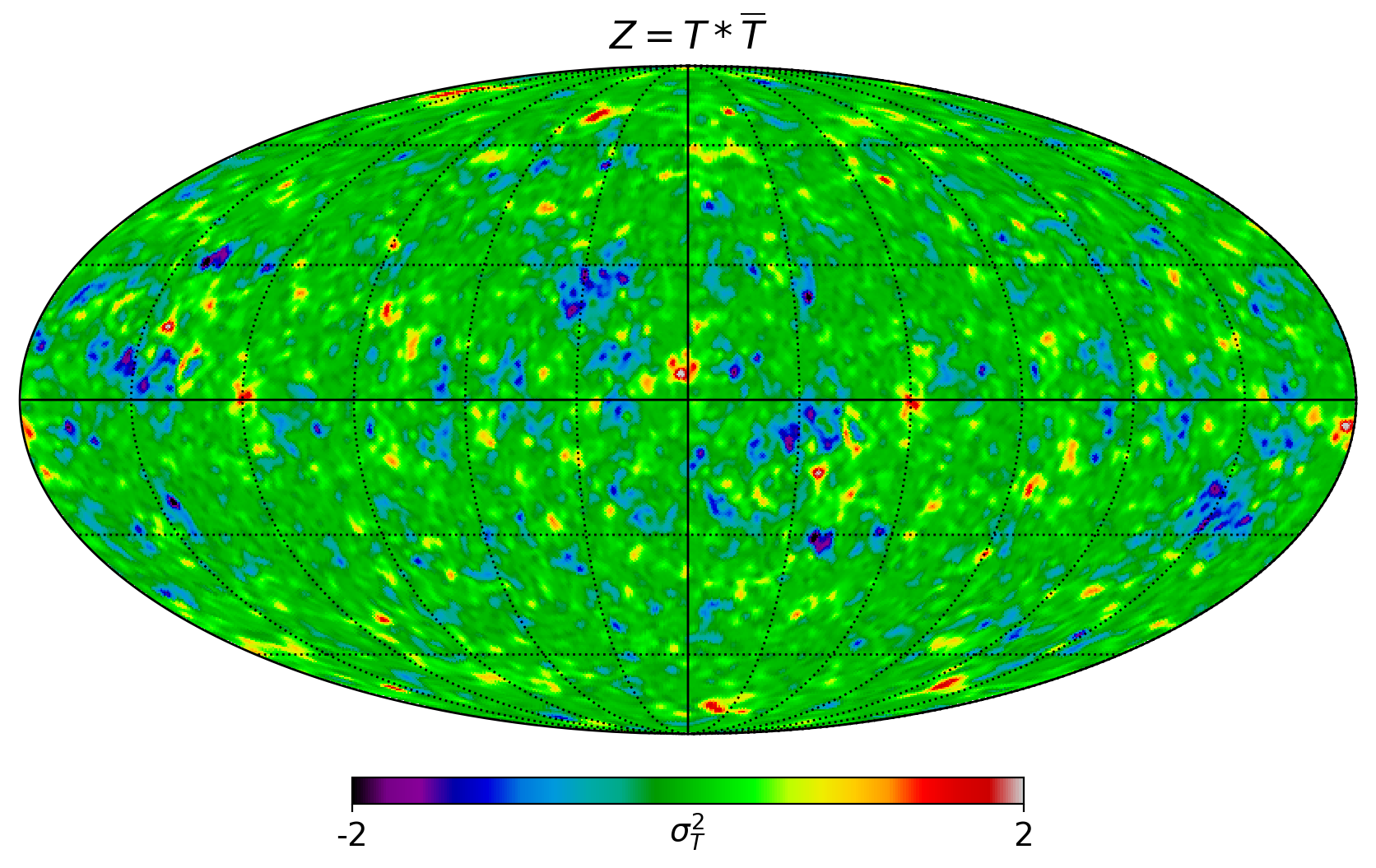}
		\includegraphics[width=0.49\linewidth]{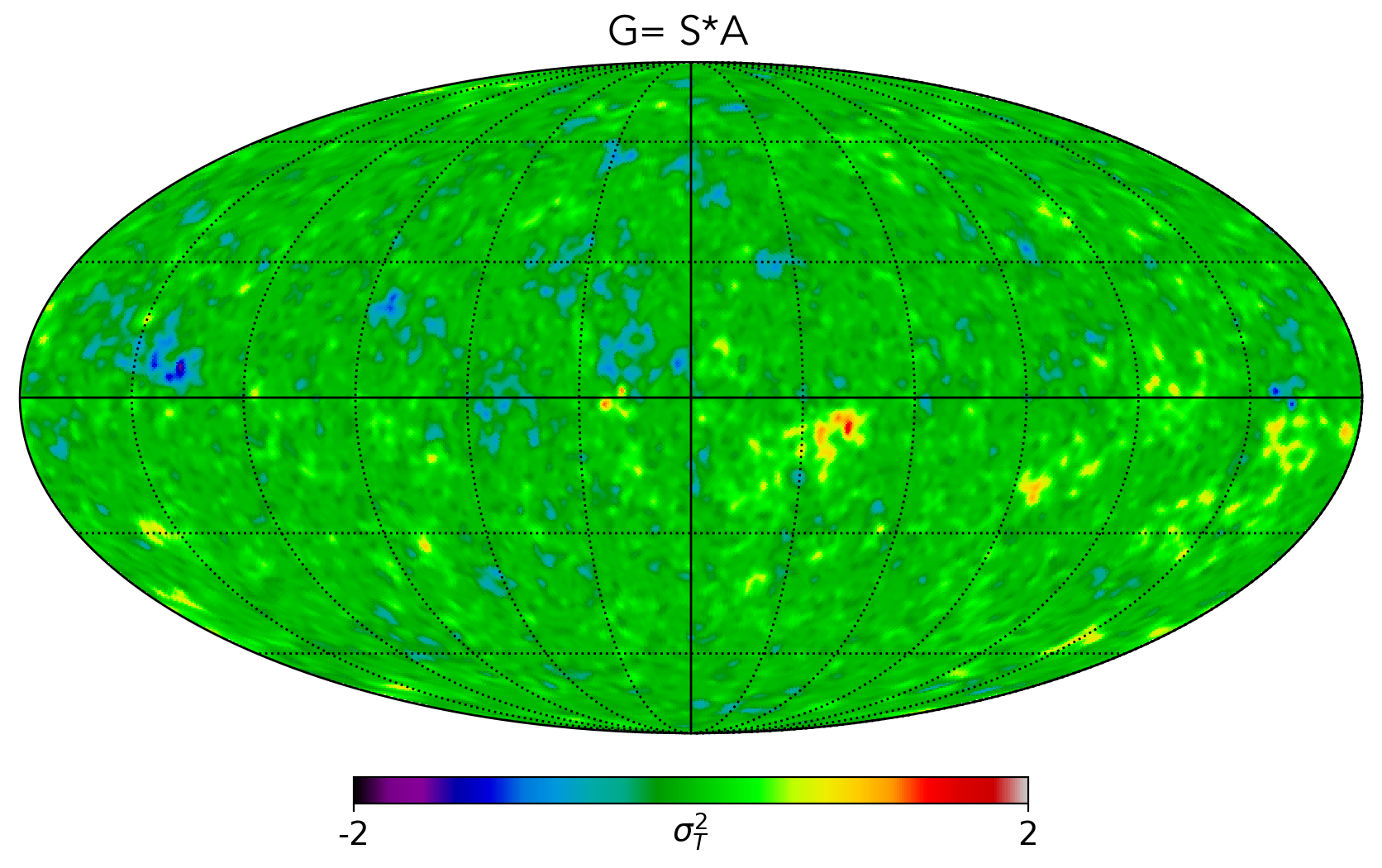}
		\caption{
			In these Mollview projections, we compare the antisymmetric ($A$ odd parity, middle left) and symmetric ($S$ even parity, middle right) decompositions of the Planck 2018 map (SMICA, top left). Furthermore, we showcase its $\Pc$ conjugate image $\overline{T}(\hat{n})=T(-\hat{n})$ (top right). The odd parity map resembles the original more closely than the even parity 
			counterpart. This can also be seen in the $Z=T *\overline{T}$ map which is characterized by a prevalence of negative (blue) values.
			The cross-map $G=S*A$ is predominantly near zero, except in the proximity of the cold structure close to the Galactic Center. }
		\label{fig:smicaParity1}
	\end{figure}

 \section{Homgeneity Index and lack of correlations at $\theta>60^\circ$}
	\label{sec:characterization}


	Both Eq.~\ref{eq:cl} and Eq.~\ref{eq:C_theta} assume local statistical isotropy. But even when a fluctuating field $\Tc(\hat{n})$ is found to be locally isotropic, we can also test if, on the largest scales, the amplitude of such fluctuations tend towards global isotropy, which we will also call Homogeneity (to distinguish it from the local isotropy, but not to be confused with 3D translation invariance) and will be better defined in the next paragraph. COBE \cite{COBEw2}, WMAP  (\cite{Bennett-wmap,Bennett-wmap,gazta2003}) and latter Planck (\cite{Planck:2013lks}), found an anomalous (lack of) amplitude of large-scale fluctuations in the CMB sky (see our Fig.~\ref{fig:Ctheta}).
	The integrated value of the square 2-point temperature correlation  $S_{1/2}$ for separations above $\theta>60^\circ$ has been used as a measure of how much power there is on such large scales \cite{Bennett-wmap}: 
	\begin{equation}
		S_{1/2}= \int_{-1}^{1/2} w^2[\theta]  d(\cos{\theta}).
		\label{eq:S12}    
	\end{equation}
	This additional test of (global) isotropy has no directional component. Because 
	$S_{1/2}$  is quadratic in $w$ (and in $C_\ell$) , it results in a non-linear response to rare statistics which can be hard to interpret.
	
	A better measurement of such global isotropy is given by the Homogeneity index  $\mathcal{H}$ (\cite{Camacho-Quevedo:2021bvt}).
	This is a Hausdorff or fractal dimension \cite{Mandelbrot,Mandel}, which is a rigorous way to characterize geometrical irregularities and chaos in Nature. The idea of the
	homogeneity index, $\mathcal{H}$, is to test if the (2D) metric associated with a given map of fluctuations tends towards a homogeneous (2D) metric on the largest scales. The index  $\mathcal{H}$  is independent of the amplitude of fluctuations. It only depends on the slope (how the amplitude of fluctuations  changes as we change the angular separation $\theta$):
	\begin{equation}
		\mathcal{H}(\theta) = 2\frac{d\ln{\mathcal{P}}}{d\ln{\Omega}} 
		\quad ; \quad
		\mathcal{P} = \frac{1}{1-\cos{\theta}} \int_0^{\theta} \left(1+ w[\theta] \right) d(\cos{\theta})
		\label{eq:Hausdorff}
	\end{equation}
	For a global isotropic (i.e. homogeneous) metric we expect $\mathcal{H} \simeq 0$ on the largest scales. Standard Primordial Inflation (SI) is near scale-invariant Eq.~\ref{HZpower} and therefore has $\mathcal{H} \neq 0$ also on the largest scales. This is at high odds with the actual CMB measurements (\cite{Camacho-Quevedo:2021bvt}), which shows  $\mathcal{H} \simeq 0$ for $\theta>60^\circ$, which corresponds to $\ell<4$ (i.e. a lower than expected quadrupole $C_2$ and octopole $C_3$). We have checked here that the p-values of $\mathcal{H}$ at $\theta=180^\circ$ follow a very similar trend to the ones of $Z^1$ shown in Table \ref{tab1:annomalies}-\ref{tab2:annomalies}, which confirms the odd parity in the 2-point function using a more robust measurement of global isotropy.

	\section{Visual comparison of the models to data} 

 \label{sec:visualcom}
	
	In this section, we'll demonstrate how one can easily discern odd parity with a simple visual inspection of the right CMB maps. 
	In the upper-left panel of Figure \ref{fig:smicaParity1}, we present the Cosmic Microwave Background (CMB) temperature, denoted as $T=T(\hat{n})$, extracted from the Planck map using the SMICA component separation method.
	
\begin{figure}
		\centering
		\includegraphics[width=0.49\linewidth]{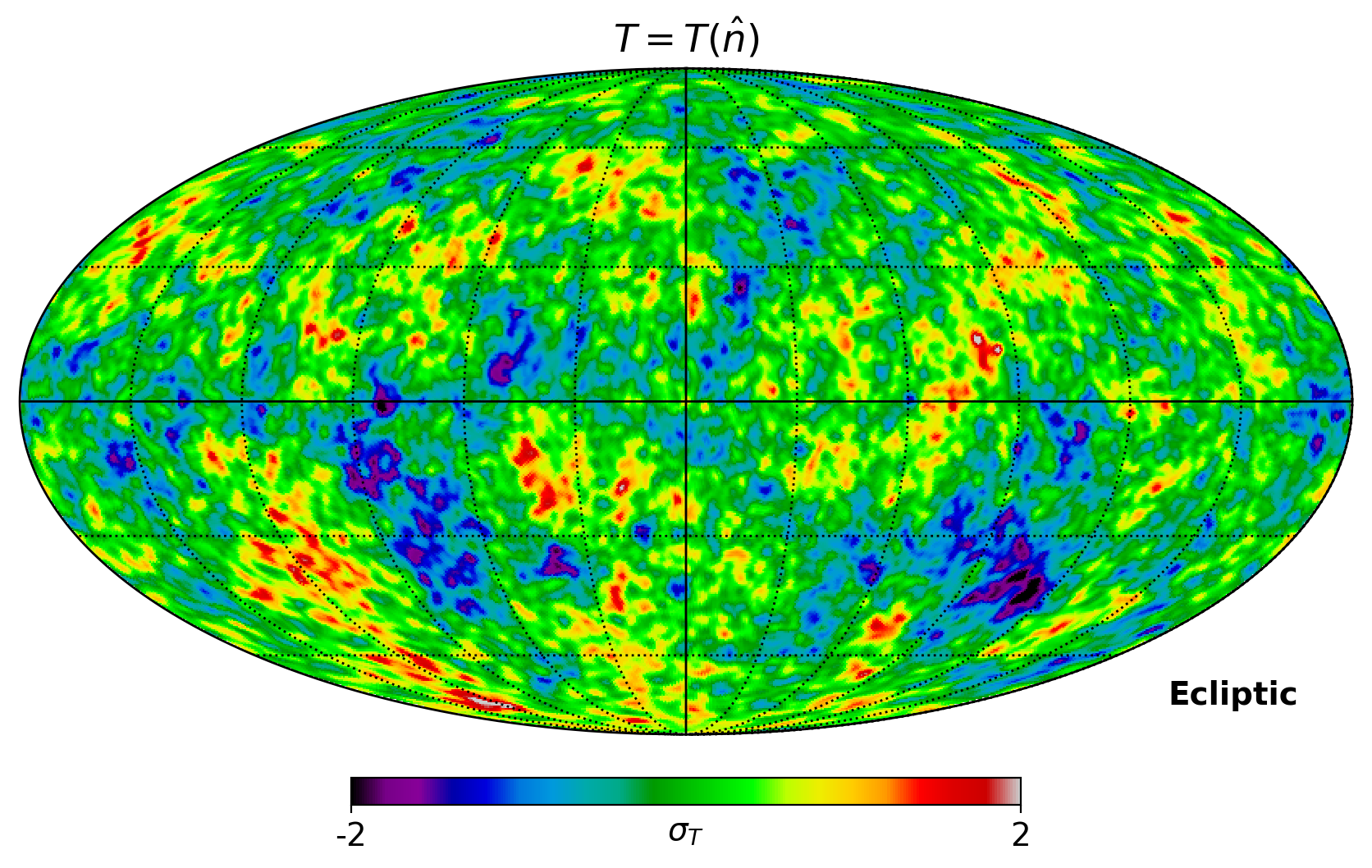}
		\includegraphics[width=0.49\linewidth]{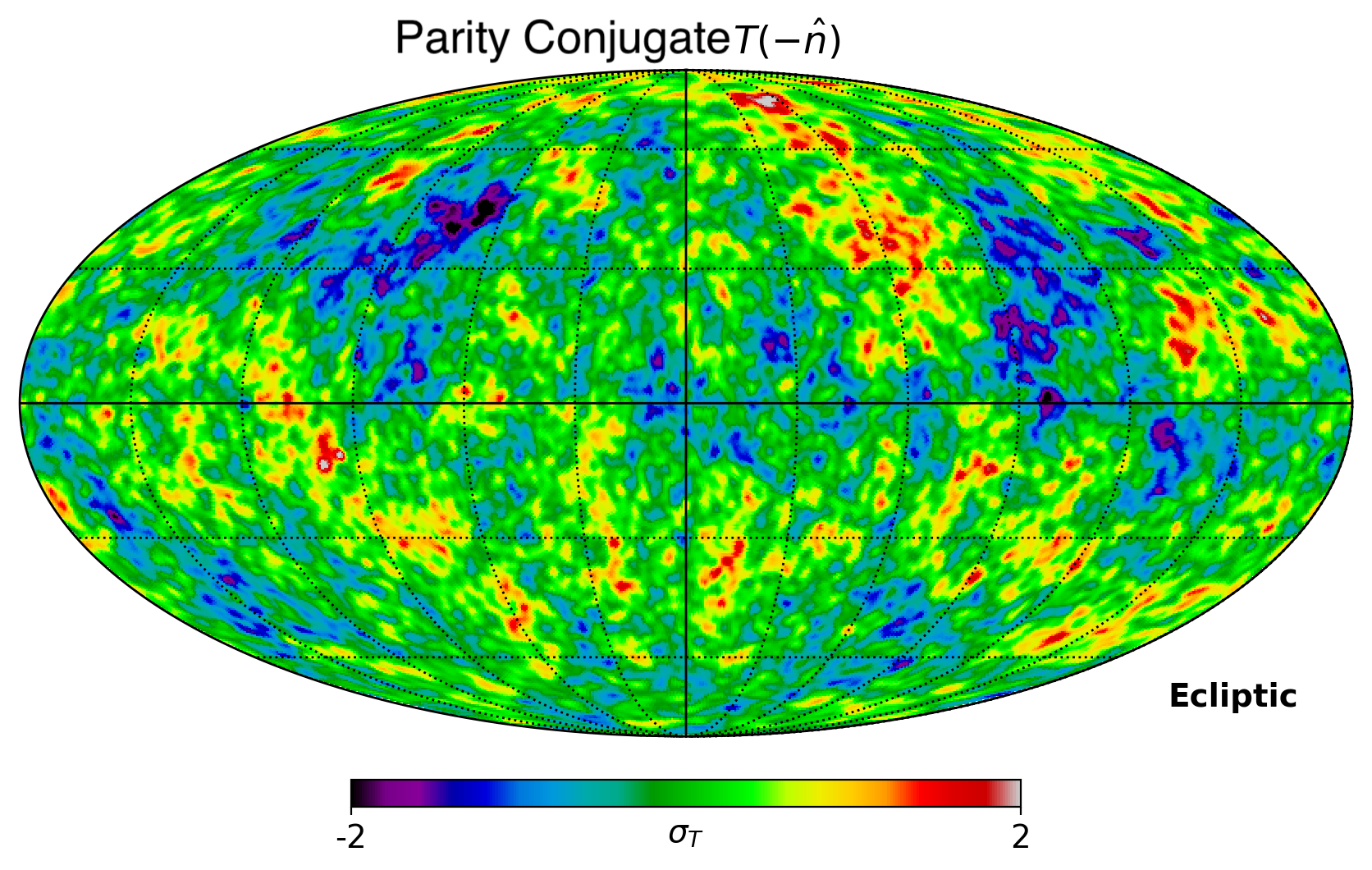} 
		\vskip -0.1cm
		\includegraphics[width=0.49\linewidth]{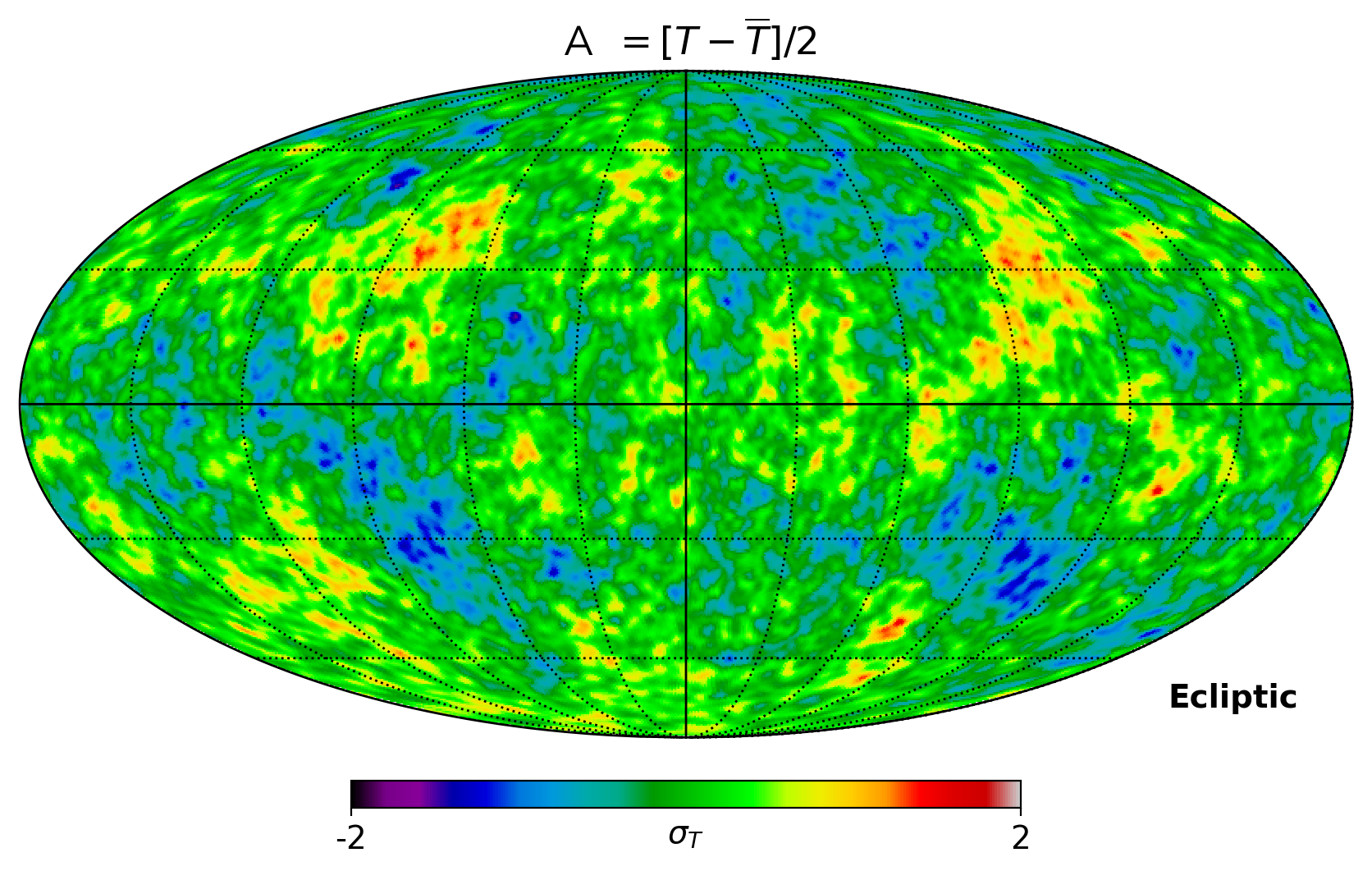}
		\includegraphics[width=0.49\linewidth]{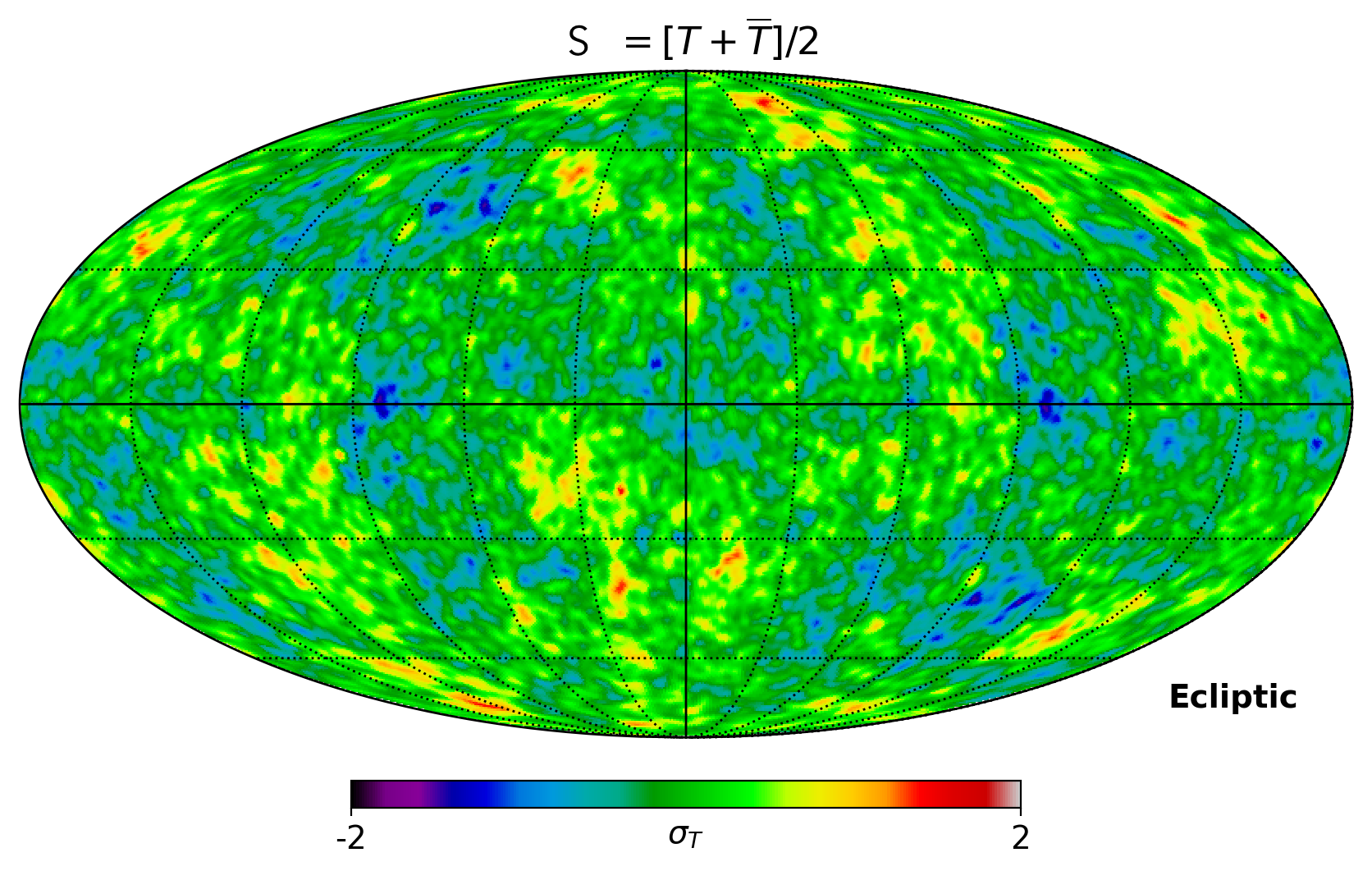}
		\vskip -0.1cm
		\includegraphics[width=0.49\linewidth]{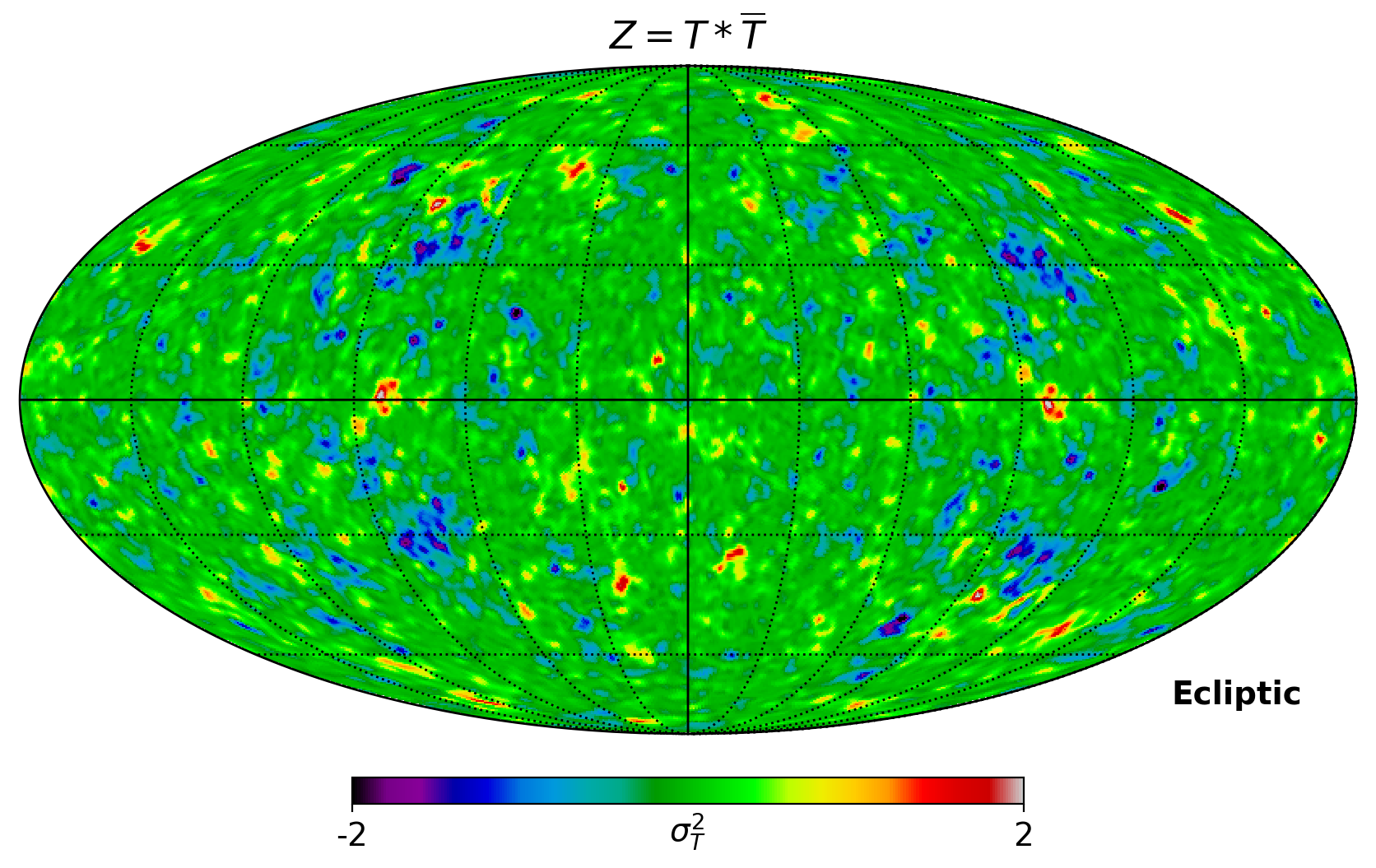}
		\includegraphics[width=0.49\linewidth]{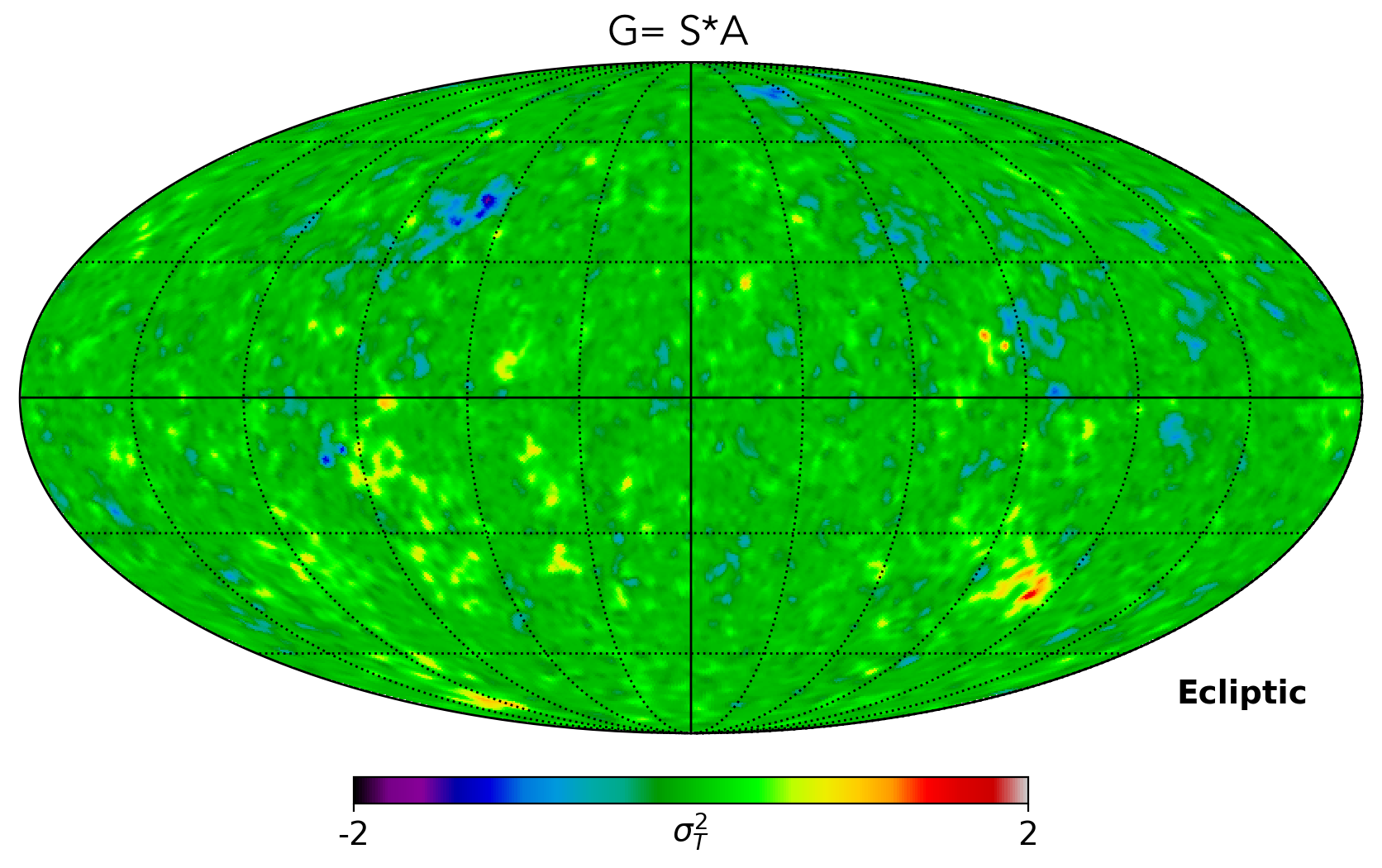}
		\caption{
			Same as Fig.~\ref{fig:smicaParity1} in Ecliptic coordinates, where there is the maximal hemispherical power difference.}
		\label{fig:smicaParity2}
	\end{figure}

	In the top-right panel, we compare it to its spherical $\Pc$ conjugate image, denoted as $\bar{T}=T(-\hat{n})$. Moving to the middle panels, we explore the odd and even parity components: $A=(T-\bar{T})/2$ and $S=(T+\bar{T})/2$. Further, we examine their multiplication, represented by $Z= T *\bar{T}$ and $G= S *A$ in the bottom panels. All these maps are presented in Galactic coordinates. For the corresponding maps in Ecliptic coordinates, refer to Figure \ref{fig:smicaParity2}.
	
	If you squint at the middle maps, you'll observe that the original $T$ map bears a closer resemblance to the odd $A$ component than the even $S$ component.  This holds in both coordinate systems. 
	This visual impression agrees very well with the 
	more quantitative split  $w^+/w^-$ shown for $w(\theta)$ in the left panel of Fig.~\ref{fig:Ctheta} or the even/odd oscillations of $C_\ell$ in Fig.~\ref{fig:clfit}.
	Another indicator is that the $Z$ map is predominantly dominated by blue (i.e., negative) values, which matches with the asymmetry evident in $P(Z)$ distribution in Fig.~\ref{fig:ZhistSMICA}.
	
\begin{figure}
		\centering
		\includegraphics[width=0.49\linewidth]{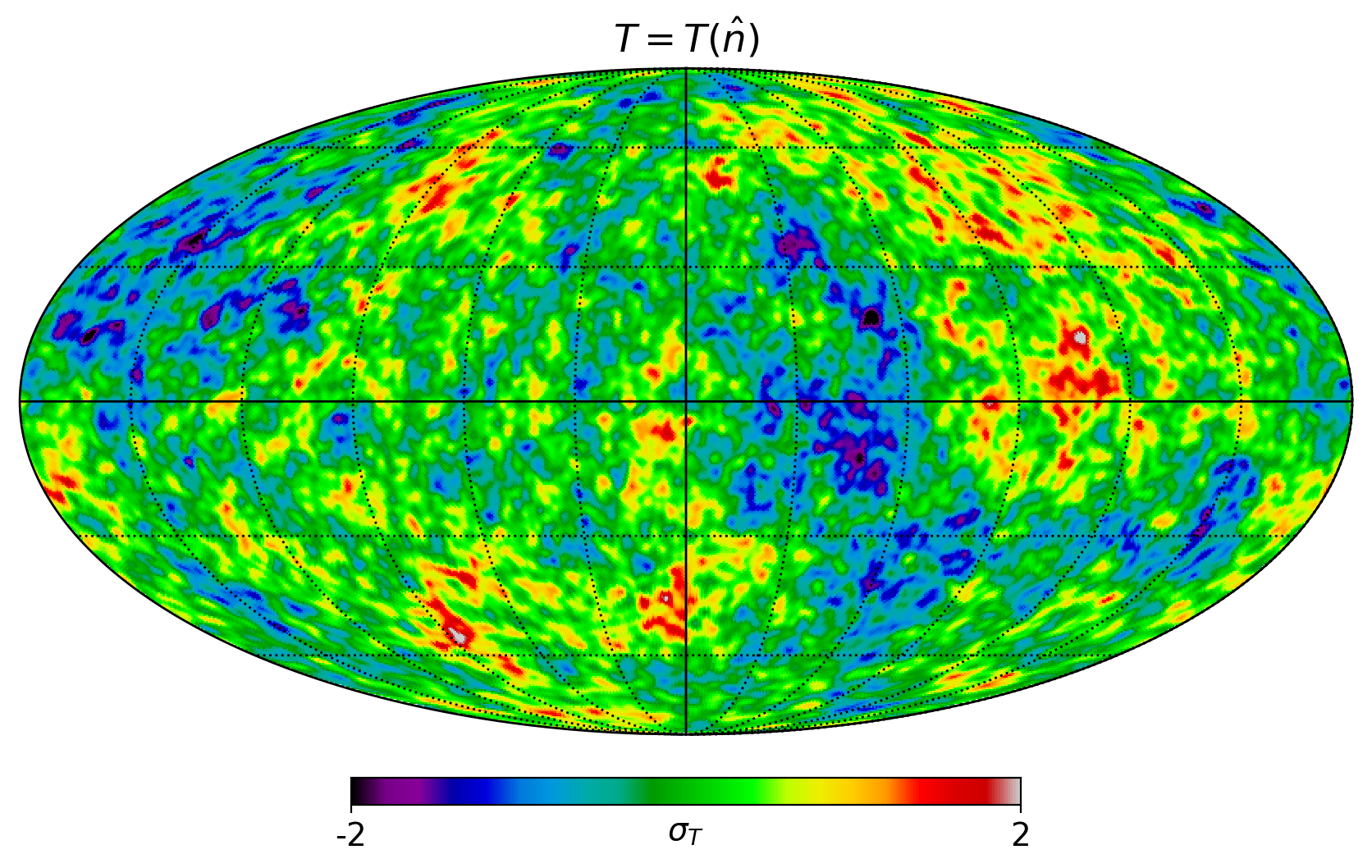}
		\includegraphics[width=0.49\linewidth]{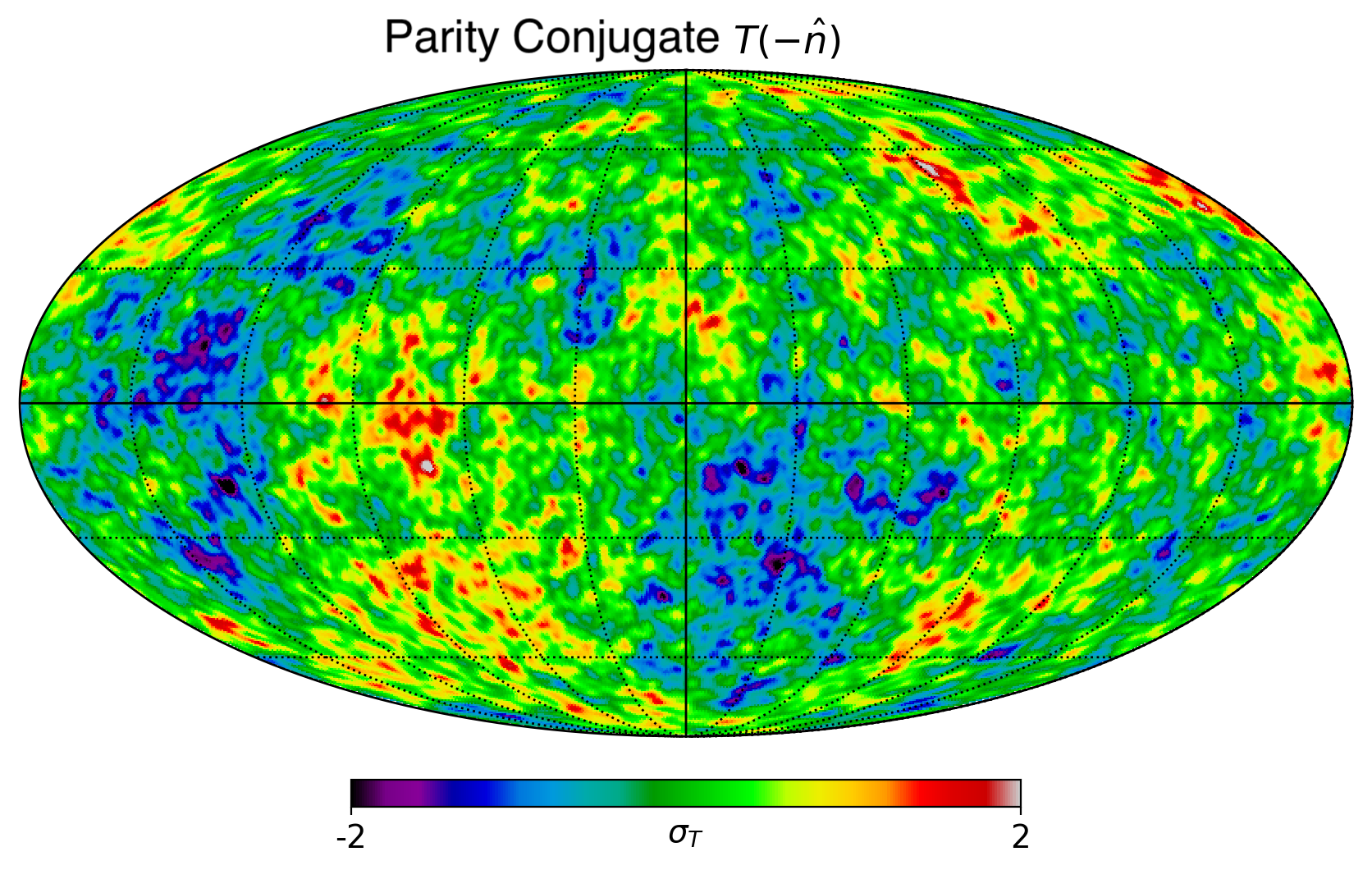} 
		\vskip -0.1cm
		\includegraphics[width=0.49\linewidth]{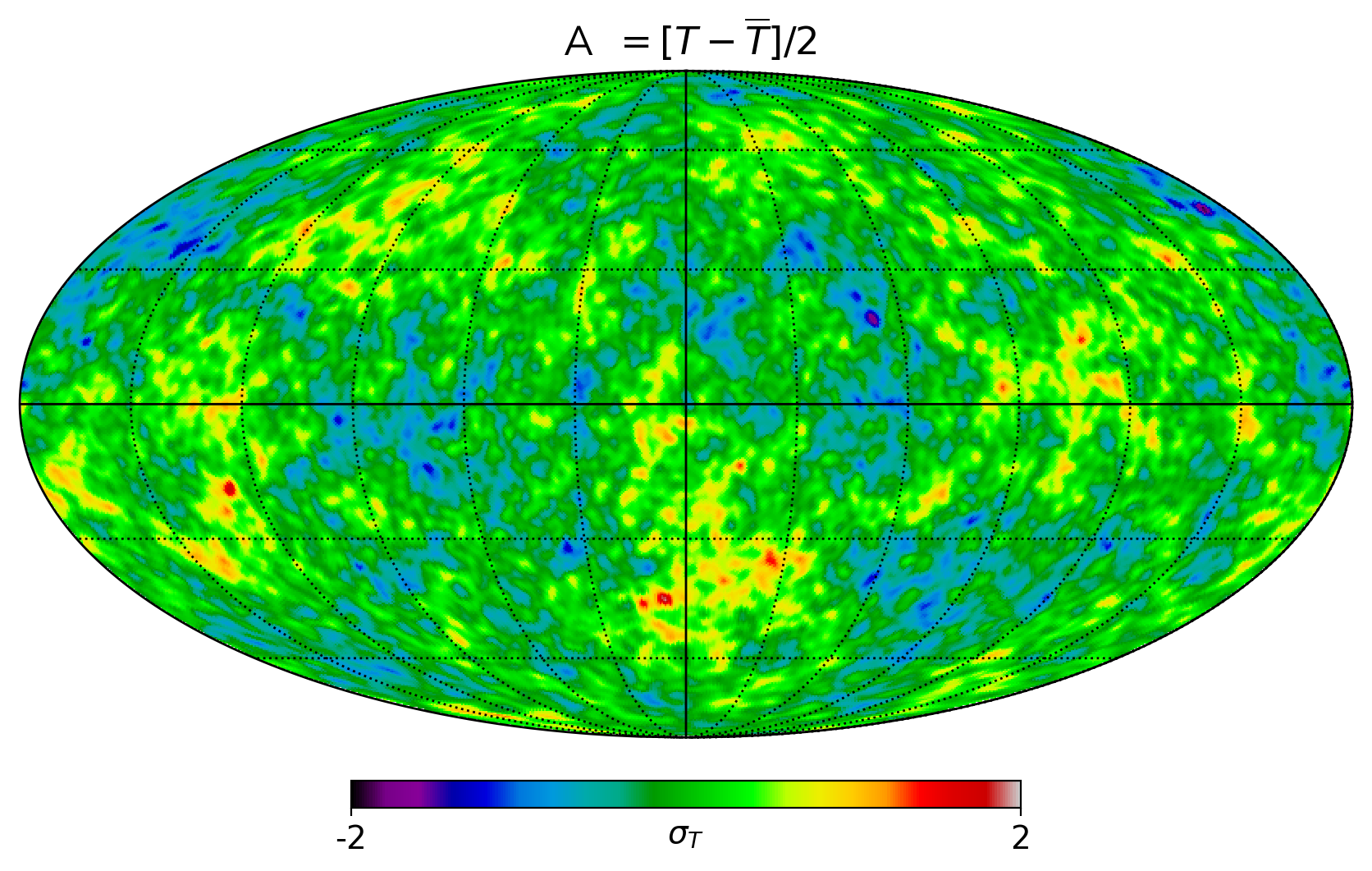}
		\includegraphics[width=0.49\linewidth]{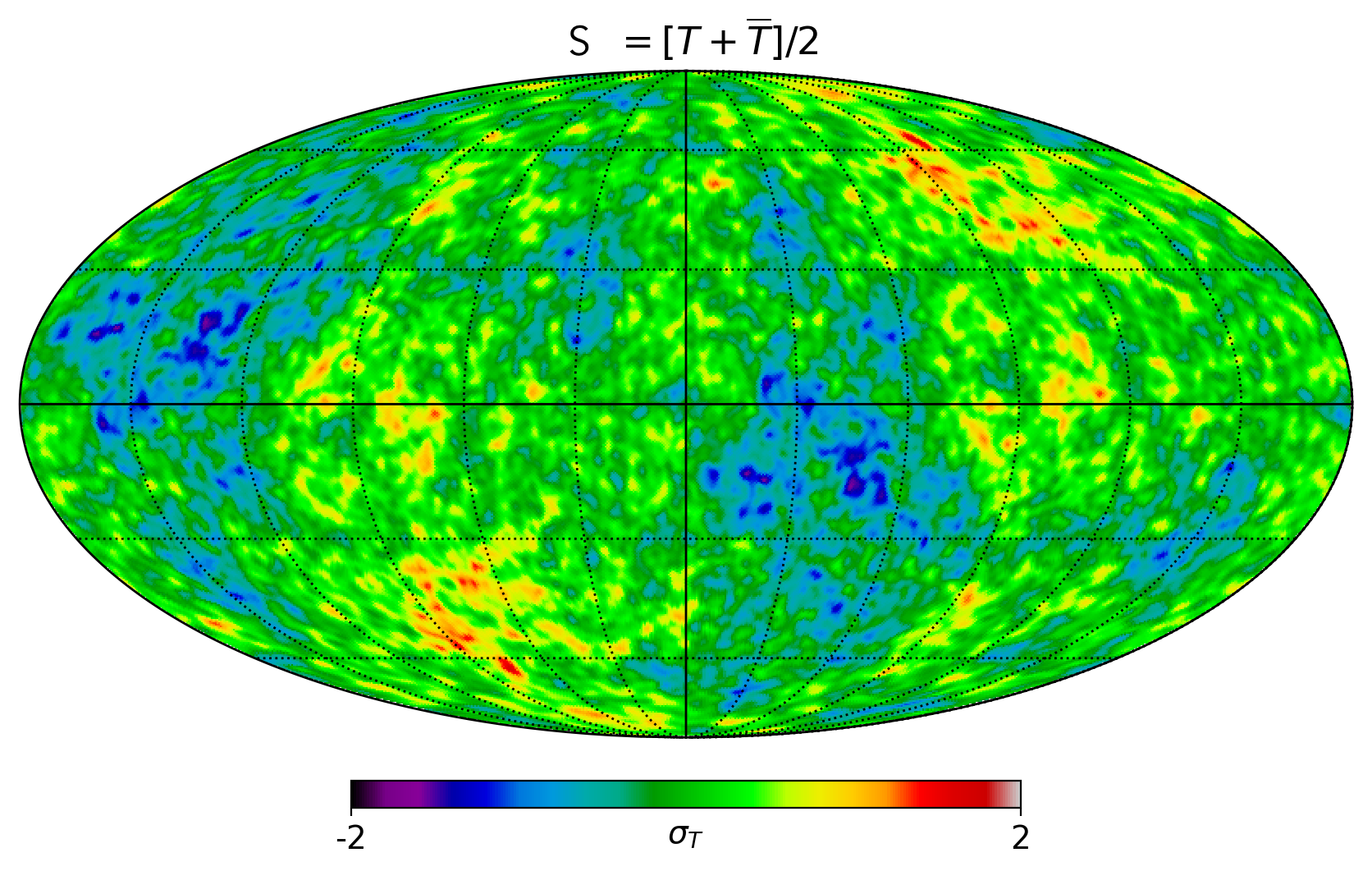}
		\vskip -0.1cm
		\includegraphics[width=0.49\linewidth]{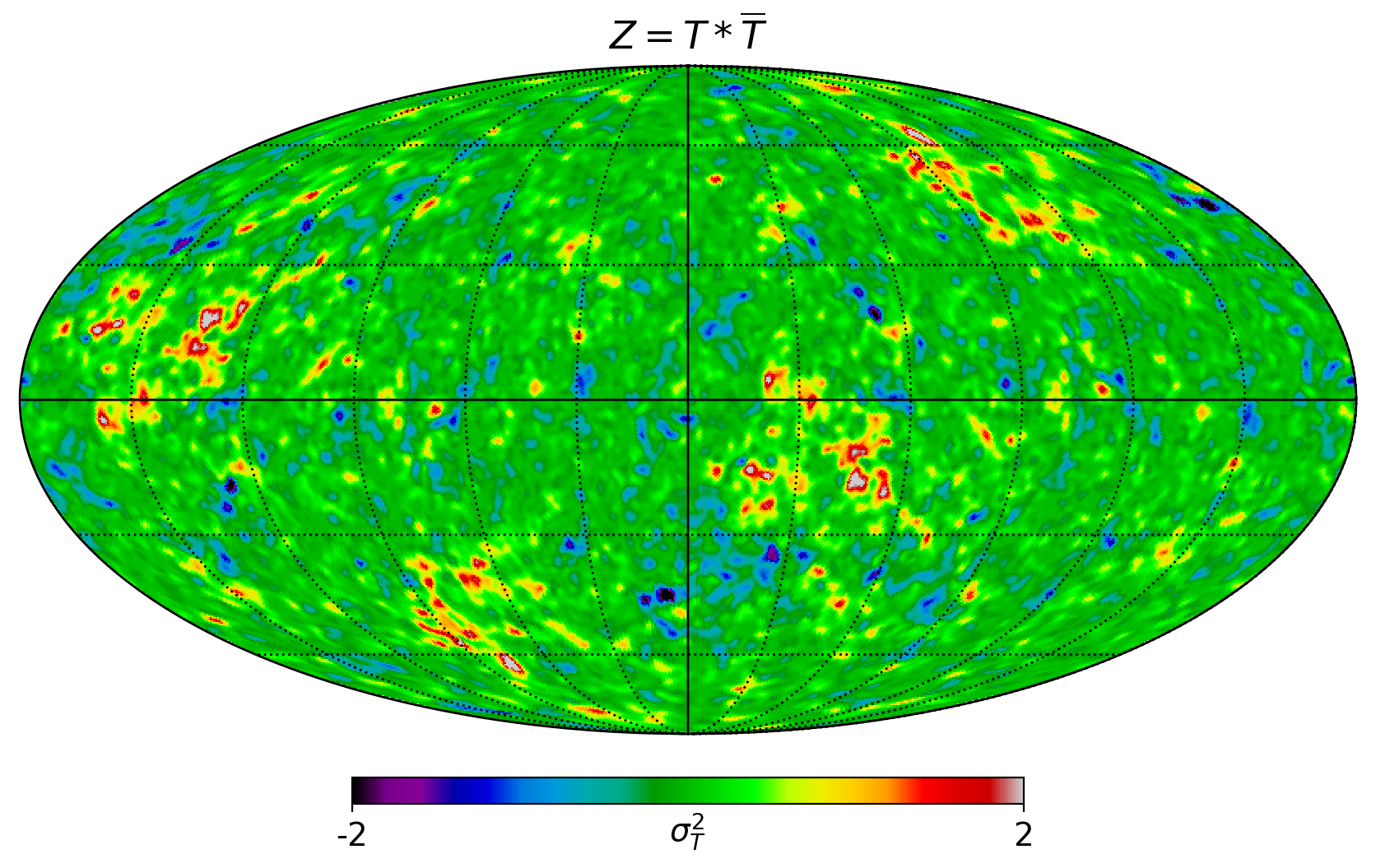}
		\includegraphics[width=0.49\linewidth]{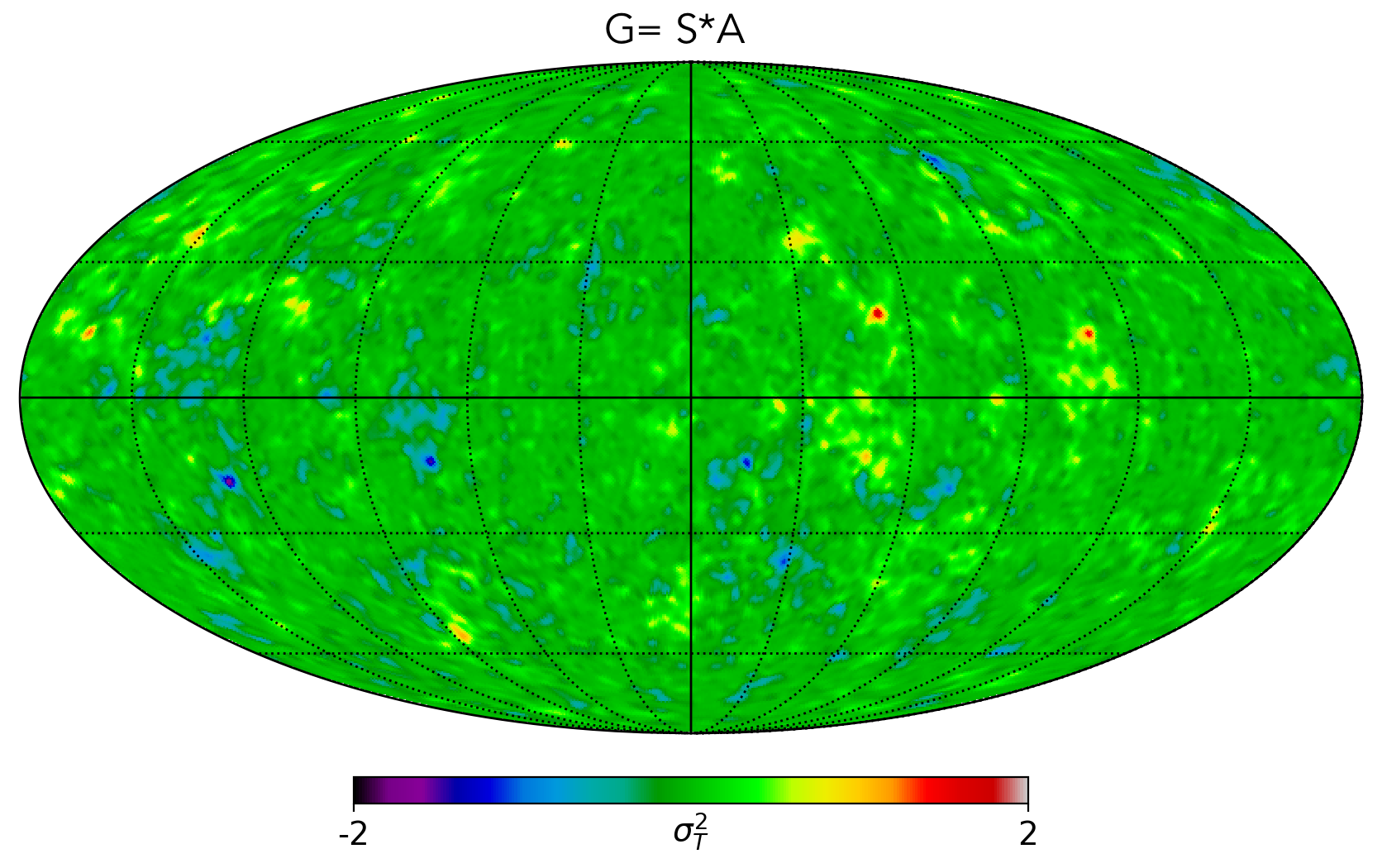}
		\caption{
			Same as Fig.~\ref{fig:smicaParity1} for a randomly selected realization of the SI+LCDM (with standard primordial inflation, SI).  On average LCDM shows more even distribution of power in the odd/even parity maps than the Planck map.}
		\label{fig:simParitySI}
	\end{figure}

\begin{figure}
		\centering
		\includegraphics[width=0.49\linewidth]{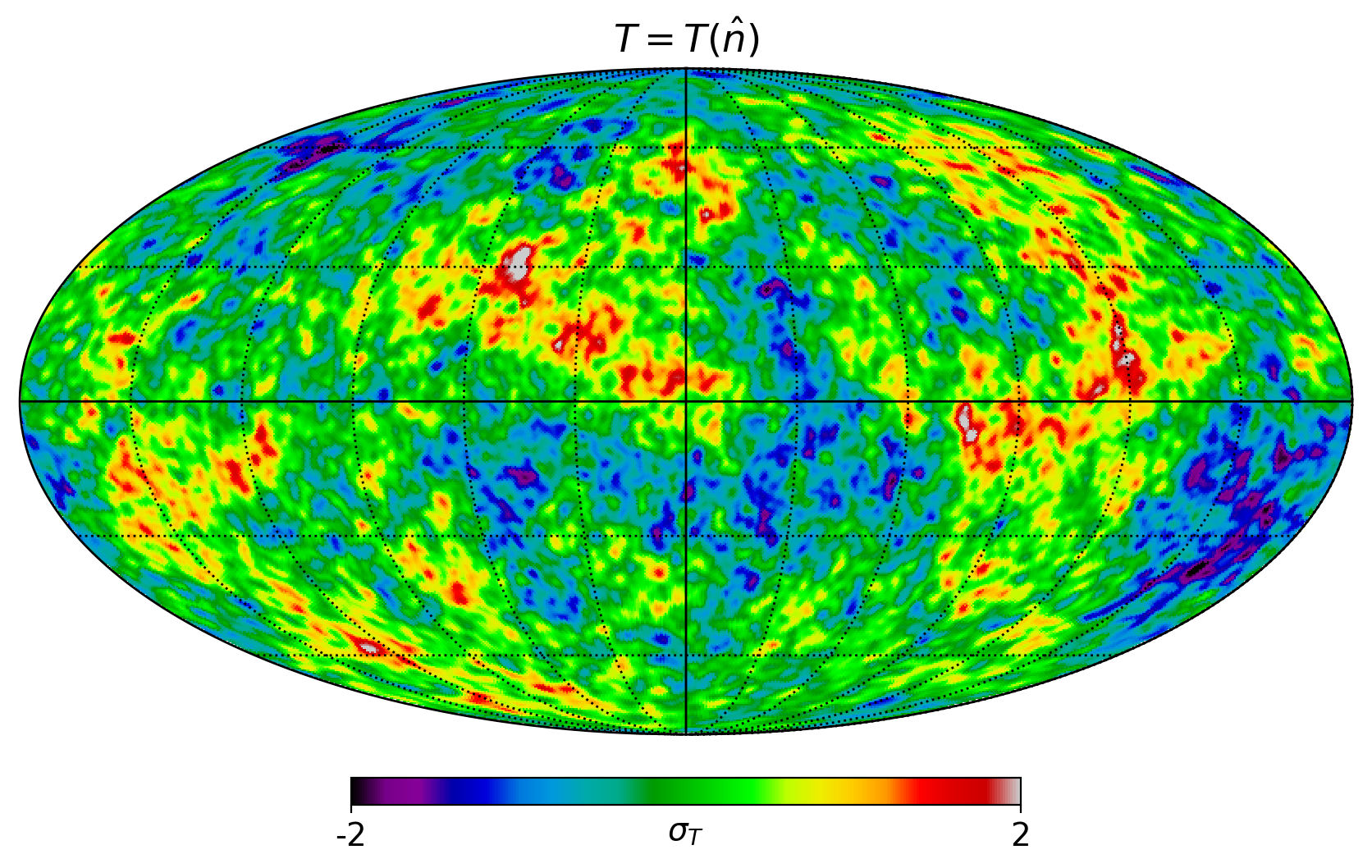}
		\includegraphics[width=0.49\linewidth]{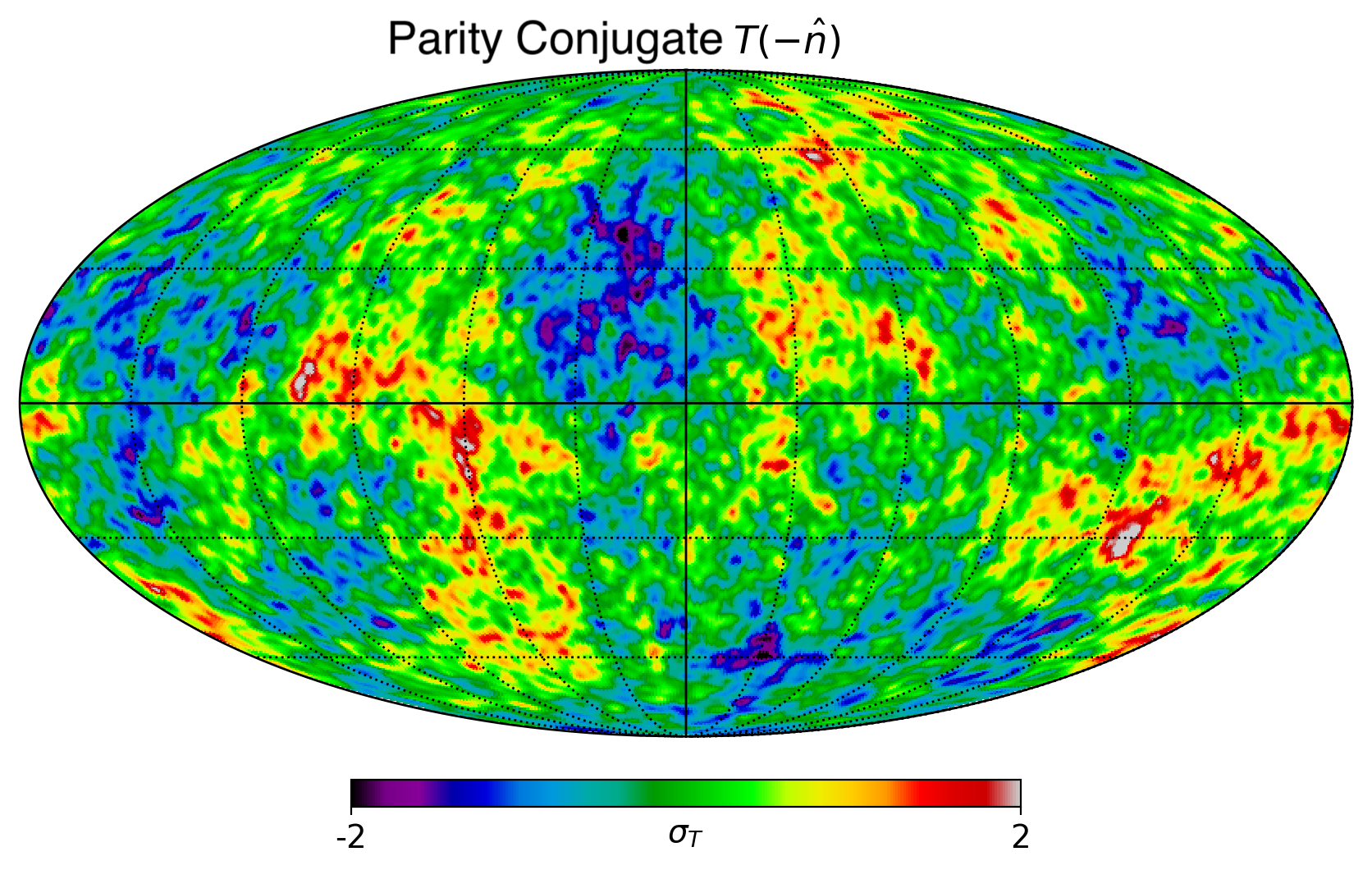} 
		\vskip -0.1cm
		\includegraphics[width=0.49\linewidth]{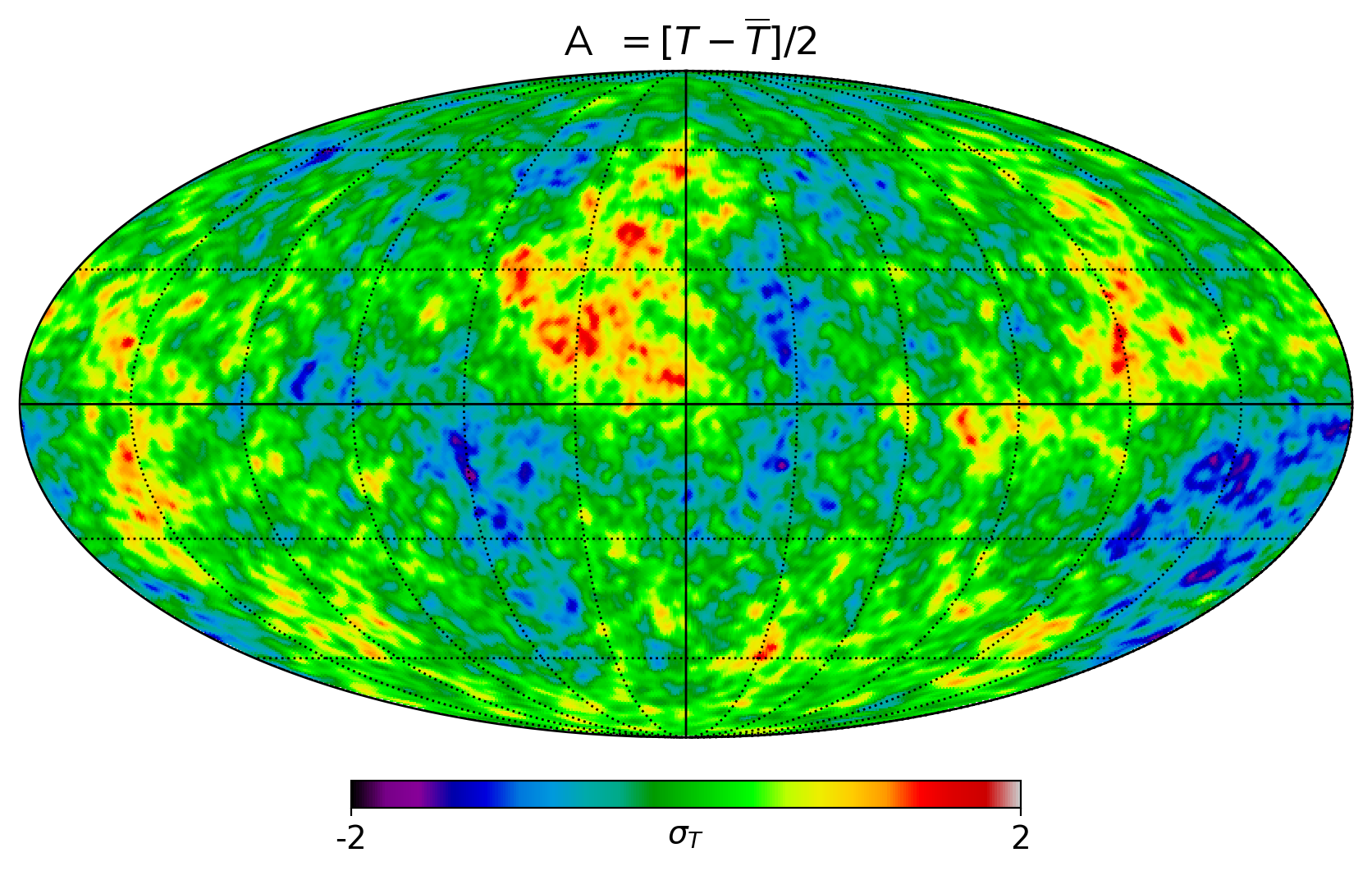}
		\includegraphics[width=0.49\linewidth]{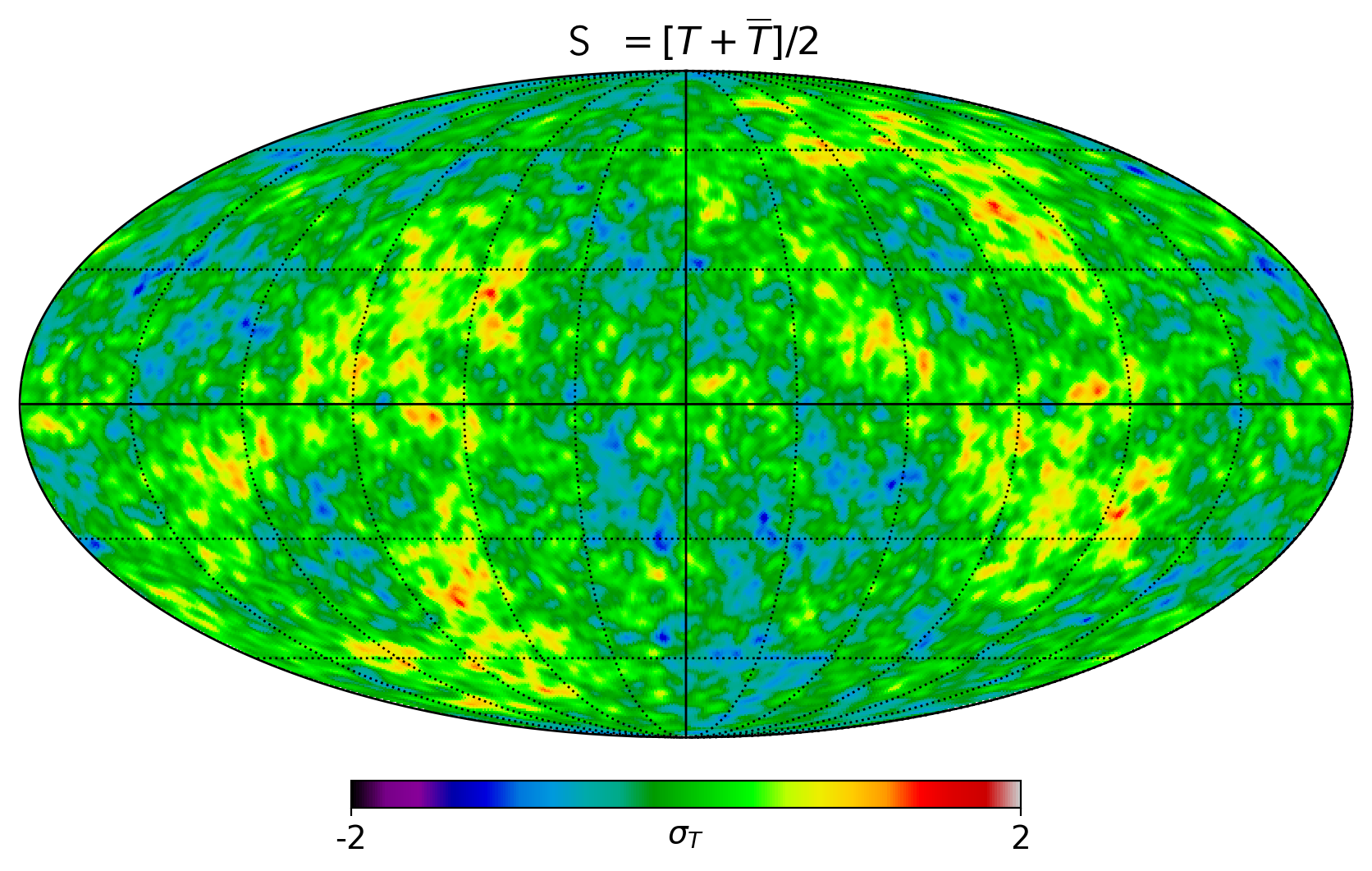}
		\vskip -0.1cm
		\includegraphics[width=0.49\linewidth]{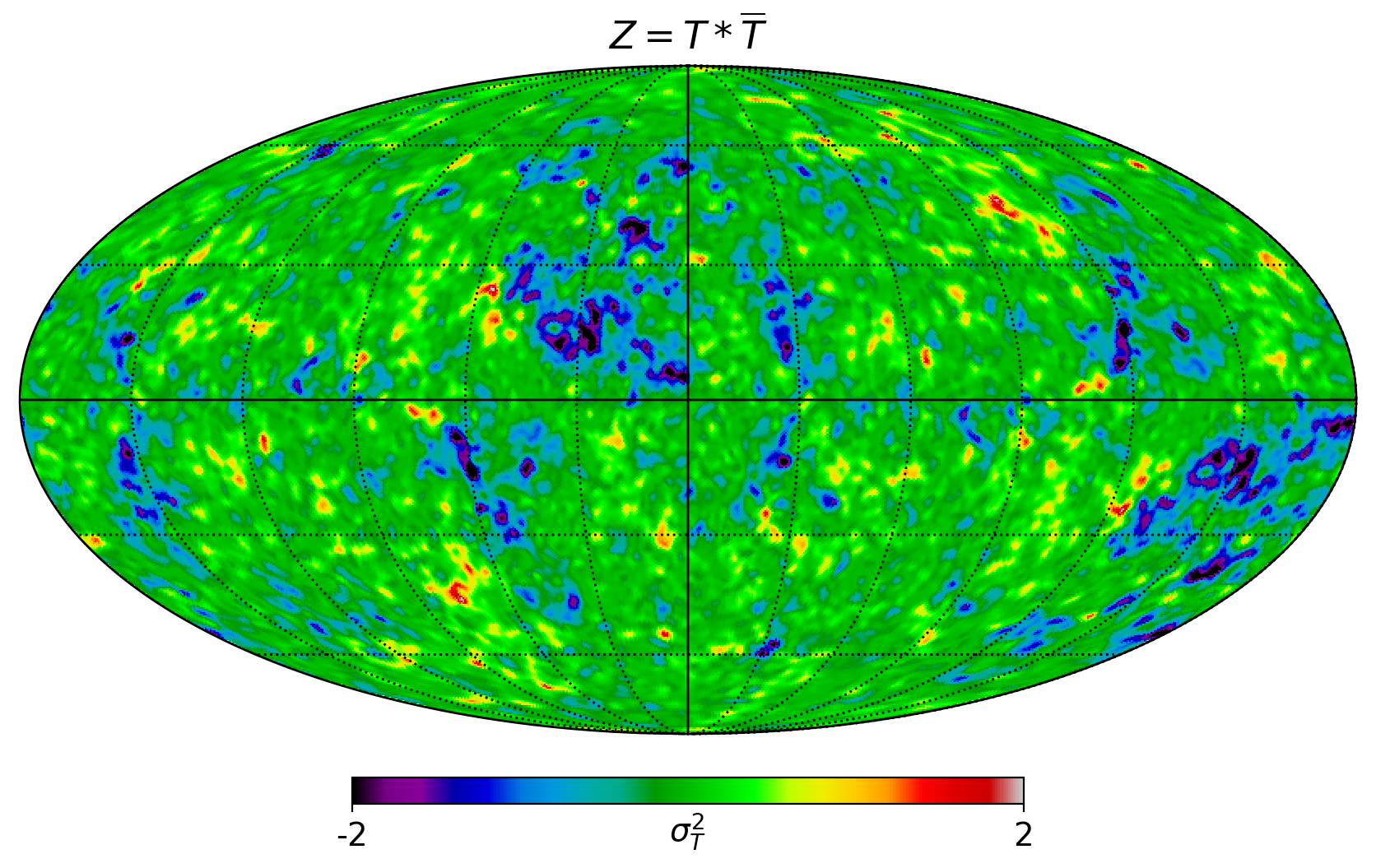}
		\includegraphics[width=0.49\linewidth]{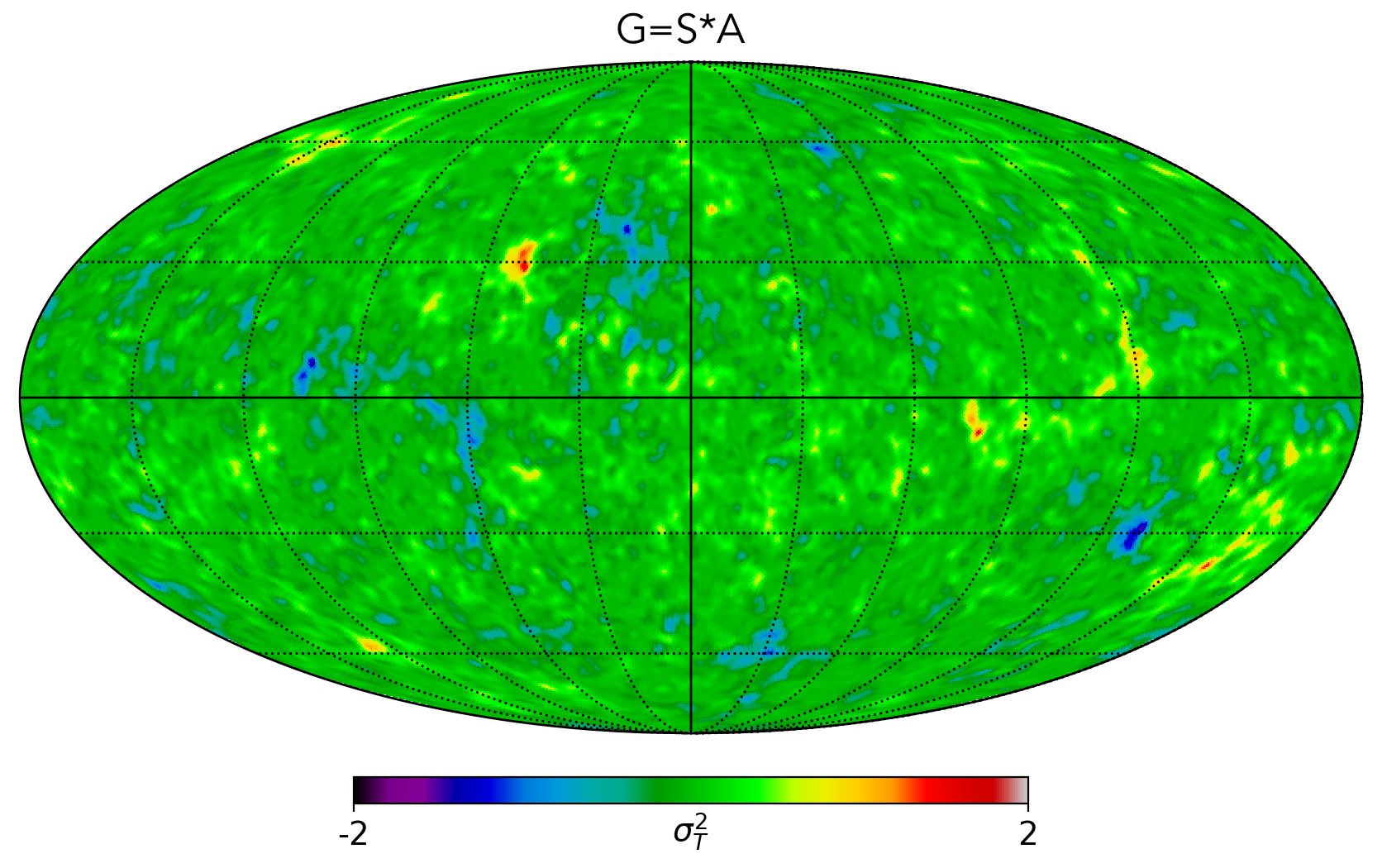}
		\caption{
			Same as Fig.~\ref{fig:smicaParity1} for a randomly selected realization of the DSI+LCDM.  It shows prominent odd parity maps $A$ and negative values of $Z$, as in the observed Planck map.}
		\label{fig:simParityDSI}
	\end{figure}

	In comparison, the corresponding maps for SI realization, as illustrated in Fig.~\ref{fig:simParitySI}, exhibit no discernible preference for even or odd parity. These visual observations can be further contrasted with the DSI maps in Fig.~\ref{fig:simParityDSI}, which exhibit similar odd parity features as observed in the Planck maps.
	
	In contrast with the other maps, the maps of $G=A*S$ (bottom right corner of each figure) are consistent with white noise in all cases. This indicates that the odd and even components are not coupled, which is what we expect from Gaussian statistics.
	This characteristic arises from the independence of each multipole, and it trivially extends to the independence between even and odd multipoles.
	
	Exceptions to this rule are observed in the form of rare bright hot spots. These spots, appearing as cold spots in the antipodal opposite direction due to $\Pc$ conjugate symmetry construction, might signify rare events or could be artifacts caused by galactic or foreground contamination in some of the maps. 
	Rare events are inevitable due to the finite size of the observable sky realization. At any resolution, this realization has a maximum number of unique draw values, with the extreme values being labeled as rare events or anomalies, as discussed in \cite{Creswell:2021eqi}. In our analysis, we mitigate the impact of these rare events by employing a 4-sigma clipping to the maps before conducting statistical measurements. This approach allows us to distinctly separate the parity study from the interpretation of rare events.
	
\begin{figure}
		\centering
		\includegraphics[width=0.49\linewidth]{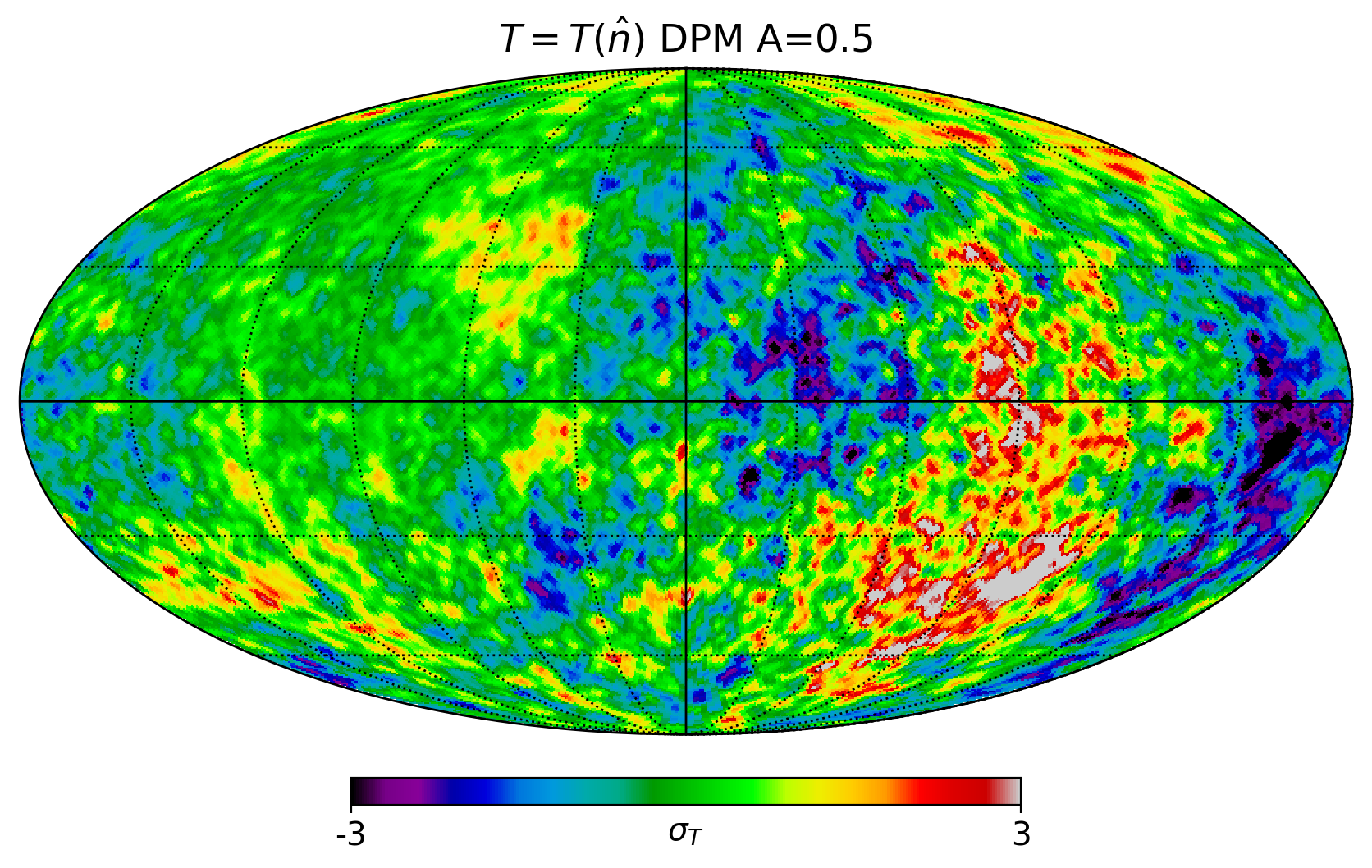}
		\includegraphics[width=0.49\linewidth]{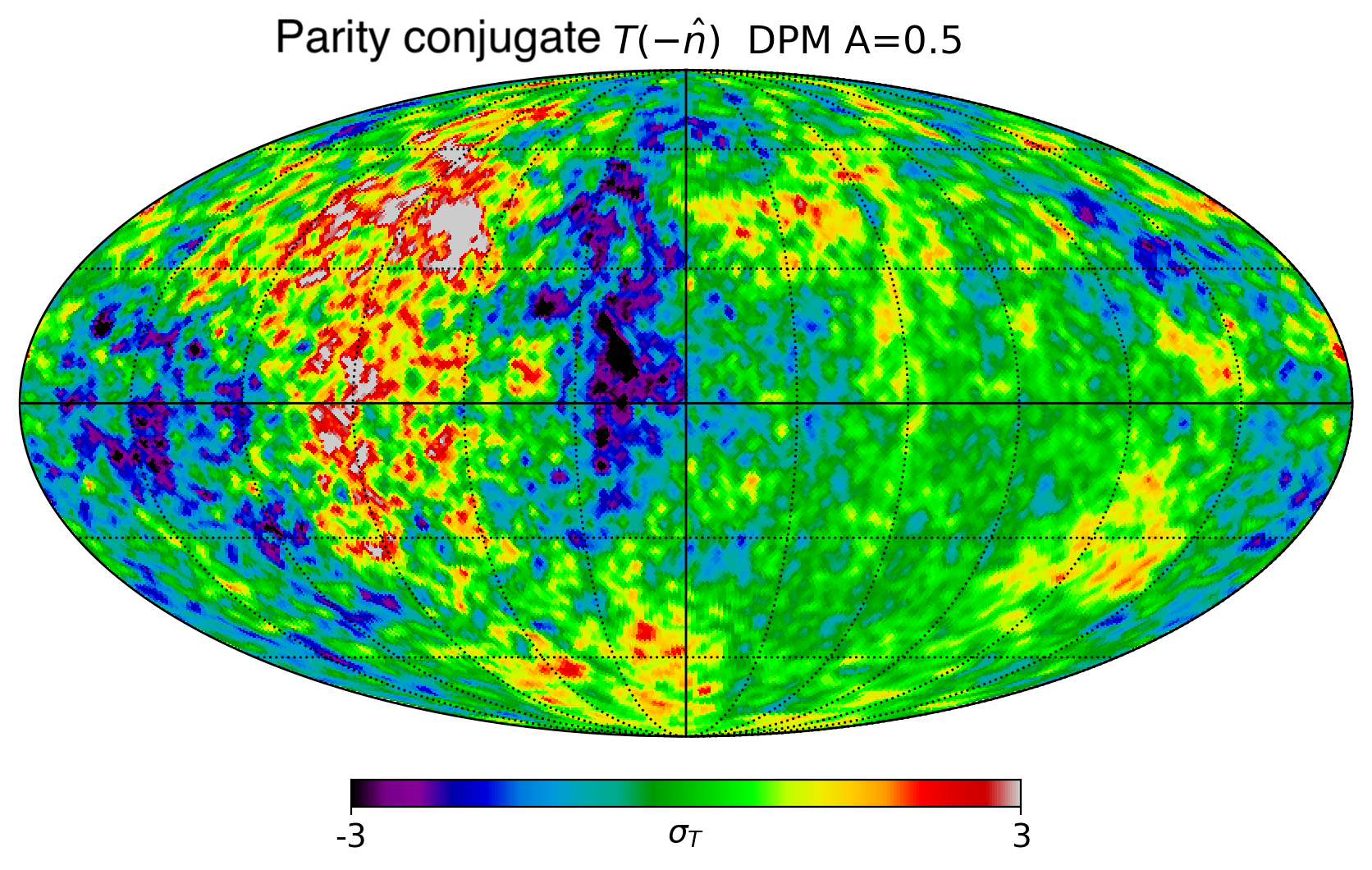} 
		\vskip -0.1cm
		\includegraphics[width=0.49\linewidth]{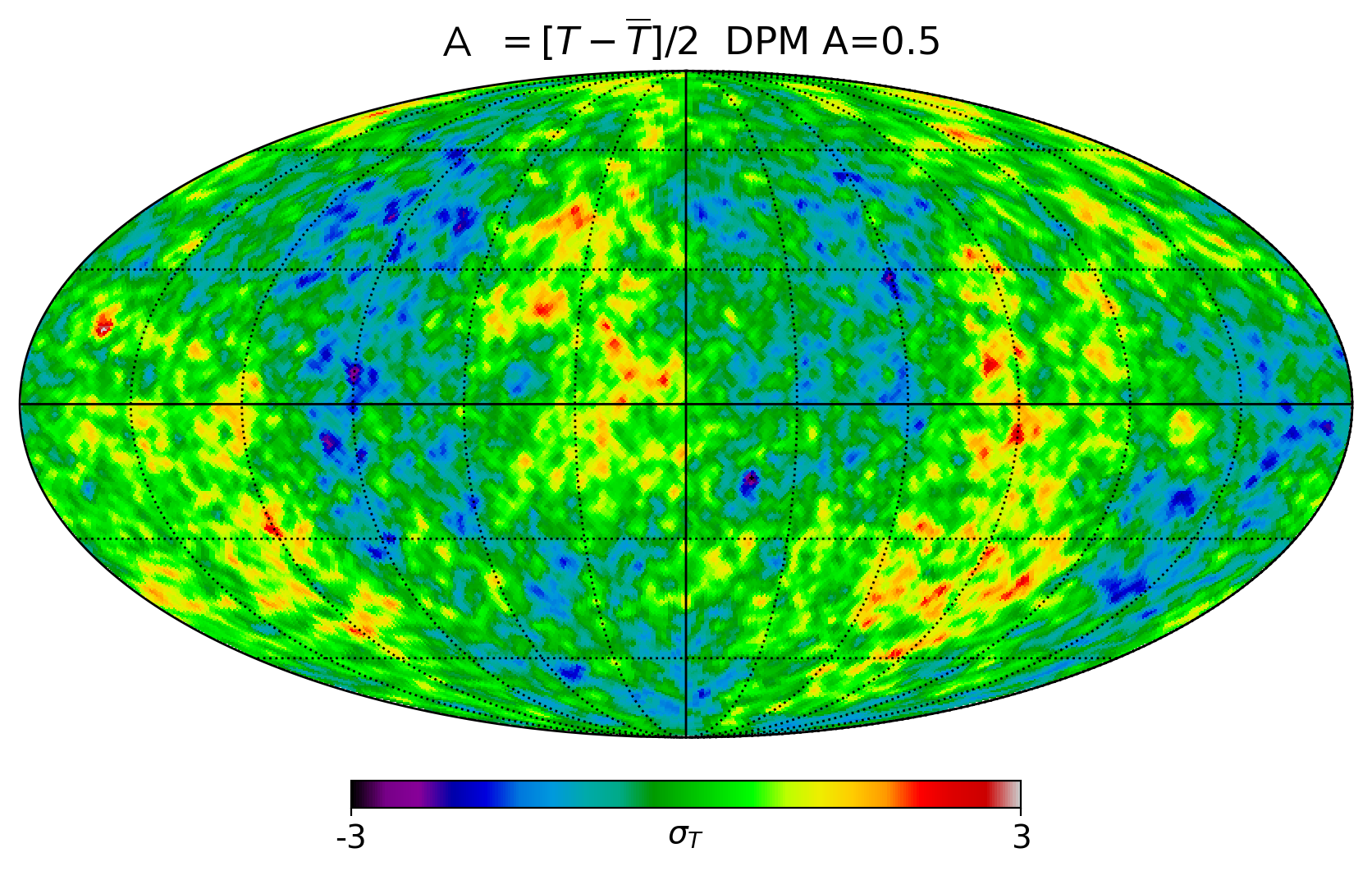}
		\includegraphics[width=0.49\linewidth]{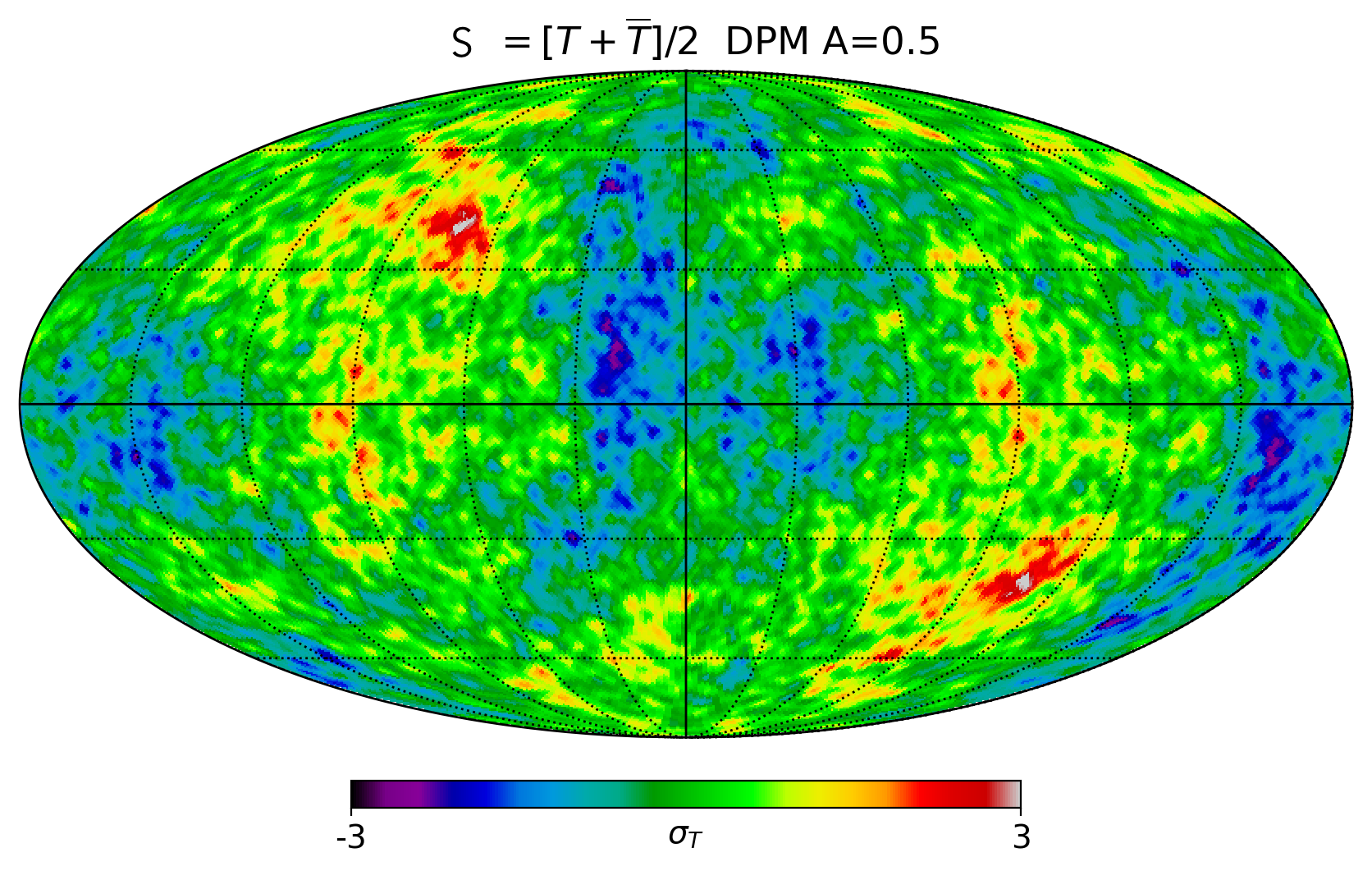}
		\vskip -0.1cm
		\includegraphics[width=0.49\linewidth]{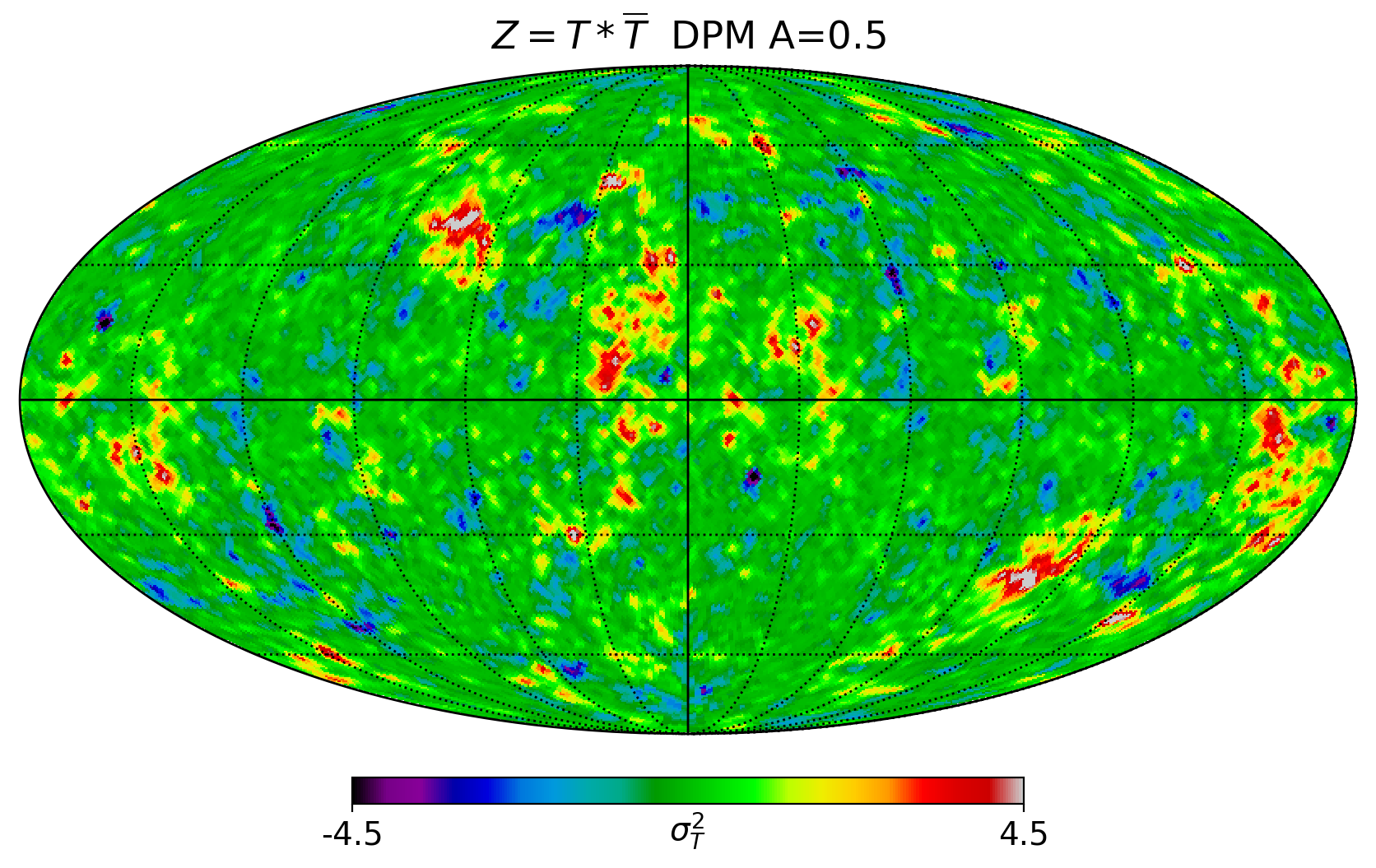}
		\includegraphics[width=0.49\linewidth]{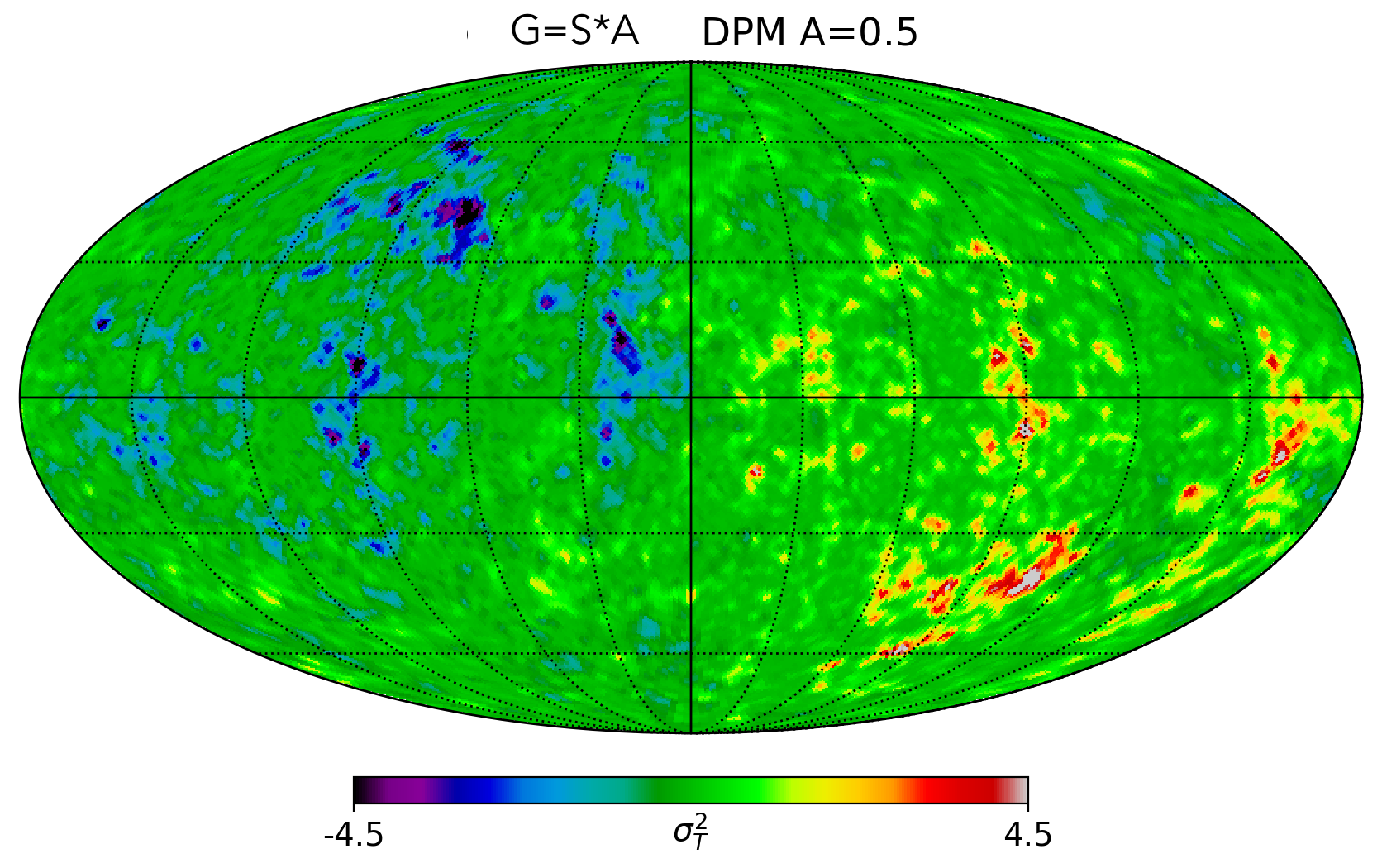}
		\caption{
			Same as Fig.~\ref{fig:smicaParity1} for a randomly selected realization of the extremely anisotropic DPM model with $A=0.5$.  Despite the very large dipolar anisotropy, the model shows no preference for odd parity in the $Z$ maps other than some rare peaks at the position of North and South Ecliptic poles, where the dipole mimics parity. 
			Such rare peaks have little impact on the overall parity statistics (especially when we include a 4-$\sigma$ clipping). }
		\label{fig:simParityDPM}
	\end{figure}

	The very extreme $A=0.5$ DPM model in Figure \ref{fig:simParityDPM}, shows a dipolar component in the $G$ map. Despite such strong anisotropy, the $Z$ map and the parity statistics, e.g. $P(Z)$ or $R^{TT}$, show no preference for parity (see Table.~\ref{tab2:annomalies}). This is a clear demonstration of how the $P(Z)$ distribution allows us to separate isotropy from parity.
	
		\section{Reviewing Standard Primordial Inflation (SI)}
		
		\label{sec:reviewSI}
	
	Inflation by definition, is a quasi-de Sitter (qdS) expansion of the Universe that is expected to last around 50-60 e-folds before exiting to reheating phase where particle production happens and the Universe eventually enters into a radiation-dominated era \cite{Starobinsky:1981vz}. To drive inflationary cosmic expansion we require at the minimum a new scalar degree of freedom which would result in a non-perturbative modification of GR whose action can be written as 
	\begin{equation}\label{action-S}
		S= \int d^4x\sqrt{-g} \LT \frac{1}{2} R- \frac{1}{2} (\pd_\mu \phi)( \pd^\mu \phi  )-V\LF \phi \RF\RT  \,, 
	\end{equation}
	where the potential $V(\phi)$ is essential and it require to have a plateau-like shape to explain the the current CMB contraints \cite{Planck:2018jri}. The Starobinsky potential 
	\begin{equation}
		V(\phi) = \frac{3}{4}M^2 \LF 1-e^{-\sqrt{\frac{2}{3}}\frac{\phi}{M_p}} \RF^2\,. 
	\end{equation}
	which is based on the quadratic scalar curvature modification of gravity is the most consistent theory with observations so far. Moreover, the Starobinsky-like inflationary scenarios are found to occur in multiple frameworks of quantum gravity that are currently the active field of research \cite{Koshelev:2017tvv,Koshelev:2022olc,Koshelev:2023elc,Linde:2014nna,Ellis:2013nxa}.  Even though inflationary quantum fluctuations are largely understood to be the origin of density fluctuations observed through the CMB, still there are fundamental questions related to the quantum effects in curved spacetime which we shall discuss in the next section. Before that, it is vital here to recall the minute details of SI quantum fluctuations \cite{Mukhanov:1990me}.

	The scalar perturbations arise from metric fluctuations ($g_{\mu\nu}=\bar{g}_{\mu\nu}+\delta g_{\mu\nu}$) and also the scalar field fluctuations ($\phi = \bar{\phi}+\delta\phi$). The metric fluctuations can be represented by the Arnowitt-Deser-Misner (ADM) metric of the form 
	\begin{equation}\label{ADMmetric}
		ds^2 =  a^2\LF \tau \RF \Big(-\Nc^2d\tau^2+  \gamma_{ij} \LF dx^i+\Nc^i d\tau \RF \LF dx^j+\Nc^j d\tau \RF \Big)\,.
	\end{equation}
	where $d\tau = \frac{dt}{a}$ is conformal time, $ \Nc$ and $\Nc_i$ are the lapse and shift functions respectively.  In the unitary gauge, we fix 
	$\delta\phi=0$ then we linearly expand the ADM metric  as
	\begin{equation}\label{eq:ADMfluctuation}
		\Nc= 1+\delta \Nc,\quad \Nc_i = a\pd_i\chi,\quad \gamma_{ij} =  \LF 1+2\zeta \RF\delta_{ij}+h_{ij}\,, 
	\end{equation}
	where $\chi$ is a scalar function of spacetime, $\zeta $ is the curvature perturbation , $h_{ij}$ is the transverse and traceless spin-2 fluctuation. In this paper, we only focus on scalar flucutations as our study is only about temperature fluctuations in the CMB. 
	From the linear perturbed equations of motion, we obtain the following constraints 
	\begin{equation}
		\begin{aligned}
			\delta \Nc & = \frac{\dot{\zeta}}{H} \\ 
			\pd^2\chi & = -\frac{\pd^2 \zeta}{H}+ \epsilon a^2\dot{\zeta}
		\end{aligned}
	\end{equation}
	{Substituting Eq.~\ref{ADMmetric} into the action Eq.~\ref{action-S} and expanding to second order in the fluctuations, we find the second order action for the scalar perturbation}:
	\begin{equation}\label{scalar}
		\delta^{(2)}S_{s} = \frac{1}{2}\int d\tau d^3x a^2\frac{\phi^{\prime 2}}{\Hc^2} \Bigg[ \zeta^{\prime 2} -\LF \pd\zeta \RF^2 \Bigg]
		\,\text{,}
	\end{equation}
	where $\Hc = \frac{a^\prime}{a}= aH$ and $\tau = \int \frac{dt}{a} = - \frac{1}{aH}$ is the conformal time.

	To quantize the scalar fluctuations we perform a field redefinition to define a canonical variable 
	\begin{equation}
		v=a\zeta \dot{\phi}/H
		\label{canvar}
	\end{equation}
	given by 
	\begin{equation}
		\delta^{(2)}S_s = \frac{1}{2} \int d\tau d^3x \Big[v^{\prime 2}-\LF\pd_iv\RF^2-\LF \frac{\nu_s^2-\frac{1}{4}}{\tau^2} \RF v^2\Big],
		\label{screfa}
	\end{equation}
	where $\nu_s$ is a function of the slow-roll parameters 
	\begin{equation}
		\epsilon = -\frac{\dot{H}}{H^2},\quad \eta = \frac{\dot{\epsilon}}{H\epsilon}
		\label{slp}
	\end{equation}
	as
	\begin{equation}
		\nu_s = \frac{3}{2}+ 2\epsilon+\eta\,. 
	\end{equation}
	
	\subsection{Standard inflationary quantum fluctuations}
	
	By inspection, we can recognize that Eq.~\ref{screfa} $v$ is a Klein-Gordan field with time-dependent mass $m_{eff}^2= \frac{\nu_s^2-\frac{1}{4}}{\tau^2}$. Quantization means we promote the canonical variable to an operator and express it in terms of creation and annihilation operators as 
	\begin{equation}
		\hat{v}\LF \tau,\,\textbf{x} \RF = \int \frac{d^3k}{\LF 2\pi \RF^{3/2}} \Big[\hat{a}_\textbf{k}v_k\LF \tau \RF e^{i\textbf{k}\cdot x}+ \hat{a}_{\textbf{k}}^\dagger v^\ast_{k}\LF \tau \RF e^{-i\textbf{k}\cdot x}\Big]\,. 
		\label{qunfluc}
	\end{equation}
	The crux of quantum theory lies in the non-commutativity of the fields and its conjugate momenta that results in 
	\begin{equation}
		\Big[	\hat{v}\LF \tau,\, \textbf{x} \RF,\, \hat{v}^\prime\LF \tau,\, \textbf{y} \RF\Big] = i\delta \LF \textbf{x}-\textbf{y}  \RF
		\label{comq}
	\end{equation}
	supplemented by 
	\begin{equation}
		\Big[a_\textbf{k},\,a_\textbf{p}^\dagger\Big] = \delta\LF \textbf{k}-\textbf{p} \RF\quad 	\Big[a_\textbf{k},\,a_\textbf{p}\Big] = 0
	\end{equation}
	The mode function $v_{k}$ satisfies the Mukhanov-Sasaki (MS) equation 
	\begin{equation}
		v_{k}^{\prime\prime}+ \LF k^2-\frac{\nu_s^2-\frac{1}{4}}{\tau^2} \RF v_k = 0\,. 
		\label{MSeq}
	\end{equation}
	whose solution in the limit $\nu_s\to 3/2$ (i.e., neglecting slow-roll contributions since $\epsilon,\,\eta\ll 1$ during inflation) is given by 
	\begin{equation}
		\begin{aligned}
			v_k(\tau)  & = A_k \frac{\sqrt{\pi \vert \tau \vert}}{2} H_{3/2}^{(1)}\LF -k\tau \RF+ B_k \frac{\sqrt{\pi}\vert \tau \vert}{2} H_{3/2}^{(2)}\LF -k\tau \RF \\ 
			& =  \frac{A_k}{\sqrt{2k}} e^{-ik\tau } \LF 1-\frac{i}{k\tau} \RF + \frac{B_k}{\sqrt{2k}}  e^{ik\tau } \LF 1+\frac{i}{k\tau} \RF\,
			\label{modeg}
		\end{aligned}
	\end{equation}
	where $A_k$ and $B_k$ are the Bogoliubov coefficients which can are in principle functions of $k$ satisfying the following constraint which is obtained by substituting Eq.~\ref{modeg} in Eq.~\ref{comq}  
	\begin{equation}
		\vert A_k\vert^2 - \vert B_k\vert^2 = 1 \implies \Big\{A= \cosh\alpha_k,\, B_k=\sinh\alpha_k\Bigg\}\,. 
	\end{equation}
	From Eq.~\ref{modeg} we can learn that in the limit $\vert k\tau \vert \gg 1$ we end up with 
	\begin{equation}
		v_{k}^{\rm{MS}}	= v_{k} \Bigg\vert_{\vert k\tau\vert \gg 1} \approx \frac{A_k}{\sqrt{2k}} e^{-ik\tau }  + \frac{B_k}{\sqrt{2k}}  e^{ik\tau }
	\end{equation}
	Considering $A_k=1$ and $B_k=0$ the mode function behaves as a positive energy state that is defined according to the Schr\"{o}dinger equation
	\begin{equation}
		i\frac{\pd}{\pd t_p}	\vert \Psi\rangle  = \hat{H} \vert\Psi\rangle \implies \Psi_t = e^{-i\hat{H} t_p } \vert \Psi\rangle_0\,, 
	\end{equation}
	where $t_p$ is the parametric time of quantum mechanics and $\hat{H}$ is the time-independent Hamiltonian and $\vert \Psi\rangle $ state vector of Heisenberg representation. The vacuum 
	\begin{equation}
		a_\textbf{k} \vert 0\rangle_{\rm{BD}} = 0\,. 
	\end{equation}
	corresponds to the choice   $A=1$ and $B=0$ is also called the Bunch-Davies or adiabatic vacuum. The power spectrum of curvature perturbation is obtained by
	\begin{equation}
		\begin{aligned}
			P_\zeta & = \frac{k^3}{2\pi^2a^2}\LF\frac{1}{\sqrt{2\epsilon}}\RF^2 \Big\vert v_k\Big\vert^2\, \\ 
			& = \frac{H^2}{8\pi^2\epsilon} \LF 1+\frac{k^2}{a^2H^2} \RF
			\label{powersup}
		\end{aligned}
	\end{equation}
	where in the second line we substituted $\tau^2= \frac{1}{a^2H^2}$. For $k^2\ll a^2H^2$ the power spectrum just depends on background quantities $H,\, \epsilon$. Therefore, we can evaluate the power spectrum at a moment $k=aH$ by normalizing it with a factor of $\sqrt{2}$ for every mode that crosses the horizon \cite{Kinney:2009vz}. This means the observed power spectrum is $\Pc_\zeta\vert_{k=aH} = 2\times \Pc_\zeta\vert_{k\ll aH}$
	Thus the (near) scale-invariant power spectrum is just given by the background quantities as 
	\begin{equation}
		P_\zeta \approx \frac{H^2}{8\pi^2\epsilon} \,. 
		\label{pz}
	\end{equation}
	This power spectrum has a non-zero tilt because the background quantities $H,\, \epsilon$ are time-dependent and every time a mode exits the horizon $k= aH$ the value of the power spectrum is slightly changed by
	\begin{equation}
		n_s-1 = \frac{d\ln P_\zeta}{d\ln k}\approx -\frac{d\ln P_\zeta}{dN} \approx -2\epsilon-\eta\,. 
		\label{pzr}
	\end{equation}
	where $N$ is the number of e-foldings counted from the end of inflation. Combining Eq.~\ref{pz} and Eq.~\ref{pzr} we get the near-scale invariant power spectrum in the well-known power-law form 
	\begin{equation}
		P_\zeta \approx A_s \LF \frac{k}{k_\ast} \RF^{n_s-1}\,, 
		\label{HZpowe}
	\end{equation}
	where $k_\ast = 0.05\, {\rm Mpc}^{-1}$ at which $A_s= 2.2\times 10^{-9}$ from the latest Planck data \cite{Planck:2018jri}. The angular power spectrum of SI is computed following Eq.~\ref{eq:Cl_DQFT}. 
	
	A caveat in the calculation of power spectrum Eq.~\ref{HZpowe} is that slow-roll corrections in the MS equation Eq.~\ref{MSeq} are neglected by taking $\nu_s\approx \frac{3}{2}$. On the other hand, the full actual calculation including slow-roll corrections has to deal with applying approximations on the  Hankel functions in the super-horizon limit and eventually estimating the quantities at Horizon exit \cite{Kinney:2009vz,Powell:2006yg,vennin:tel-01094199}. The approximate mode functions Eq.~\ref{modeg} are devoid of information related to slow-roll parameters and thus we do not capture correctly the quantum nature of fluctuations subject to the inflationary background.\footnote{In fact, including accurately the slow-roll corrections to the mode functions in the SI is expected to give further enhancement to the angular power spectra at the low-multipoles. This would reduce even further the chances of SI matching with the data. Since this is not the primary goal of this paper, we leave it for future investigation.} In other words, with $\nu_s = \frac{3}{2}$, the quantum fluctuations Eq.~\ref{qunfluc} described by Eq.~\ref{modeg} are identical to those of de Sitter space. The power spectrum of curvature perturbation Eq.~\ref{HZpowe} that follows from the super-horizon limit of Eq.~\ref{powersup} is a consequence of classically rescaling back the canonical variable to curvature perturbation using inflationary background quantities $H,\,\epsilon$ which can be read from the first line of Eq.~\ref{powersup}. In the next section, we continue discussing further the fundamental questions associated with SI. 
	
	\subsection{The fundamental questions associated with inflationary quantum fluctuations}
	
	There are key fundamental questions on standard inflationary quantum fluctuations we discussed in the previous section. We discuss them briefly here following the investigations in \cite{Kumar:2022zff, Kumar:2023ctp}
	
	\subsubsection{Problem of time}
	
	The conundrum between gravity and quantum mechanics is the concept of time. According to GR \say{time} is a coordinate and it plays a dynamical role. Whereas in quantum mechanics \say{time} is a parameter which only characterizes whether a state is positive energy or not as per the Schr\"{o}dinger equation. In the case of inflationary quantum fluctuations, we have to deal with combining these two concepts. One of the prominent attempts to combine GR and QM is by the formulation of the Wheeler-DeWitt equation \cite{Kiefer:2007ria,Rovelli:2004tv} which gives us a surprise as the wave-function of the Universe only becomes a function of scale factor and the matter fields that satisfy a timeless equation 
	\begin{equation}
		\Hc_g \Psi\LF g_{ij},\, \phi \RF =0 \,, 
		\label{Wdeq}
	\end{equation}
	where $\Hc_g$ is the gravitational Hamiltonian and $\Psi$ is the wavefunction of the Universe \cite{Kiefer:2007ria} which is a function of spatial metric and the matter content which is assumed here to be the inflaton field $\phi$. The fact that there is no explicit time in Wheeler-de Wit equation Eq.~\ref{Wdeq} indicates that the merge of gravity and quantum mechanics requires us to think differently about time. Since only scale factor and matter content determine the Wave function of the Universe, it conveys that the arrow of time in cosmology is dictated by scale factor and dynamics of matter rather than the \say{time} coordinate that appears in the metric. It is also worth pointing out here that the stochastic inflationary framework which addresses the inflationary quantum fluctuations in a non-perturbative way also indicates the scale factor (or the Number of e-folds) should be considered as a \say{time} rather than the co-ordinate time that appears in the metric (See Appendix A of \cite{Vennin:2015hra}).

	\bibliographystyle{utphys.bst}
	\bibliography{Anoqft.bib}
	
\end{document}